\documentclass[preprint]{aastex62}

\usepackage{natbib}
\usepackage{amsmath}
\received{March 4, 2025}
\revised{April 20, 2025}
\accepted{May 1, 2025}
\submitjournal{\apj, March 3, 2025}

\shorttitle{Narrow Eccentric Ringlets}
\shortauthors{Hahn et al.}

\begin{document}

%title page
\title{N-body simulations of the Self--Confinement of 
Viscous Self--Gravitating Narrow Eccentric Planetary Ringlets}

\correspondingauthor{Joseph M. Hahn}
\email{jhahn@spacescience.org}

\author{Joseph M. Hahn}
\affiliation{Space Science Institute}

\author{Douglas P. Hamilton}
\affiliation{University of Maryland}

\author{Thomas Rimlinger}
\affiliation{University of Maryland}

%\author{Lucy Luu}
%\affiliation{University of Maryland}
\author[0000-0003-4769-3273]{Yuxi(Lucy) Lu}
\affiliation{Department of Astronomy, The Ohio State University, Columbus, 140 W 18th Ave, OH 43210, USA}
\affiliation{Center for Cosmology and Astroparticle Physics (CCAPP), 
The Ohio State University, 191 W. Woodruff Ave., Columbus, OH 43210, USA}
\affiliation{American Museum of Natural History, Central Park West, Manhattan, NY, USA}

%abstract
\begin{abstract}

Narrow eccentric planetary ringlets have sharp edges, sizable eccentricity gradients, and a 
confinement mechanism that prevents radial spreading due to ring viscosity. 
Most proposed ringlet confinement mechanisms presume that there are one or more 
shepherd satellites whose gravitational perturbations keeps the ringlet confined radially, 
but the absence of such shepherds in Cassini observations of Saturn's rings
casts doubt upon those ringlet confinement mechanisms.
The following uses a suite of N-body simulations to explore an alternate scenario,
whereby ringlet self-gravity drives a narrow eccentric ringlet into a 
self--confining state. These simulations show that, under a wide variety of initial conditions,
an eccentric ringlet's secular perturbations of itself causes the eccentricity of its outer
edge to grow at the expense of its inner edge. This causes the
ringlet's nonlinearity parameter $q$
to grow over time until it exceeds the $q\simeq\sqrt{3}/2$ threshold where 
the ringlet's orbit-averaged angular momentum flux due to viscosity + self-gravity is zero.
The absence of any net radial angular momentum transfer through the ringlet 
means that the ringlet has settled into a self-confining state, {\it i.e.} it does 
not spread radially due to its viscosity, and simulations also show that such ringlets have sharp edges.
Nonetheless, viscosity still circularizes the ringlet in time $\tau_e\sim10^6$ orbits $\sim1000$ years, 
which will cause the ringlet's nonlinearity parameter to shrink below the 
$q\simeq\sqrt{3}/2$ threshold and allows radial spreading to resume. Either sharp-edged
narrow eccentric ringlets are transient phenomena, or exterior perturbations 
are also sustaining the ringlet's eccentricity.
We then speculate about how such ringlets might come to be.

\end{abstract}

%keywords

\section{Introduction}
\label{sec:intro}

Narrow eccentric planetary ringlets have properties both interesting and
not well understood: sharp edges,
sizable eccentricity gradients, and a confinement mechanism that
opposes radial spreading due to ring viscosity. To date, nearly all of the
prevailing ringlet confinement mechanisms assume that there also exists a pair of 
unseen shepherd satellites that straddle the ringlet, with those shepherds' gravities
also torquing the ringlet's edges' in a way that keeps them radially confined 
(\citealt{GT79, GT79c, GT81, CG00, ME02}), with \cite{Getal95} showing
that a single shepherd satellite can provide temporary confinement.
However the local gravitational perturbations exerted by 
the shepherds on the nearby ring material also excites prominent kinks (\citealt{Metal05}) and 
scalloping (\citealt{Wetal09}) of the ring edges,
which should invite detection of the hypothetical satellites.
That the Cassini spacecraft did not detect any shepherds near Saturn's
well-studied narrow ringlets casts doubt upon this ringlet confinement 
mechanism \citep{L18}. 

Note though that \cite{BGT82} showed that a viscous ringlet having a
a sufficiently high eccentricity gradient can in fact be self-confining,
due to a reversal of its viscous angular momentum flux, which in turn would
cause the ringlet to get narrower over time. 
That suggestion also motivates this study,
which uses the epi\_int\_lite N-body integrator to investigate whether a 
viscous and self--gravitating ringlet might evolve into a self-confining state.

We also note that Uranus hosts several narrow eccentric ringlets, but Section \ref{subsec:variations}
uses results from a recent study of the Uranian ringlets (\citealt{Fetal24})
to argue that these are not candidates for the self-confinement
mechanism that is considered here; see Section \ref{subsec:variations} for details.

\section{epi\_int\_lite}
\label{sec:epi_int_lite}

Epi\_int\_lite is a child of the epi\_int N-body integrator that was used to
simulate the outer edge of Saturn's B ring while it is sculpted by satellite perturbations
\citep{HS13}. The new code is very similar to its parent but differs in two significant ways:
({\it i}.) epi\_int\_lite is written in python and is recoded for more efficient execution, and
({\it ii}.) epi\_int\_lite uses a more reliable drift step to handle
unperturbed motion around an oblate planet (detailed in Appendix \ref{sec:Appendix A}).
Otherwise epi\_int\_lite's treatment of ring self--gravity and viscosity are identical
to that used by the parent code, and see \cite{HS13} for additional details. The epi\_int\_lite 
source code is available at https://github.com/joehahn/epi\_int\_lite, and the
code's numerical quality is benchmarked in Appendix \ref{sec:Appendix B}
where the output of several numerical experiments are compared against theoretical expectations.

Calculations by epi\_int\_lite use natural units with gravitation constant $G=1$, 
central primary mass $M=1$, and the ringlet's inner edge has initial radius
$r_0=1$, and so the ringlet masses $m_r$ and radii $r$ quoted below are in units of $M$ and $r_0$.
Converting code output from natural units to physical units requires choosing	
physical values for $M$ and $r_0$ and multiplying accordingly, and when this text does so
it assumes Saturn's mass $M=5.68\times10^{29}$ gm and a characteristic
ring radius $r_0=1.0\times10^{10}$ cm. Simulation time $t$ is in units of $T_{\text{orb}}/2\pi$
where $T_{\text{orb}} = 2\pi\sqrt{r_0^3/GM}$ is the orbit period at $r_0$, 
so multiply time $t$ by $T_{\text{orb}}/2\pi$ 
to convert from natural to physical units.
The simulated particles' motions during the drift step are also
sensitive to the $J_2$ portion of the primary's non-spherical gravity component 
(see Appendix \ref{sec:Appendix A}), and all simulations
adopt Saturn-like values of $J_2=0.01$ and $R_p=r_0/2$ where $R_p$ is the planet's
mean radius.

\subsection{streamlines}
\label{subsec:streamlines}

Initially all particles are assigned to various streamlines across the simulated ringlet. A streamline
is a closed eccentric path around the primary, and each streamline is populated by $N_p$ particles that are
initially assigned a common semimajor axis $a$ and eccentricity $e$ while
distributed uniformly in longitude. Most of the simulations described below
employ only $N_s=2$ streamlines, so that the model output can be compared against
theoretical treatments that also treat the ringlet as two gravitating streamlines
(e.g.\ \citealt{BGT83}). But the following also performs a few higher-resolution simulations
using $N_s=3-14$ streamlines, to demonstrate that the $N_s=2$ treatment is perfectly
adequate and reproduces all the relevant dynamics. All simulations use $N_p=241$ particles 
per streamline, and the total number of particles is $N_sN_p$.
Note that the assignment of particles to a given streamline is merely
for labeling purposes, as particles are still free to wander in response
to the ring’s internal forces, namely, ring gravity and viscosity. But as \cite{HS13} as well
as this work shows, the simulated ring stays coherent and highly organized throughout the 
simulation such that particles on the same streamline do not pass each other longitudinally,
nor do they cross adjacent streamlines. Because the simulated ringlet stays highly organized,
there is no radial or longitudinal mixing of the ring particles, and simulated particles preserve
memory of their streamline membership over time. 

The epi\_int\_lite code also monitors all particles and checks whether any have crossed adjacent streamlines.
If that happens the simulation is then terminated since the particles' subsequent evolution
would no longer be computed reliably.

\subsection{N-body method}
\label{subsec:N-body method}

The epi\_int\_lite N-body integrator uses the same second-order sympletic drift-kick
scheme used by the MERCURY Nbody algorithm \citep{C99}, except that
epi\_int\_lite particles that do not interact with each other directly.
Rather, epi\_int\_lite particles
are only perturbed by the accelerations exerted by the ringlet's individual streamlines. 
Those accelerations are sensitive to the streamline's relative separations and orientations, 
which are inferred from the particles' positions
and velocities. Epi\_int\_lite particles are thus tracer particles
that indicate the streamlines' locations and orientations, which the N-body
integrator uses to compute the orbital evolution of those tracer particles
due to the perturbations exerted by those streamlines. This streamline approach 
is widely used in theoretical studies of planetary rings (c.f.\ \citealt{GT79, BGT83, BGT85})
as well as in N-body studies of rings \citep{HS13, RHH16}. The great benefit of the streamline concept
in numerical work is that it allows one to swiftly track the 
global evolution of the ringlet's streamlines numerically
using only a modest numbers of tracer particles, typically $N_sN_p\sim500$.

The simulations reported on here account for streamline gravity
and ringlet viscosity. Because a ringlet is narrow, all particles
are in close proximity to the nearby portions of all streamlines,
which allows us to approximate a streamline as an infinitely
long wire of matter having linear density $\lambda$.
Consequently the gravity of each perturbing streamline draws a particle
towards that streamline with acceleration
\begin{equation}
\label{eqn:gravity}
    A_g=\frac{2G\lambda}{\Delta },
\end{equation}
where $\Delta$ is the particle's distance from the streamline.

The streamline's linear density $\lambda$ is inferred from a streamline's total
mass, $m_1=m_r/N_s=\int_0^{2\pi}\lambda d\ell$
where the integration is about the streamline's circumference. Replacing
$d\ell$ with $vdt$ where $v$ is the velocity of the streamline's particles
allows replacing the spatial integral with a timewise integral,
$m_1=\int_0^{T}\lambda vdt = \lambda v T$ where the streamline's
orbital period $T=2\pi/\Omega$ and $\Omega$ its mean orbital frequency, hence
\begin{equation}
\label{eqn:lambda}
    \lambda=\frac{m_1\Omega}{2\pi v}.
\end{equation}

The hydrodynamic approximation is used here to account for
the dissipation that occurs as particles in adjacent particle streamlines
shear past and collide with the perturbed particle,
without having to monitor individual particle-particle collisions.
The particle's Eulerian acceleration due to the ring particles' shear viscosity is
\begin{equation}
\label{eqn:shear_viscosity}
    A_{\nu,\parallel}=-\frac{1}{\sigma r}\frac{\partial {\cal F}_L}{\partial r},
\end{equation}
where $r$ is the particle's radial coordinate, $\sigma$ is the surface density
\begin{equation}
    \label{eqn:sigma}
    \sigma = \frac{\lambda}{\Delta r}
\end{equation}
where $\Delta r$ is the radial separation between the particle
and the perturbing streamline (and see Section 2.3.3 of \cite{HS13} for details
about how that is evaluated numerically),
and ${\cal F}_{L,\nu}$ is the flux of angular momentum
that is transported radially across the particle's streamline
due to its collisions with particles in adjacent streamlines, {\it i.e.}\
\begin{equation}
    \label{eqn:F_nu_theta}
    {\cal F}_{L,\nu} = -\nu_s \sigma r^2\frac{\partial\omega}{\partial r}
\end{equation}
where $\nu_s$ is the ringlet's kinematic shear viscosity
and $\omega=v_\theta/r$ is the particle's angular velocity (Appendix A of \citealt{HS13}).
The acceleration $A_{\nu,\parallel }$ is parallel to the perturbed particle's streamline
{\it i.e.}\ parallel to particle's velocity vector
$\mathbf{v}=\dot{\mathbf{r}}=v_r\hat{r} + v_\theta\hat{\theta}$ 
where $\mathbf{r} = r\hat{\mathbf{r}}$ is the particle's position vector.

Dissipative collisions also transmits linear momentum in the perpendicular direction,
which results in the additional acceleration
\begin{equation}
\label{eqn:bulk_viscosity}
    A_{\nu, \perp}=-\frac{1}{\sigma}\frac{\partial {\cal G}}{\partial r},
\end{equation}
where the radial flux of linear momentum due to ringlet viscosity is
\begin{equation}
    \label{eqn:G}
    {\cal G} = -\left(\frac{4}{3}\nu_s + \nu_b\right)\sigma\frac{\partial v_r}{\partial r}
        -\left(\nu_b - \frac{2}{3}\nu_s\right)\frac{\sigma v_r}{r},
\end{equation}
$\nu_b$ is the ringlet's kinematic bulk viscosity, and $v_r$ is the paricle's radial velocity
\citep{HS13}.

In the hydrodynamic approximation there is also the acceleration due to ringlet pressure $p$
that is due to particle-particle collisions,
\begin{equation}
\label{eqn:pressure}
    A_p=-\frac{1}{\sigma }\frac{\partial p}{\partial r}.
\end{equation}
Epi\_int\_lite treats the particle ring as a dilute gas of
colliding particles for which the 1D pressure is $p=c^2\sigma$ where
c is the particles dispersion velocity. However \cite{HS13} found ring pressure
to be inconsequential in N-body simulations of Saturn's A ring, and the ringlet simulation
examined in great detail in Section \ref{subsec:nominal} 
also showed no sensitivity to pressure effects, so all other
simulations reported on here have $c=0$.

\section{N-body simulations of viscous gravitating ringlets}
\label{sec:nbody}

The folowing describes a suite of N-body simulations of narrow viscous gravitating planetary
ringlets, to highlight the range of initial ringlet conditions that do evolve into
a self-confining state, and those that do not.

\subsection{nominal model}
\label{subsec:nominal}

\begin{figure}
\plotone{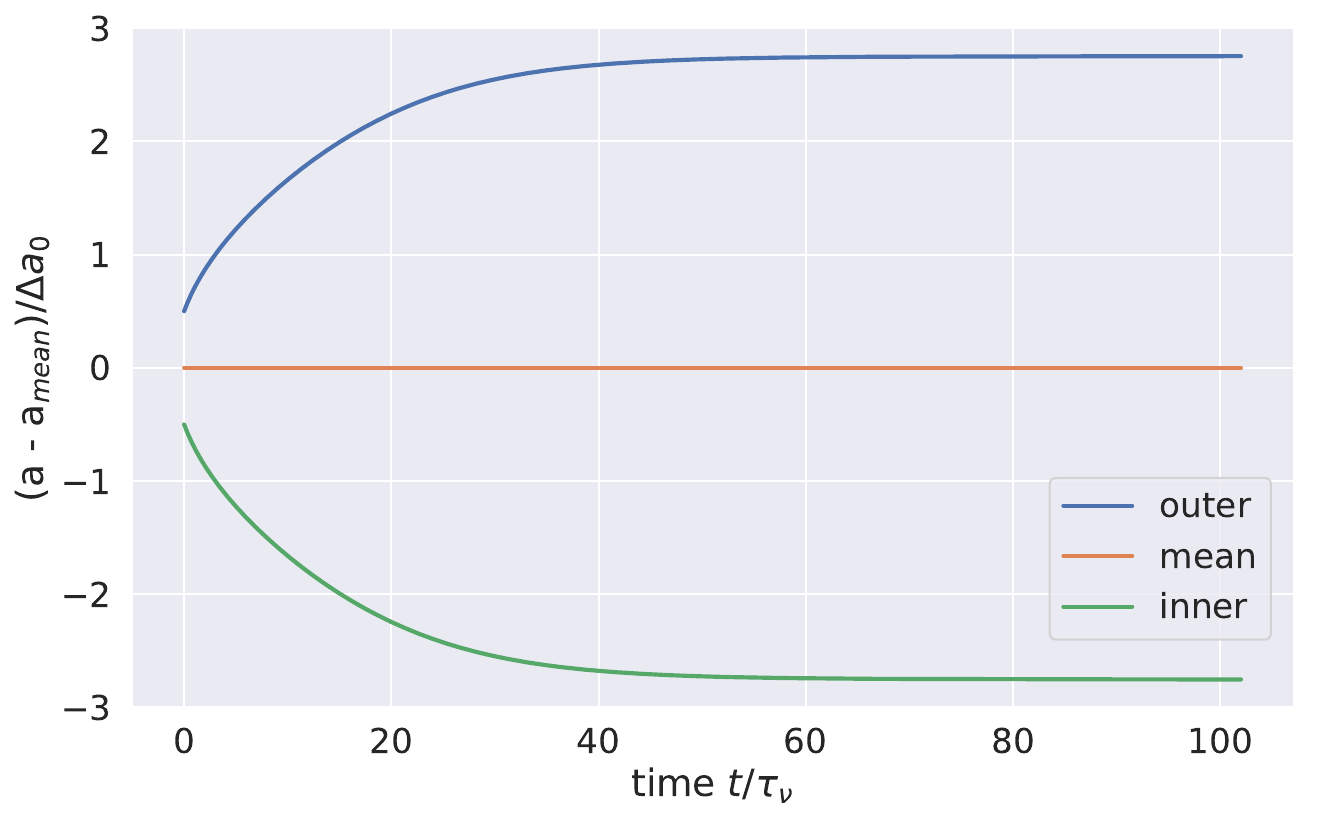}
\caption{
\label{fig:a_nominal}
Evolution of the nominal ringlet's semimajor axes $a$
versus time $t$ in units of the ringlet's viscous timescale
$\tau_\nu=1.3\times10^3$ orbital periods. This ringlet is composed of $N_s=2$ streamlines,
and the outer (blue) and inner (green) streamlines' semimajor axes are plotted relative
to their mean $a_{\text{mean}}$, and displayed in units of the ringlet's
initial width $\Delta a_0 = 10^{-4}$ in natural units ({\it i.e.}\ $G=M=r_0=1$).
The simulated ringlet has total mass $m_r=10^{-10}$, shear viscosity $\nu_s=10^{-13}$,
and initial eccentricity $e=0.01$. See Section \ref{subsec:nominal} to convert
$m_r$, $a$ and $\nu_s$ from natural units to physical units.}
\end{figure}

\begin{figure}
\plotone{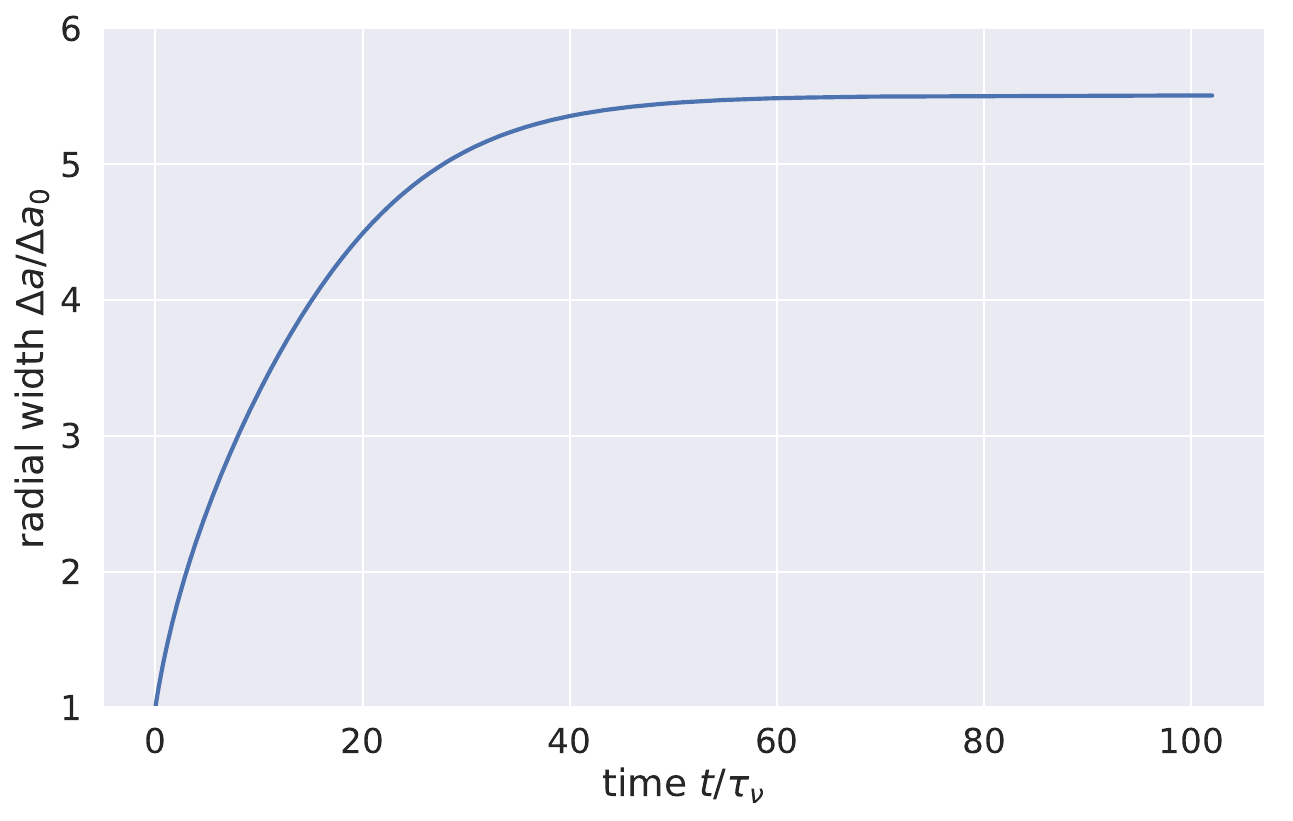}
\caption{
\label{fig:da_nominal}
The nominal ringlet's semimajor axis width $\Delta a = a_{\text{outer}} - a_{\text{inner}}$ over time
and in units of its initial radial width $\Delta a_0$.}
\end{figure}

Figure \ref{fig:a_nominal} shows the semimajor axis evolution of what is referred to
as the nominal model since this ringlet readily evolves into a self-confining state.
The simulated ringlet is composed of $N_s=2$ streamlines having $N_p=241$ particles
per streamline, and the integrator timestep is $\Delta t=0.5$ in natural units, so
the integrator samples the particles' orbits $2\pi/\Delta t\simeq13$ times per orbit, and this
ringlet is evolved for $1.4\times10^5$ orbits, which requires 50 minutes execution time
on a ten year old laptop. The ringlet's mass is
$m_r=10^{-10}$, its shear viscosity is $\nu_s=10^{-13}$, and its
bulk viscosity is $\nu_b=\nu_s$. The ringet's initial radial width is
$\Delta a_0 = 10^{-4}$, its initial eccentricity is $e=0.01$, and its
eccentricity gradient is initially zero. A convenient measure of time is the ringlet's
viscous radial spreading timescale
\begin{equation}
\label{eqn:viscous-timesscale}
    \tau_\nu=\frac{\Delta a_0^2}{12\nu_s} 
\end{equation}
which can be inferred from Eqn.\ (2.13) of \cite{P81}. 
This simulation's viscous timescale is $\tau_\nu=8.3\times10^3$ in natural units
or $\tau_\nu/2\pi=1.3\times10^3$ orbital periods. If this ringlet were orbiting Saturn
at $r_0=1.0\times10^{10}$ cm then the simulated ringlet's physical mass
would be $m_r=5.7\times10^{19}$ gm which is equivalent to the mass of a $24$ km radius iceball assuming
a volume density $\rho=1$ gm/cm$^3$, and the ringlet's initial physical radial width would be
$\Delta a_0 = 10^{-4}r_0=10$ km. This ringlet's
orbit period would be $T_{orb}=2\pi\sqrt{r_0^3/GM}=9.0$ hours in physical units, so 
the ringlet's viscous timescale is $\tau_\nu=1.4$ years, and
so its shear viscosity is $\nu_s=\Delta a_0^2/12\tau_\nu = 1.9\times10^3$ cm$^2$/sec
when evaluated in physical units. 
This ringlet's initial surface density would be $\sigma=m_r/2\pi r_0\Delta a_0=900$ gm/cm$^2$, but
Figs.\ \ref{fig:a_nominal}--\ref{fig:da_nominal} show that shrinks by a factor of about 5 as the 
ringlet's sememajor axis width $\Delta a$ grows via viscous spreading until it settles into
the self-confining state at time $t\sim40\tau_\nu$.
This so-called nominal ringlet is probably somewhat overdense and overly viscous compared to known 
planetary ringlets,
but that is by design so that the simulated ringlet quickly settles into the self-confining state.
Section \ref{subsec:variations} also shows how outcomes vary 
when a wide variety of alternate initial masses, widths, and viscosities are also considered.

\begin{figure}
    \plotone{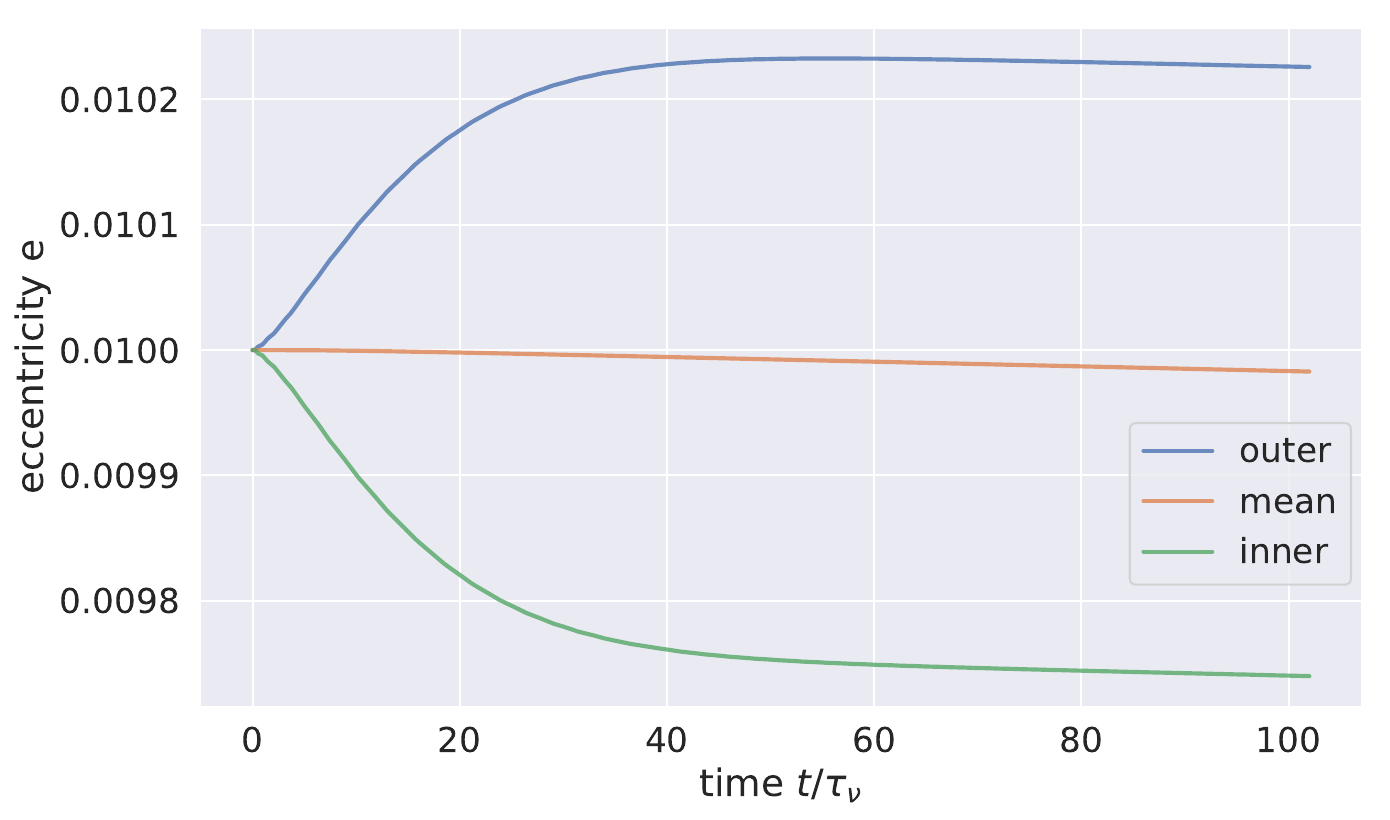}
    \caption{
        \label{fig:e_nominal}
        The nominal ringlet's eccentricity evolution.
    }
\end{figure}

Figure \ref{fig:e_nominal} shows that the outer streamline's eccentricity initially grows at the
expense of the inner streamline's, and that is a consequence the self-gravitating ringlet's
secular perturbations of itself, which is also demonstrated in Appendix \ref{sec:Appendix C}. 
Figure \ref{fig:de_nominal} shows
the ringlet's eccentricity difference $\Delta e = e_{\text{outer}} - e_{\text{inner}}$
and longitude of periapse difference
$\Delta\tilde{\omega} = \tilde{\omega}_{\text{outer}} - \tilde{\omega}_{\text{inner}}$,
which both settle into equilibrium values after the ringlet arrives at the self-confining
state. 

In all self-confing ringlet simulations examined here, the ringlet's periapse twist 
$\Delta\tilde{\omega} = \tilde{\omega}_{\text{outer}} - \tilde{\omega}_{\text{inner}}$ is negative,
so the outer streamline's longitude of periapse $\tilde{\omega}$ trails
the inner streamline's, which in turn causes the streamlines' separations along
the ringlet's pre-periapse side (where $\varphi = \theta - \tilde{\omega} < 0$) 
to differ slightly from the post-periapse ($\varphi>0$) side.
Which in turn makes the ringlet's surface density asymmetric {\it i.e.}\ is
maximal just prior to periapse, see
Figs.\ \ref{fig:nominal_sigma_vs_longitude}--\ref{fig:radial_sigma_nominal}.

\begin{figure}
    \plotone{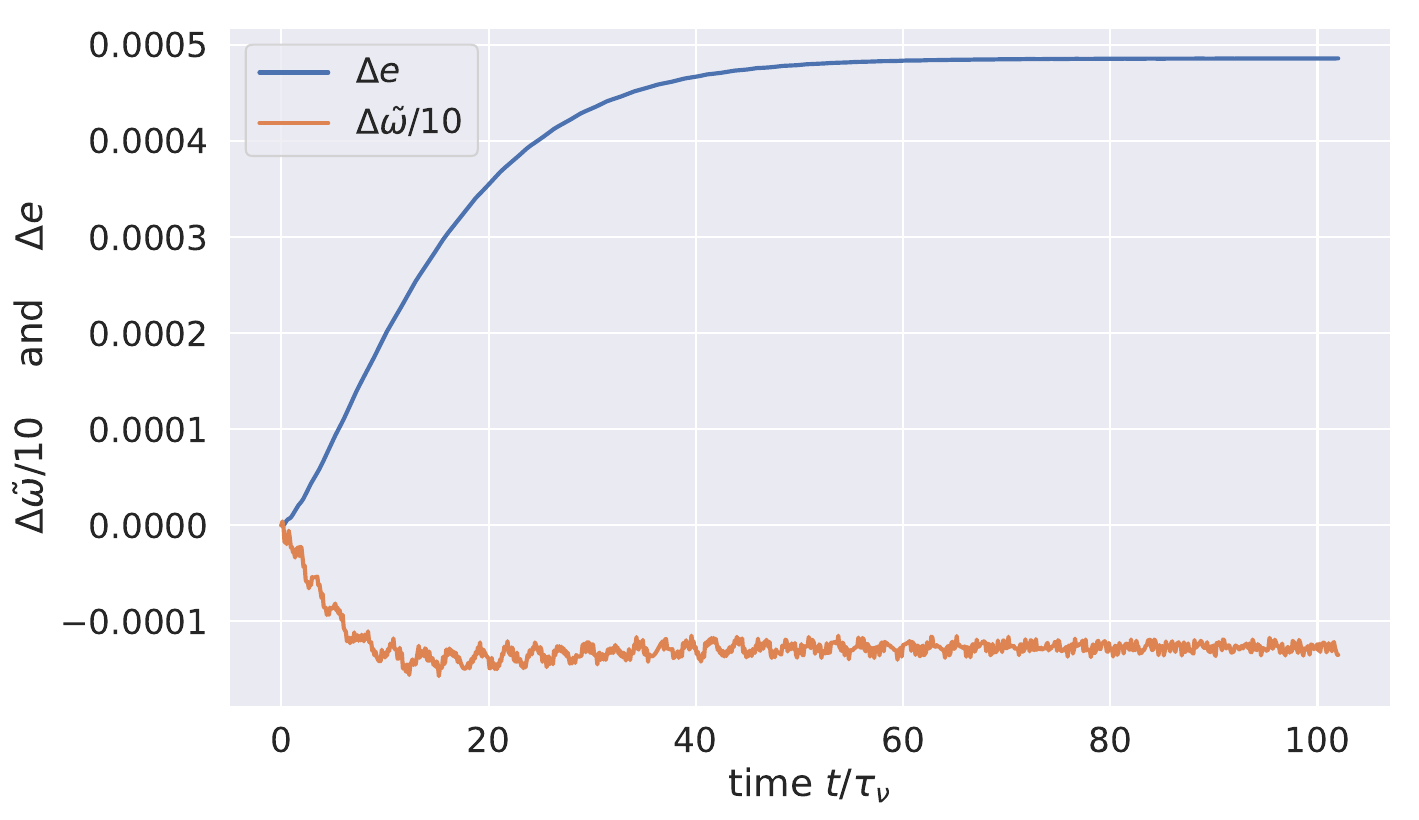}
    \caption{
        \label{fig:de_nominal}
        The nominal ringlet's eccentricity difference $\Delta e = e_{\text{outer}} - e_{\text{inner}}$
        and longitude of periapse difference
        $\Delta\tilde{\omega} = \tilde{\omega}_{\text{outer}} - \tilde{\omega}_{\text{inner}}$
        in radians divided by 10.
    }
\end{figure}

\begin{figure}
    \plotone{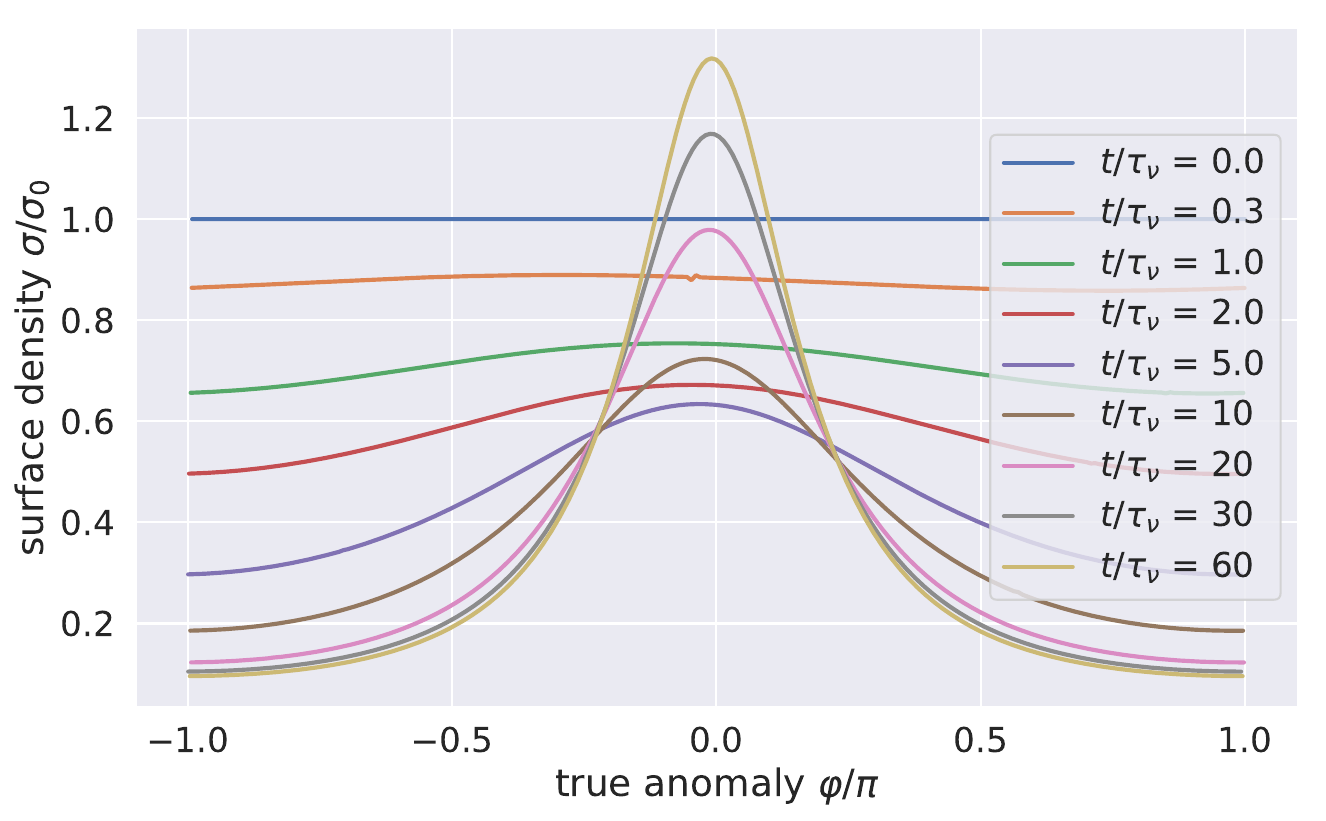}
    \caption{
        \label{fig:nominal_sigma_vs_longitude}
        Nominal ringlet's surface density $\sigma(\varphi)$ versus true anomaly 
        $\varphi$ at selected times $t$ and plotted in units of ringlet's initial
        mean surface density $\sigma_0$. Note that the ringlet's surface
        density maxima occurs just before peripase,  and is due to the ringlet's
        negative periapse twist 
        $\Delta\tilde{\omega} = \tilde{\omega}_{\text{outer}} - \tilde{\omega}_{\text{inner}} < 0$.
    }
\end{figure}

\begin{figure}
    \plotone{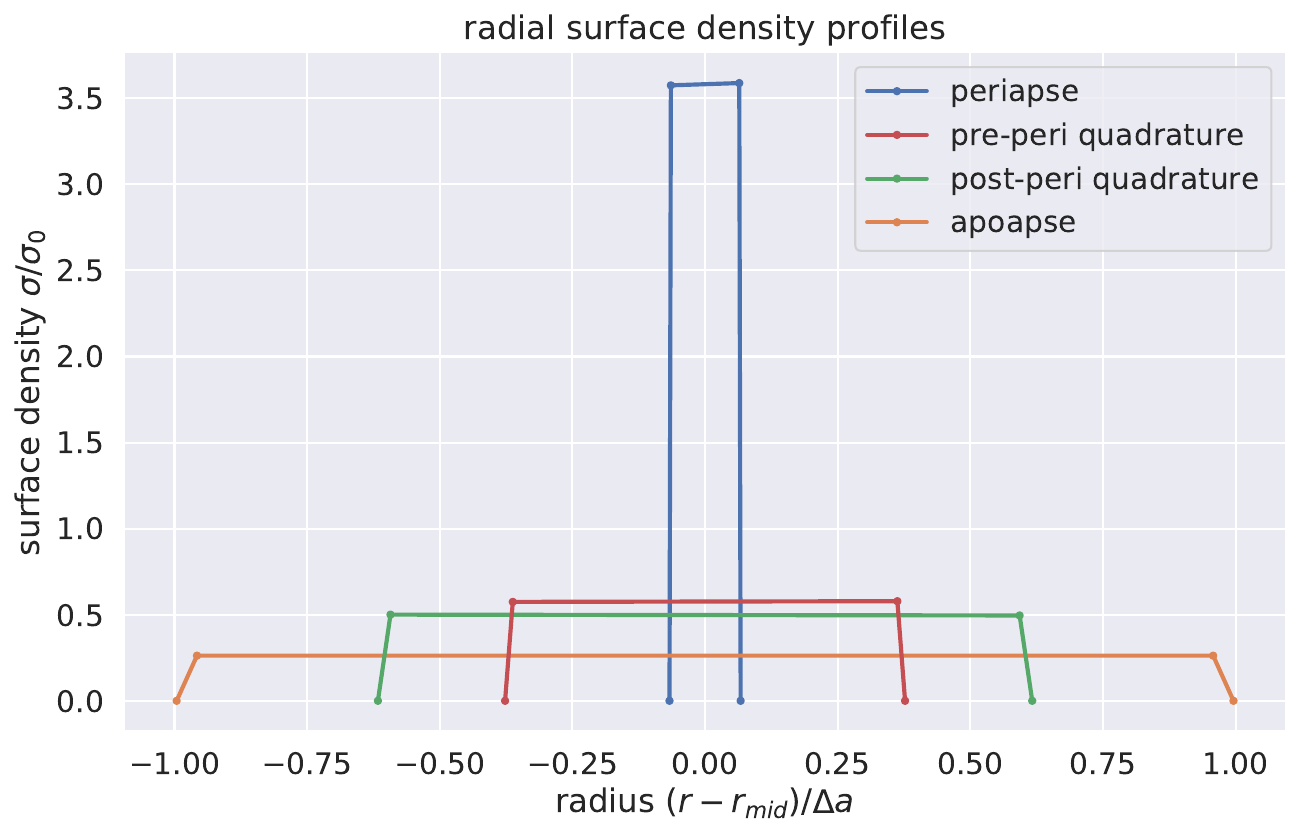}
    \caption{
        \label{fig:radial_sigma_nominal}
        Radial profiles of the nominal ringlet's surface density $\sigma(\varphi)$ at time $t=100\tau_\nu$
        when the ringlet is self-confining. Each surface density profile is plotted versus radial distance $r$ 
        relative to $r_{mid}$, which is the ringlet's midpoint along true anomaly $\varphi = \theta-\tilde{\omega}$,
        with those radial distances $r - r_{mid}$ measured in units of the ringlet's final semimajor axis width $\Delta a$,
        and surface density is shown in units of the ringlet's longitudinally-averaged surface density $\sigma_0$.
        Radial surface density profiles are plotted along the ringlet's periapse ($\varphi=0$, blue curve), which is 
        where the ringlet's streamlines are most concentrated and surface denisity $\sigma$ is
        greatest due to the ringlet's eccentricity gradient $e'$, at the pre-periapse
        quadrature ($\varphi=-\pi/2$, red curve), post-periapse
        quadrature ($\varphi=\pi/2$, green curve)
        and at apoapse ($|\varphi|=\pi$, orange curve) where streamlines have their greatest separation
        and ringlet surface density is lowest. This ringlet's surface density contrast between periapse and
        apoapse is about 14.
    }
\end{figure}

\begin{figure}
    \plotone{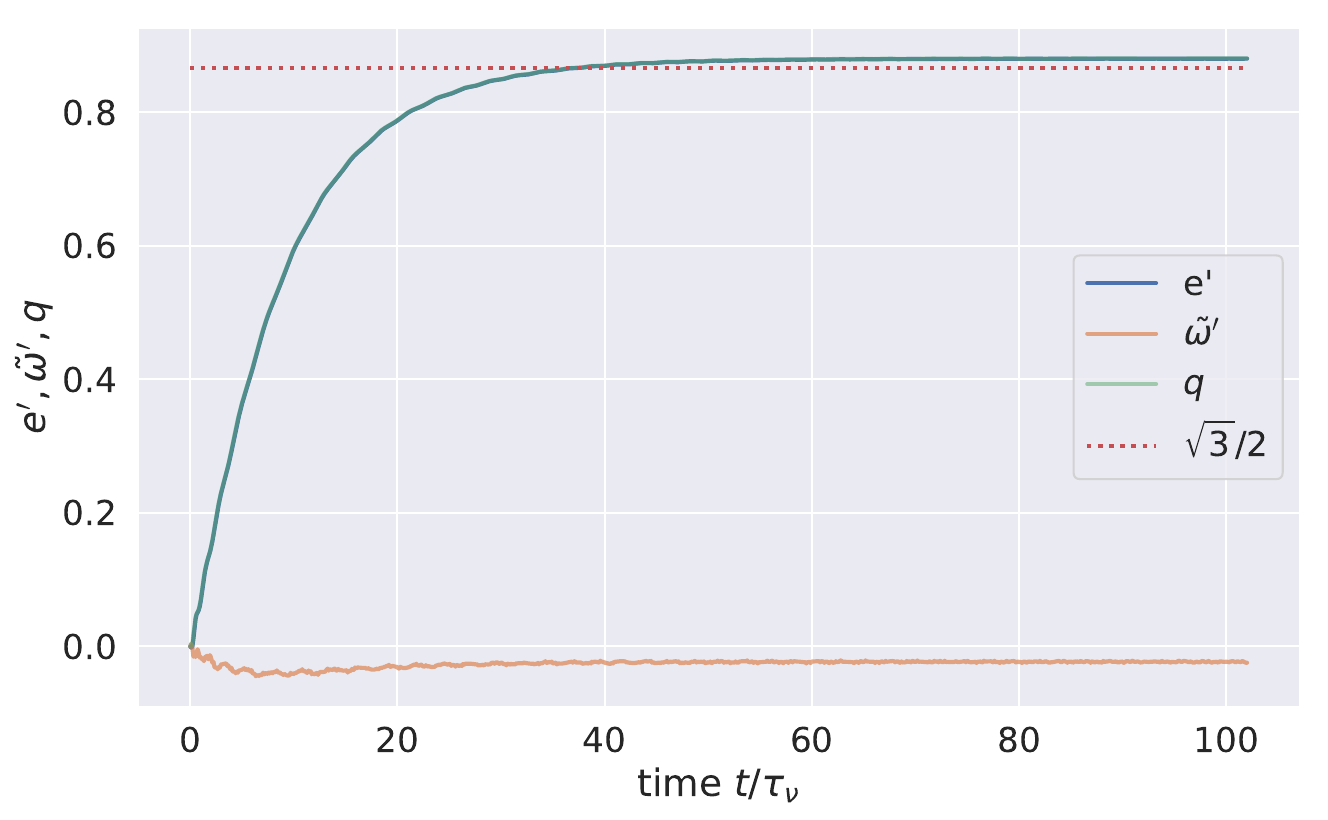}
    \caption{
        \label{fig:de_prime_nominal}
        The nominal ringlet's dimensionless eccentricity gradient $e' = a\Delta e/\Delta a$
        (blue curve), dimensionless periapse twist $\tilde{\omega}' = ea\Delta\tilde{\omega}/\Delta a$
        (orange), and nonlinearity parameter $q=\sqrt{e'^2 + \tilde{\omega}'^2}$
        (green curve which overlaps blue) versus time $t/\tau_\nu$. Dotted red line 
        is the threshold for self-confinement in a non-gravitating ringlet, $e'=\sqrt{3}/2\simeq0.866$
    }
\end{figure}

It is convenient to recast these orbit element differences as dimensionless gradients
\begin{equation}
    \label{eqn:e_prime}
    e' = a\frac{de}{da}
    \qquad\mbox{and}\qquad
    \tilde{\omega}' = ea\frac{d\tilde{\omega}}{da}
\end{equation}
as these are the terms that contribute to the nonlinearity parameter of \cite{BGT83}:
\begin{equation}
    \label{eqn:q}
    q = \sqrt{e'^2 + \tilde{\omega}'^2}.
\end{equation}
See also Fig.\ \ref{fig:de_prime_nominal} which plot's the nominal
ringlet's dimensionless eccentricity gradient $e'$, dimensionless periapse twist $\tilde{\omega}'$,
and nonlinearity parameter $q$ versus time. All simulated self-confining ringlets
have a positive eccenticity gradient and a negative periapse twist such that
the outer ringlet's periapse trails the inner ringlet's, consistent with the findings of
\cite{BGT83}.\vfil

\section{angular momentum and energy fluxes \& luminosities}
\label{sec:fluxes}

The nominal ringlet's evolution is readily understood when the ringlet's 
radial flux of angular momentum and energy are considered. 

\subsection{angular momentum and energy fluxes}
\label{subsec:fluxes}

The torque that is exerted on a small streamline segment of mass $\delta m$
at location $\mathbf{r}=r\hat{\mathbf{r}}$
due to the streamlines orbiting interior to it is $\delta T=\delta m\mathbf{r}\times\mathbf{A}^1$ where 
$\mathbf{A}^1=A^1_r\hat{\mathbf{r}} + A^1_\theta\hat{\boldsymbol\theta}$
is the so-called one-sided acceleration that is exerted on $\delta m$ by all other streamlines
interior to it. Since $\delta m=\lambda\delta\ell$ where $\lambda$ is the streamline's linear mass density,
and $\delta\ell$ is the segment's length, then the radial flux of angular momentum flowing into
that segment due to the accelerations that are exerted by streamlines interior to that
segment is
\begin{equation}
    \label{eqn:F_L}
    {\cal F}_L(r, \theta) = \frac{\delta T}{\delta\ell} = \lambda r A^1_\theta,
\end{equation}
where $A^1_\theta$ is the tangential component of the one-sided acceleration,
and the streamline's linear mass density $\lambda$ is computed using Eqn.\ (\ref{eqn:lambda}).
The radial angular momentum flux, Eqn.\ (\ref{eqn:F_L}), is due to the ringlet's viscosity and self-gravity,
so ${\cal F}_L = {\cal F}_{L,\nu} + {\cal F}_{L,g}$ where
\begin{align}
    {\cal F}_{L,\nu} = \lambda r A^1_{\nu,\theta} \label{eqn:viscous_ang_mom_flux}\\
    \mbox{and} \quad {\cal F}_{L,g} = \lambda r A^1_{g,\theta}     \label{eqn:grav_ang_mom_flux}
\end{align}
are the viscous and gravitational angular momentum fluxes at a particle.
The particle's one-sided gravitational acceleration $A^1_{g,\theta}$ is straightforward to compute,
it is merely the tangential component of the gravitational accelerations exerted by
all streamlines orbiting interior to the particle, while the viscous angular momentum
flux is Eqn.\ (\ref{eqn:F_nu_theta}) derived in \cite{HS13}.

The work that the interior streamlines exert on $\delta m$ as that segment travels a small distance
$\delta\mathbf{r}=\mathbf{v}\delta t$ in time $\delta t$
is $\delta W=\delta m\mathbf{A}^1\cdot\delta\mathbf{r}$ where 
$\mathbf{v}=v_r\hat{\mathbf{r}} + v_\theta\hat{\boldsymbol\theta}$ is the segment's velocity, and
that work accrues at $\delta m$ at the rate 
$\delta W/\delta t=\lambda\mathbf{A}^1\cdot\mathbf{v}\delta\ell$,
so the radial flux of energy entering that ringlet segment due to accelerations
exerted by the interior streamlines is 
\begin{equation}
    \label{eqn:F_E}
    {\cal F}_E(r, \theta) = \frac{\delta W}{\delta\ell\delta t} = \lambda\mathbf{A}^1\cdot\mathbf{v}.
\end{equation}
The radial energy flux is due to the ringlet's viscosity and self-gravity, so
${\cal F}_E = {\cal F}_{E,\nu} + {\cal F}_{E,g}$. 

\subsection{luminosities}
\label{subsec:luminosities}

The streamline containing segment $\delta m$ has semimajor axis $a$, and 
integrating the radial angular momentum flux ${\cal F}_L$ about the entire streamline
then yields the radial luminosity of angular momentum entering streamline $a$,
\begin{equation}
    \label{eqn:L_L}
    {\cal L}_L(a) = \oint {\cal F}_Ld\ell , 
\end{equation}
which is the torque that is exerted on streamline $a$ by those orbiting interior to it. Similarly,
integrating the radial energy flux ${\cal F}_E$ about streamline $a$ also yields the ringlet's radial energy luminosity
\begin{equation}
    \label{eqn:L_E}
    {\cal L}_E(a) = \oint {\cal F}_Ed\ell,
\end{equation}
which is the rate that the interior streamlines communicate energy to streamline $a$.

\subsection{viscous transport of angular momentum}
\label{subsec:viscous_flux}

Angular momentum is transported radially through the ring via viscosity and self-gravity, 
so ${\cal F}_L={\cal F}_{L,\nu} + {\cal F}_{L,g}$,
where the ringlet's viscous flux of angular momentum is
\begin{equation}
    \label{eqn:F_nu_1}
    {\cal F}_{L,\nu}(r, \theta) = -\nu_s \sigma r^2\frac{\partial\omega}{\partial r}
\end{equation}
when Eqn.\ (\ref{eqn:F_nu_theta}) is written as a function of spatial coordinates and
angular velocity $\omega=\dot{\theta}$. 
The ring's surface density $\sigma$ is Eqn.\ (\ref{eqn:sigma}) where $\Delta r\simeq\Delta a(1-e'\cos\varphi)$
when it is assumed that the ringlet's eccentricity $e$ is small but its
eccentricity gradient $e'=a\partial e/\partial a$ might not be, so 
\begin{equation}
    \label{eqn:sigma2}
    \sigma \simeq \frac{\sigma_0}{1-e'\cos\varphi}
\end{equation}
where $\sigma_0 = \lambda/\Delta a$ would be the ringlet's initial surface density assuming
its initial $e'$ was zero and $\varphi=\theta-\tilde{\omega}$ is the true anomaly 
{\it i.e.}\ the longitude relative to periapse.
Now consider a small arc of  ring material of length $d\ell$, 
so ${\cal F}_{L,\nu}d\ell$ is the torque that arc exert exerts on
ring matter just exterior due to viscous friction,
and is the rate that friction transmits angular momentum radially across that arc.
And when ${\cal F}_{L,\nu}$ is evaluated along a single eccentric streamline of semimajor axis $a$, 
the above simplifies to
\begin{equation}
    \label{eqn:F_nu_varphi}
    {\cal F}_{L,\nu}(a, \varphi) = {\cal F}_{L,\nu,c}\frac{1-\frac{4}{3}e'\cos\varphi}{(1-e'\cos\varphi)^2}
\end{equation}
where the angular shear $\omega'=\partial\omega/\partial r$ in Eqn.\ (\ref{eqn:F_nu_1})
is derived in Appendix \ref{sec:Appendix A}, Eqn.\ (\ref{eqn:domega-dr}), and
${\cal F}_{L,\nu,c}=\frac{3}{2}\nu_s\sigma_0a\Omega$ is the viscous angular momentum flux through a
circular streamline of semimajor axis $a$ that has angular speed $\Omega(a)$.
Note that Eqn.\ (\ref{eqn:F_nu_varphi}) is equivalent to Eqn.\ (18) of \cite{BGT82}
provided $|\tilde{\omega}'|\ll e'$ such that $q\simeq e'$. 

Integrating the above around the streamline's circumference then yields its angular momentumum luminosity,
\begin{equation}
    \label{eqn:L_nu}
    {\cal L}_{L,\nu}(a) = \oint {\cal F}_{L,\nu}(a, \varphi)rd\varphi = {\cal L}_{L,\nu,c}\frac{1-\frac{4}{3}e'^2}{(1-e'^2)^{3/2}},
\end{equation}
which is the torque that one streamline exerts on its exterior neighbor due to viscous friction
(\citealt{BGT82}) with ${\cal L}_{L,\nu,c}=3\pi\nu_s\sigma_0a^2\Omega$
being the viscous angular momentum luminosity of a circular streamline.

\begin{figure}
    \plotone{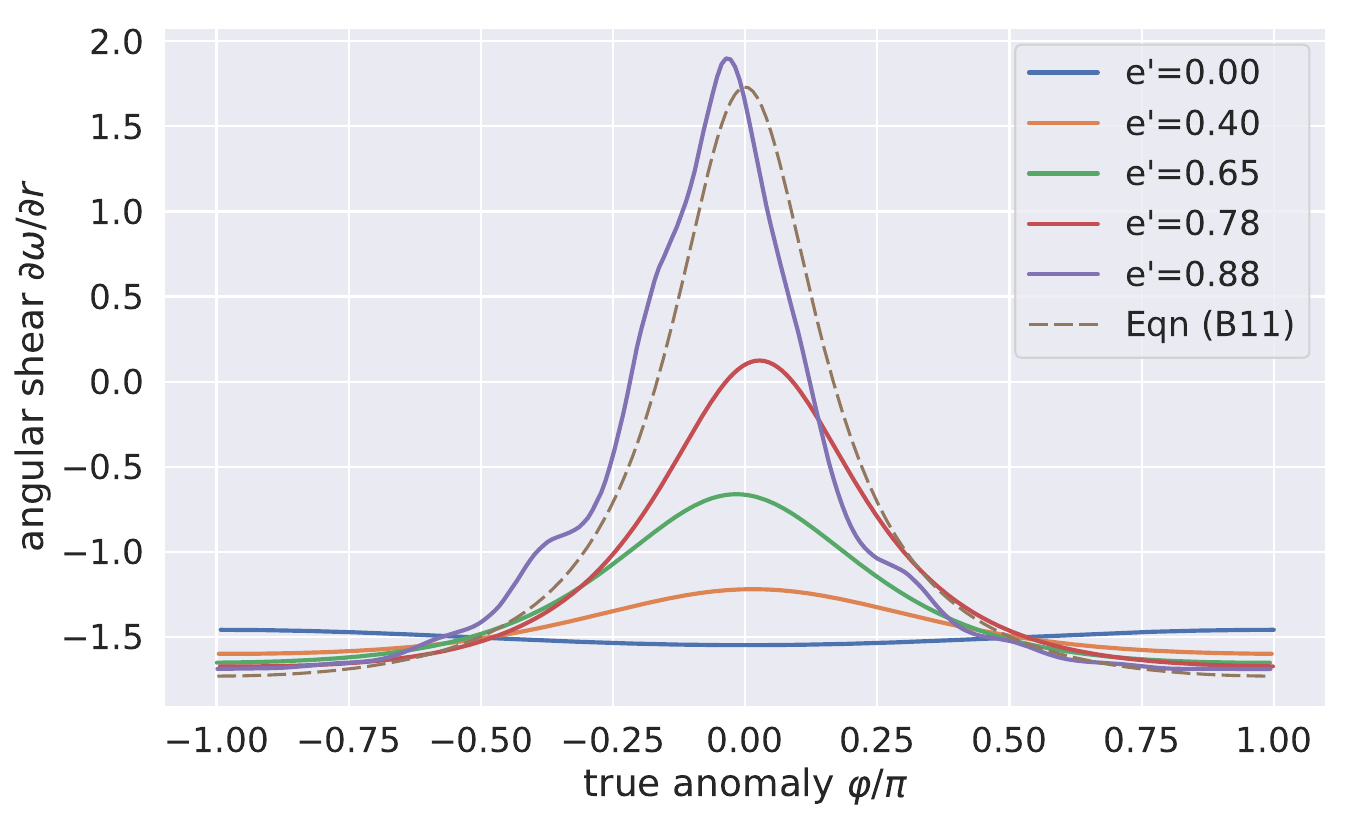}
    \caption{
        \label{fig:nominal_shear}
        The nominal ringlet's angular shear $\partial\omega/\partial r$ is plotted versus true 
        anomaly $\varphi$ at selected moments in time; this quantity is negative where the inner streamline
        has the higher angular speed $\omega=v_\theta/r$. When the simulation starts, the
        ringlet has eccentricity $e=0.01$ and eccentricity gradient $e'=0$ 
        so $\partial\omega/\partial r\simeq-3\Omega/2r\simeq-1.5$ when evaluated natural units (blue curve).
        The ringlet's $e'$ then grows over time (orange, green, red curves),
        which reverses the sign of $\partial\omega/\partial r$
        near periapse when $e'>0.75$; here the inner ringlet's angular speed
        is slower than the outer ringlet, and viscous friction causes angular momentum to instead flow inwards
        at these longitudes. Dashed curve is Eqn.\ (\ref{eqn:domega-dr}) with $e'=\sqrt{3}/2$
        and assuming $|\tilde{\omega}'|\ll1$.
    }
\end{figure}

\citet{BGT82} examine angular momentum transport through a viscous eccentric but non-gravitating ringlet, 
and use Eqns.\ (\ref{eqn:F_nu_varphi}--\ref{eqn:L_nu}) to show that this transport has three 
regimes distinguished by the ringlet's $e'$:

\begin{enumerate}

\item $e'<3/4$. The ringlet's viscous angular momentum flux ${\cal F}_{L,\nu}(\varphi)>0$
at all longitudes $\varphi$. The ringlet's viscous
angular momentum luminosity ${\cal L}_{L,\nu}>0$, so viscous friction transports angular momentum radially outwards,
and the inner ring matter evolves to smaller orbits while exterior ring matter evolves outwards, and
the ringlet spreads radially.

\item $3/4\le e'<\sqrt{3}/2$. In this regime there is a range of longitudes $\varphi$
where the viscous angular momentum flux is reversed such that ${\cal F}_{L,\nu}(\varphi)<0$. 
That angular momentum flux reversal is due to the $\partial\omega/\partial r$
term in Eqn.\ (\ref{eqn:F_nu_theta}) changing sign near periapse when $e'>0.75$, 
see Fig.\ \ref{fig:nominal_shear}.
Nonetheless ${\cal L}_{L,\nu}$, which is proportional to the orbit-average of ${\cal F}_{L,\nu}(\varphi)$,
is positive and the ringlet still spreads radially, albeit slower than when $e'<0.75$.

\item $e'\ge\sqrt{3}/2$. Viscous angular momentum flux reversal is complete such that ${\cal L}_{L,\nu}\le0$,
viscous friction transports angular momentum radially inwards, and the ringlet
shrinks radially. But if $e'=\sqrt{3}/2\simeq0.866$ then ${\cal L}_{L,\nu}=0$ and the ringlet's
radial evolution ceases, and the viscous but non-gravitating ringlet is self--confining.

\end{enumerate}
 
Note though that the nominal ringlet's eccentricity gradient
exceeds the $e'=\sqrt{3}/2\simeq0.866$ threshold 
(dotted red line in Fig.\ \ref{fig:de_prime_nominal}) when it settles into
self-confinement. This is due to the ringlet's self-gravity,
which also transports a flux of angular momentum ${\cal F}_{L,g}$ radially through the ringlet.

\begin{figure}
    \plotone{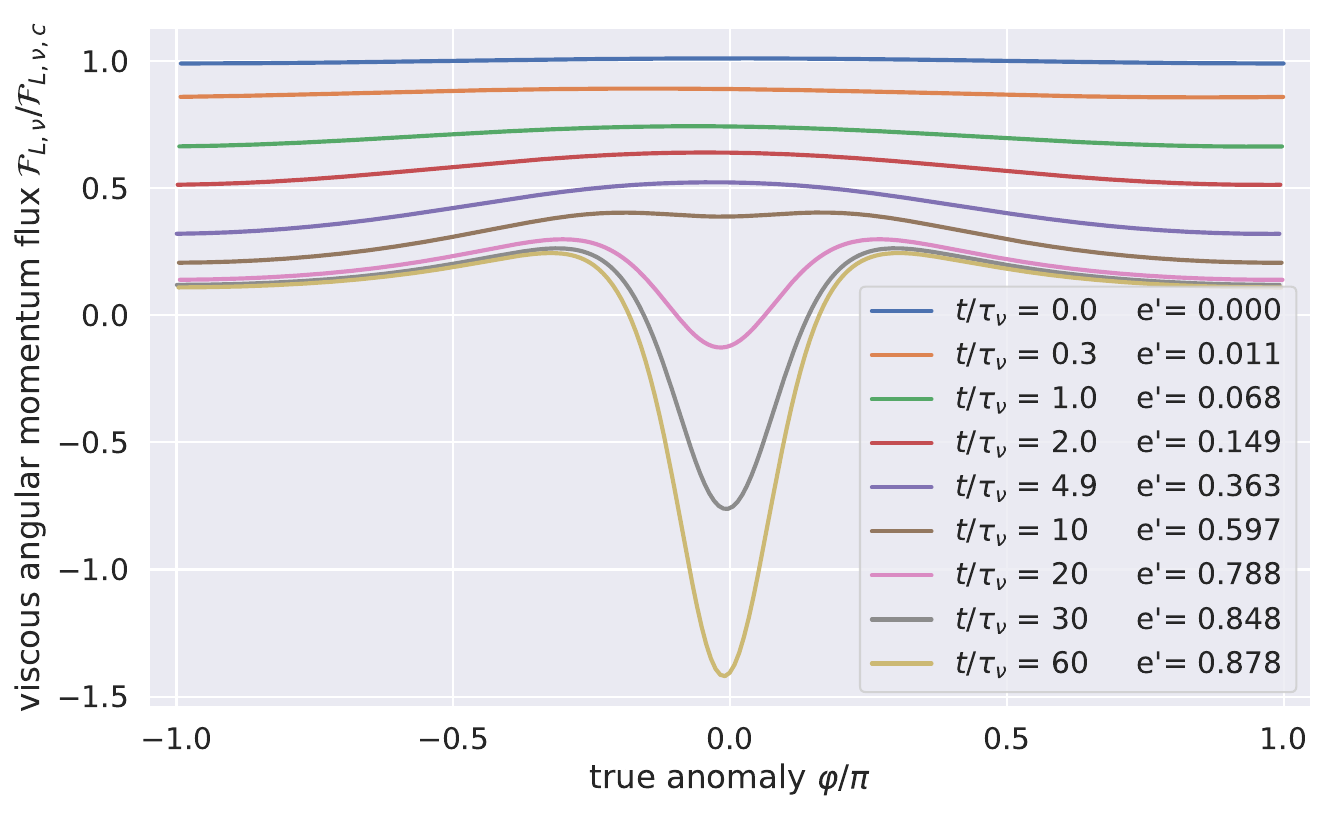}
    \caption{
        \label{fig:F_nu_nominal}
        The nominal ringlet's viscous angular momentum flux ${\cal F}_{L,\nu}(\varphi)$,
        Eqn.\ (\ref{eqn:F_nu_varphi}), is plotted 
        in units of ${\cal F}_{L,\nu,c}$ (the angular momentum flux in a circular ringlet)
        and versus the ringlet's
        true anomaly $\varphi=\theta-\tilde{\omega}$ at selected times $t/\tau_\nu$, 
        with the ringlet's eccentricy gradient $e'$ also indicated.
    }
\end{figure}

\begin{figure}
    \plotone{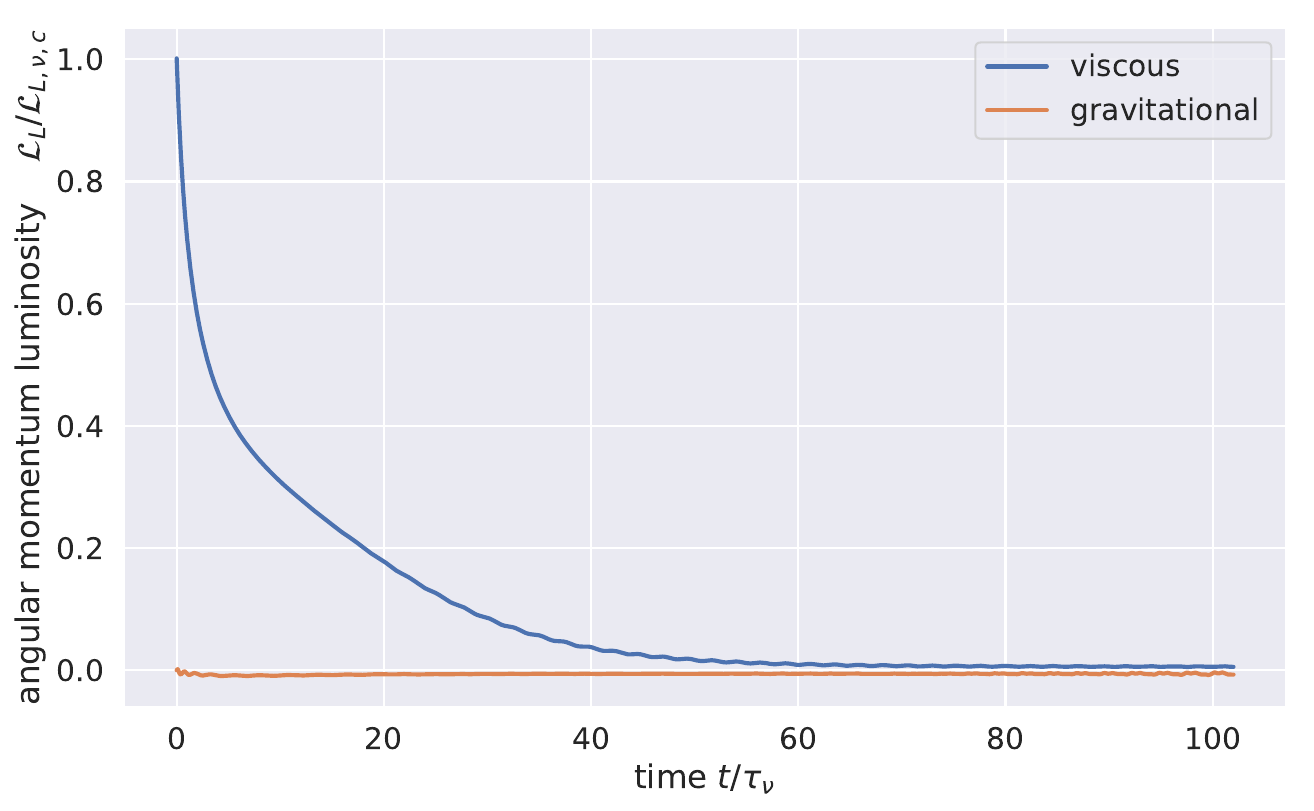}
    \caption{
        \label{fig:angular_momentum_luminosity_nominal}
        Nominal ringlet's viscous angular momentum luminosity ${\cal L}_{L,\nu}$ (blue curve) versus time $t/\tau_\nu$
        and in units of a circular ring's viscous angular momentum luminosity ${\cal L}_{L,\nu,c}$, 
        as well as the ringlet gravitational angular momentum luminosity ${\cal L}_{L,g}$ (orange curve).
    }
\end{figure}

\begin{figure}
    \plotone{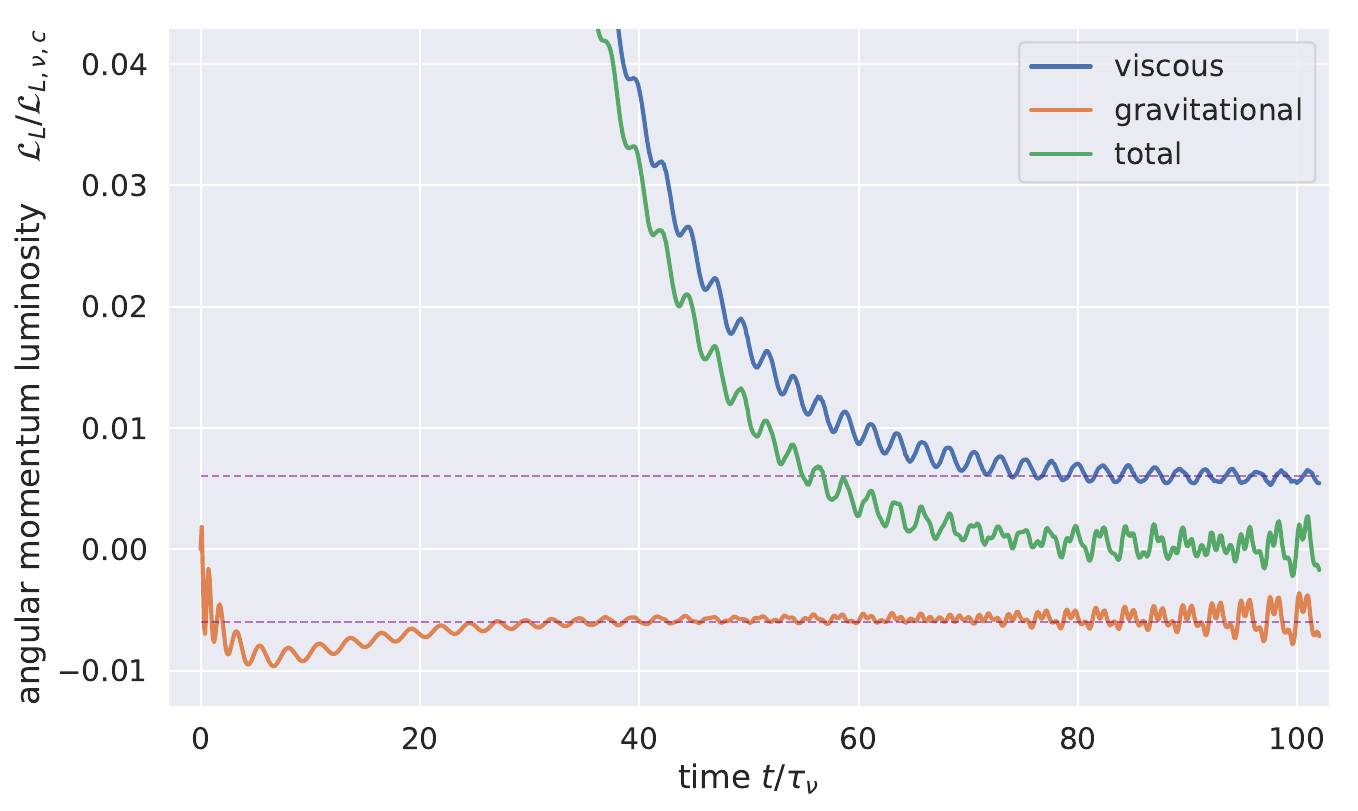}
    \caption{
        \label{fig:angular_momentum_luminosity_zoom_nominal}
        Figure \ref{fig:angular_momentum_luminosity_nominal} is replotted to show that the ringlet's 
        viscous angular momentum luminosity ${\cal L}_{L,\nu}$ (blue curve)
        always stays positive (indicating that the viscous transport of angular momentum is radially outwards)
        and is eventually balanced by the ringlet's negative ({\it i.e.}\ inwards) gravitational angular momentum luminosity 
        ${\cal L}_{L,g}$ (orange) after time $t\ge 80\tau_\nu$. Green curve is total angular momentum luminosity 
        ${\cal L}_{L,\nu} + {\cal L}_{L,g}$ whose time-average is zero when $t\ge 80\tau_\nu$.
        Dashed lines are $\pm0.006{\cal L}_{L,\nu,c}$.
    }
\end{figure}

Figure \ref{fig:F_nu_nominal} shows the nominal ringlet's viscous angular momentum flux
${\cal F}_{L,\nu}$ versus true anomaly $\varphi=\theta-\tilde{\omega}$ at selected times $t$.
Early in the ringlet's evolution when time $t \le 10\tau_\nu$ (blue, orange, green, red, purple 
and brown curves),
the ringlet is in regime 1 since $e'<0.75$ and ${\cal F}_{L,\nu}(\varphi)>0$ at all longitudes.
But by time $t \ge 20\tau_\nu$ (pink curve), this ringlet's eccentricity gradient exceeds $0.75$,
and angular momentum flux reversal ${\cal F}_{L,\nu}(\varphi)<0$ occurs near periapse where $|\varphi|\simeq0$
where the ringlet is most overdense due to its eccentricity gradient, see 
also Fig.\ \ref{fig:radial_sigma_nominal}.
This ringlet is now in regime 2 and its radial spreading is reduced by angular momentum flux reversal. 
And by time $t = 60\tau_\nu$ (yellow curve), this ringlet is seemingly in regime 3
since $e'>0.866$, so one might expect the ringlet to start contracting
now, but keep in mind that the above analysis ignores any transport
of angular momentum via ringlet self--gravity. Figure \ref{fig:da_nominal} in fact
shows that this gravitating ringlet's spreading had ceased by time $t \simeq80\tau_\nu$, 
at which point $e'=0.88$ (Fig.\ \ref{fig:F_nu_nominal} yellow curve), angular momentum flux reversal is nearly complete,
and the ringlet's total angular momentum luminosity ${\cal L}_L={\cal L}_{L,\nu}+{\cal L}_{L,g}$ is very close to zero.
Figures \ref{fig:angular_momentum_luminosity_nominal} and \ref{fig:angular_momentum_luminosity_zoom_nominal} also show that, 
when the ringlet is self-confining at times $t\ge 80\tau_\nu$, 
its small but positive viscous angular momentum luminosity ${\cal L}_{L,\nu}\simeq0.006{\cal L}_{L,\nu,c}$ 
is counterbalanced by its negative gravitational angular momentum luminosity ${\cal L}_{L,g}\simeq -0.006{\cal L}_{L,\nu,c}$,
so radial spreading has ceased and the ringlet is self-confining. 

\begin{figure}
    \plotone{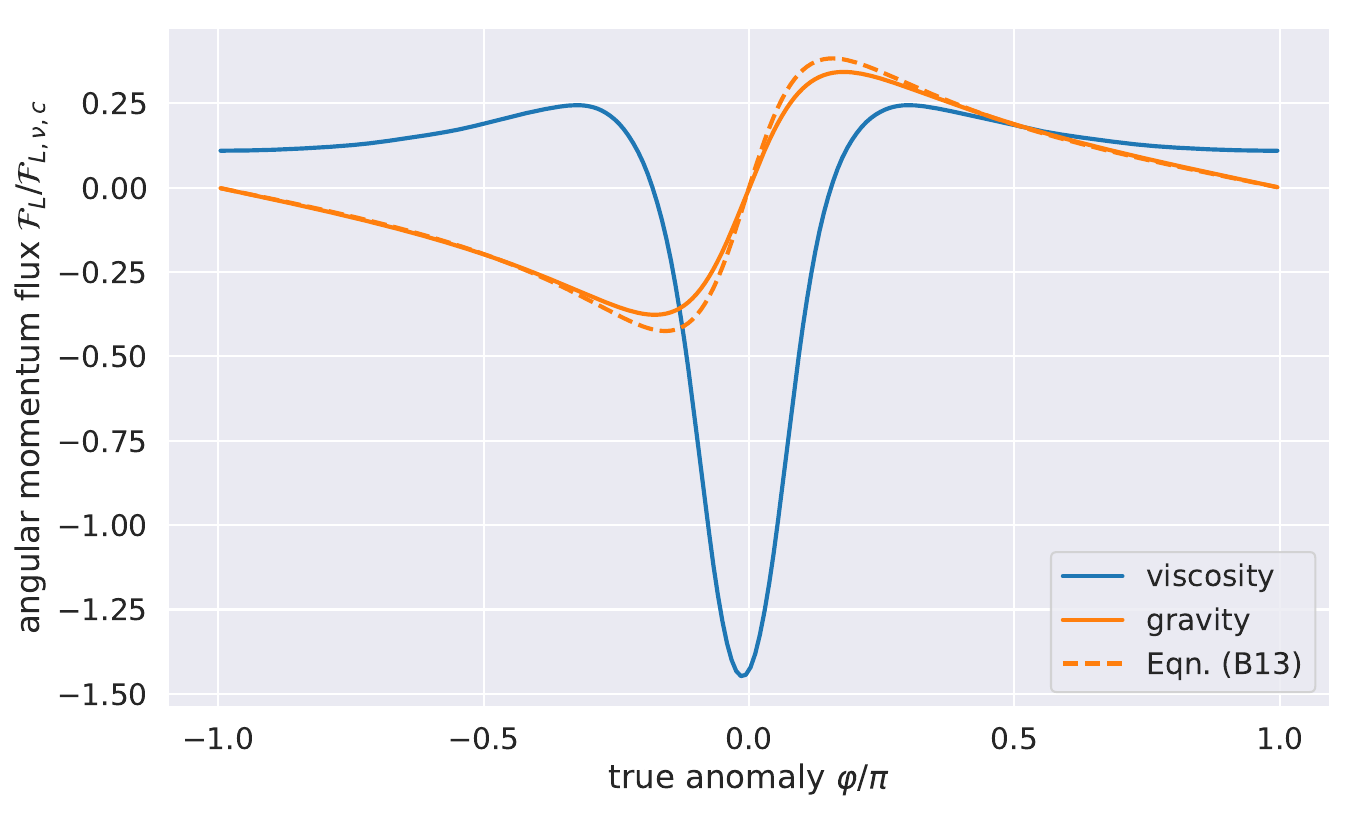}
    \caption{
        \label{fig:F_vs_longitude_nominal}
        The nominal ringlet's viscous angular momentum flux ${\cal F}_{L,\nu}(\varphi)$ (blue curve) is computed
        via Eqn.\ (\ref{eqn:F_nu_theta}) and plotted in units of a circular ringlet's flux ${\cal F}_{L,\nu,c}$
        versus true anomaly $\varphi$ at the simulation's end time $t=100\tau_\nu$, 
        as well as the ringlet's gravitational angular momentum flux ${\cal F}_{L,g}(\varphi)$
        (orange curve) computed via Eqns.\ (\ref{eqn:gravity}) and (\ref{eqn:grav_ang_mom_flux}).
        Dashed curve show the approximate gravitational flux, 
        Eqn.\ (\ref{eqn:F_L_grav_approx}), which is derived in Appendix \ref{sec:Appendix B} 
        agrees very well with the exact flux (solid curve) 
        that is derived from this simulation.
    }
\end{figure}

\subsection{gravitational transport}
\label{subsec:gravitational_flux}

The nominal ringlet's viscous ${\cal F}_{L,\nu}$ and gravitational ${\cal F}_{L,g}$
angular momentum fluxes are shown Fig.\ \ref{fig:F_vs_longitude_nominal} after it has settled into the self-confining state. 
That figure shows how viscous friction tends to transport angular momentum radially inwards, ${\cal F}_{L,\nu}(\varphi)<0$, 
at longitudes nearer periapse where $|\varphi|\sim0$, and outwards
at all other longitudes, with that flux reversal being due to the
reversal of the ringlet's angular velocity gradient, Fig.\ \ref{fig:nominal_shear}. 
Figure \ref{fig:F_vs_longitude_nominal} also shows that the ringlet's gravitational
transport of angular momentum is inwards as
ring-matter approaches periapse where $\varphi<0$, 
and is outwards ${\cal F}_{L,g}(\varphi)>0$ post-periapse, with that asymmetry being due to the ringlet's
negative periapse twist, $\tilde{\omega}'<0$ (Fig.\ \ref{fig:de_prime_nominal}).
See also Appendix \ref{sec:Appendix B}, which derives the ringlet's gravitational angular momentum flux
${\cal F}_{L,g}(\varphi)$ as a function of its eccentricity gradient $e'$. 

\begin{figure}
    \plotone{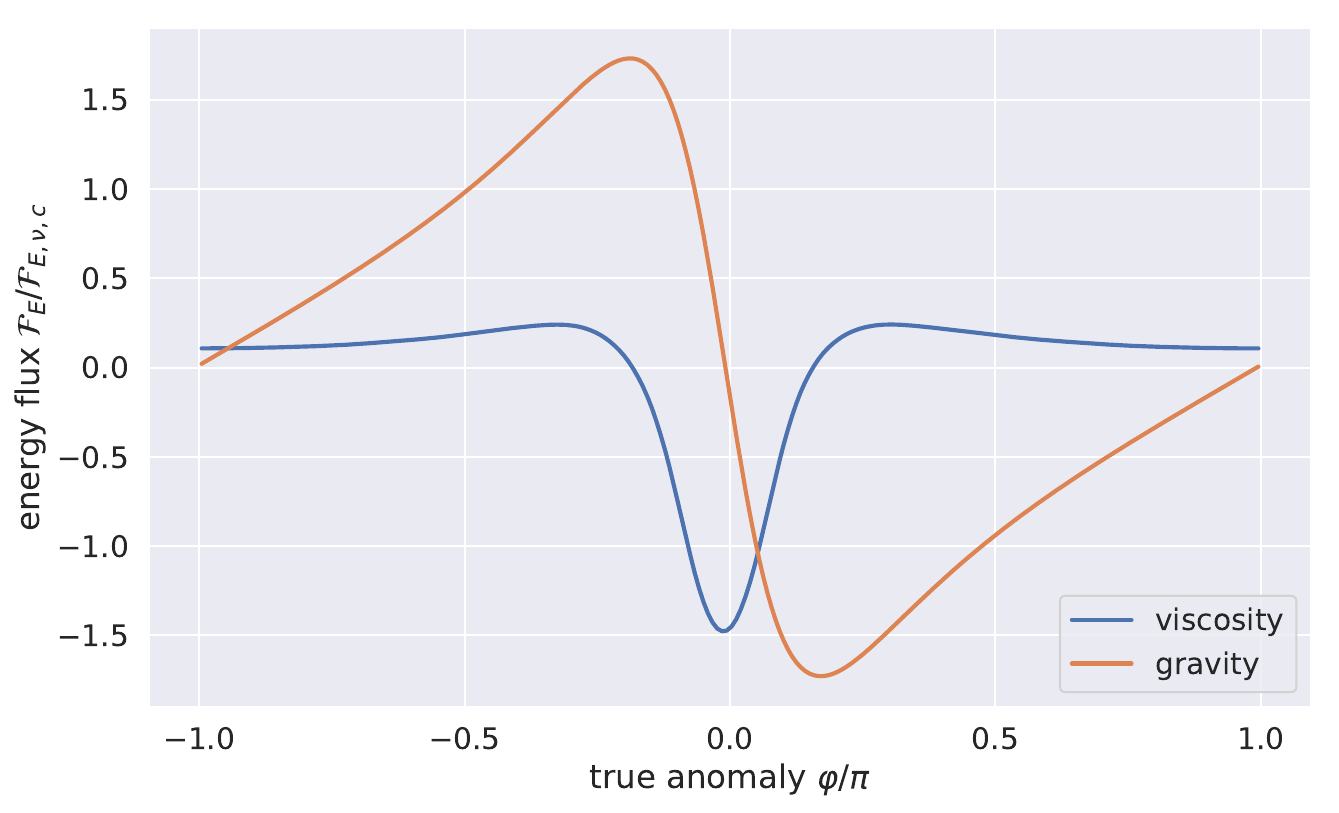}
    \caption{
        \label{fig:nominal_energy_flux}
        Blue curve is the nominal ringlet's viscous energy flux ${\cal F}_{E,\nu}(\varphi)$, plotted 
        in units of a circular ringlet's viscous energy flux ${\cal F}_{E,\nu,c}$ 
        and versus the ringlet's true anomaly $\varphi$ at the simulation's end time $t=100\tau_\nu$, 
        as well as the ringlet's gravitational energy flux ${\cal F}_{E,g}(\varphi)$.
    }
\end{figure}

\begin{figure}
    \plotone{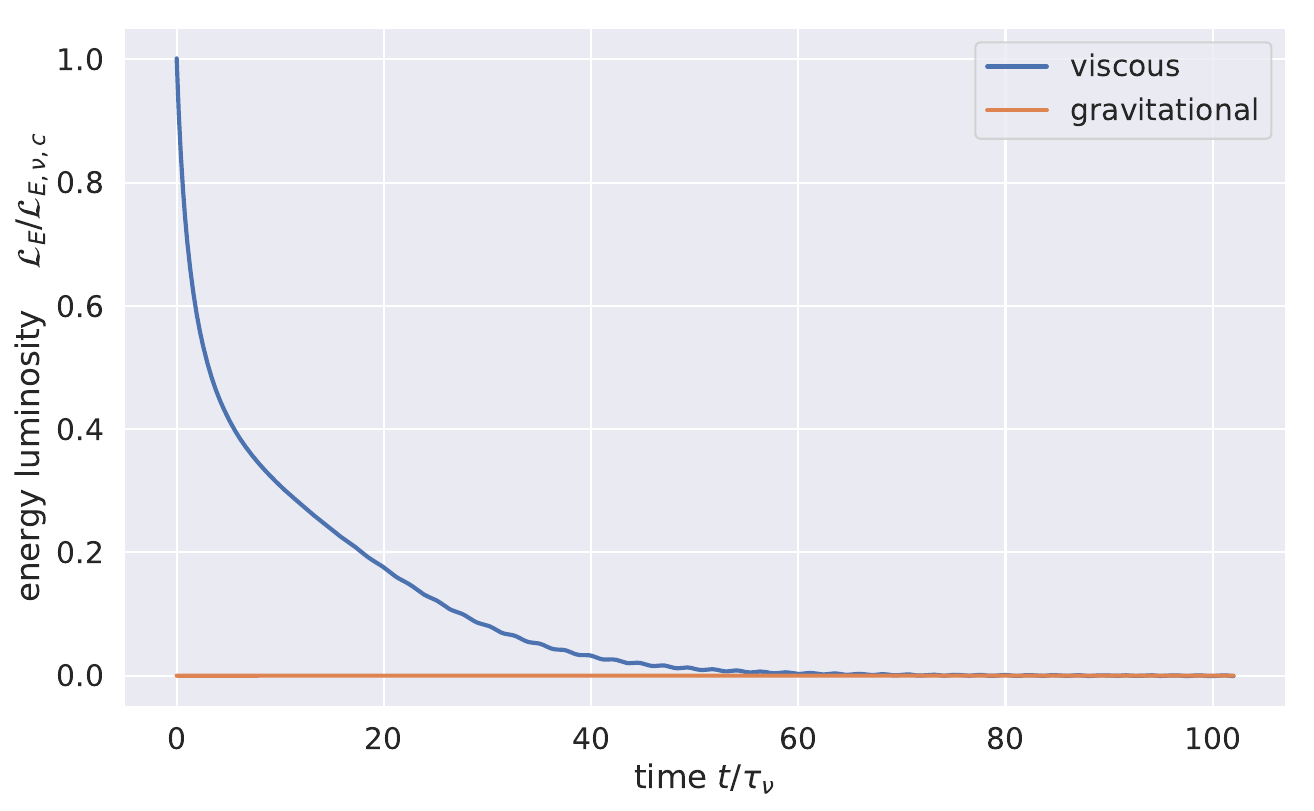}
    \caption{
        \label{fig:nominal_energy_luminosity}
        Nominal ringlet's viscous energy luminosity ${\cal L}_{E,\nu}$ (blue curve) versus time $t/\tau_\nu$
        and in units of a circular ring's viscous energy luminosity ${\cal L}_{E,\nu,c}$, 
        as well as the ringlet gravitational energy luminosity ${\cal L}_{E,g}$ (orange curve).
    }
\end{figure}

Figure \ref{fig:nominal_energy_flux} shows the ringlet's energy fluxes 
due to viscosity (blue curve) and gravity (orange) at simulation end.
Integrating these fluxes about a streamline's circumference at various times $t$ then yields the
the ringlet's viscous ${\cal L}_{E,\nu}$ and gravitational energy luminosity ${\cal L}_{E,g}$ over time,
Fig.\ \ref{fig:nominal_energy_luminosity}, where the gravitational energy luminosity is computed via
\begin{equation}
    \label{eqn:L_g}
    {\cal L}_{E,g}(a) = \oint {\cal F}_{E,g}(\varphi)rd\varphi = \oint \lambda r\mathbf{A}^1_g\cdot\mathbf{v}d\varphi
\end{equation}
where $\mathbf{A}^1_g$ is the one-sided gravitational acceleration experienced by a particle in streamline $a$.
Note that even though ${\cal F}_{E,\nu}$ and ${\cal F}_{E,g}$ have very different spatial dependances 
(see Fig.\ \ref{fig:nominal_energy_flux}),
the influence of the ringlet's viscosity and gravity still conspire such that their
orbit-integrated luminosities ${\cal L}_{E}=\oint ({\cal F}_{E,\nu}+{\cal F}_{E,g})rd\varphi$ are zero 
once the ringlet has settled into the self-confining state.

Figure \ref{fig:nominal_energy_luminosity} also shows that 
the ringlet's gravitational energy luminosity is zero at all times. Which is to be expected since 
the streamlines' gravitating ellipses only interact via their secular
perturbations, and secular perturbations do no work \citep{BC61}, hence ${\cal L}_{E,g}=0$.
That this quantity evaluates to zero within $\pm5\times10^{-24}$ (in nautural units) can also be regarded
as another test of the epi\_int\_lite integrator's numerical quality.

\subsection{variations with ringlet width, mass, and viscosity}
\label{subsec:variations}

\begin{figure}
    \gridline{\fig{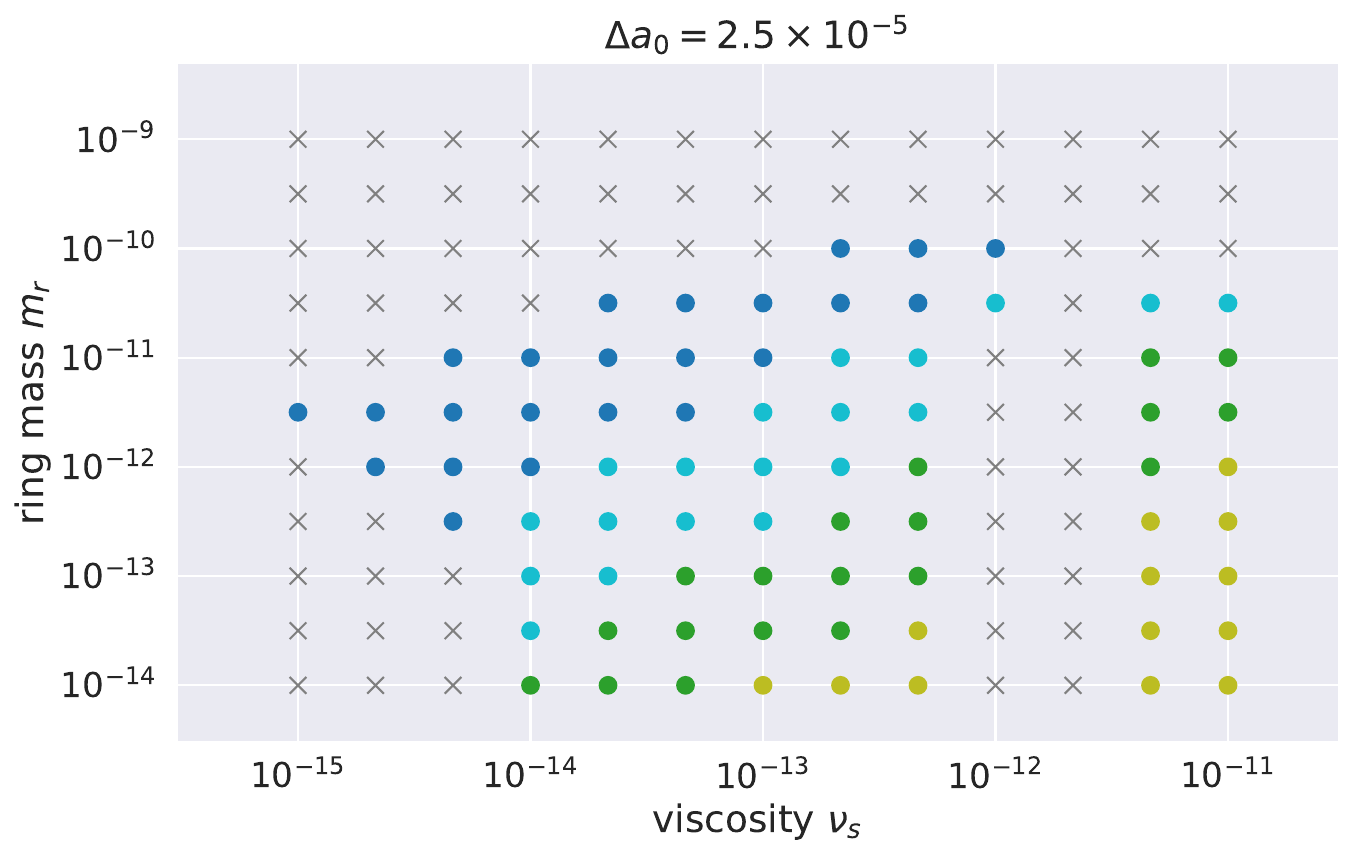}{0.5\textwidth}{}\fig{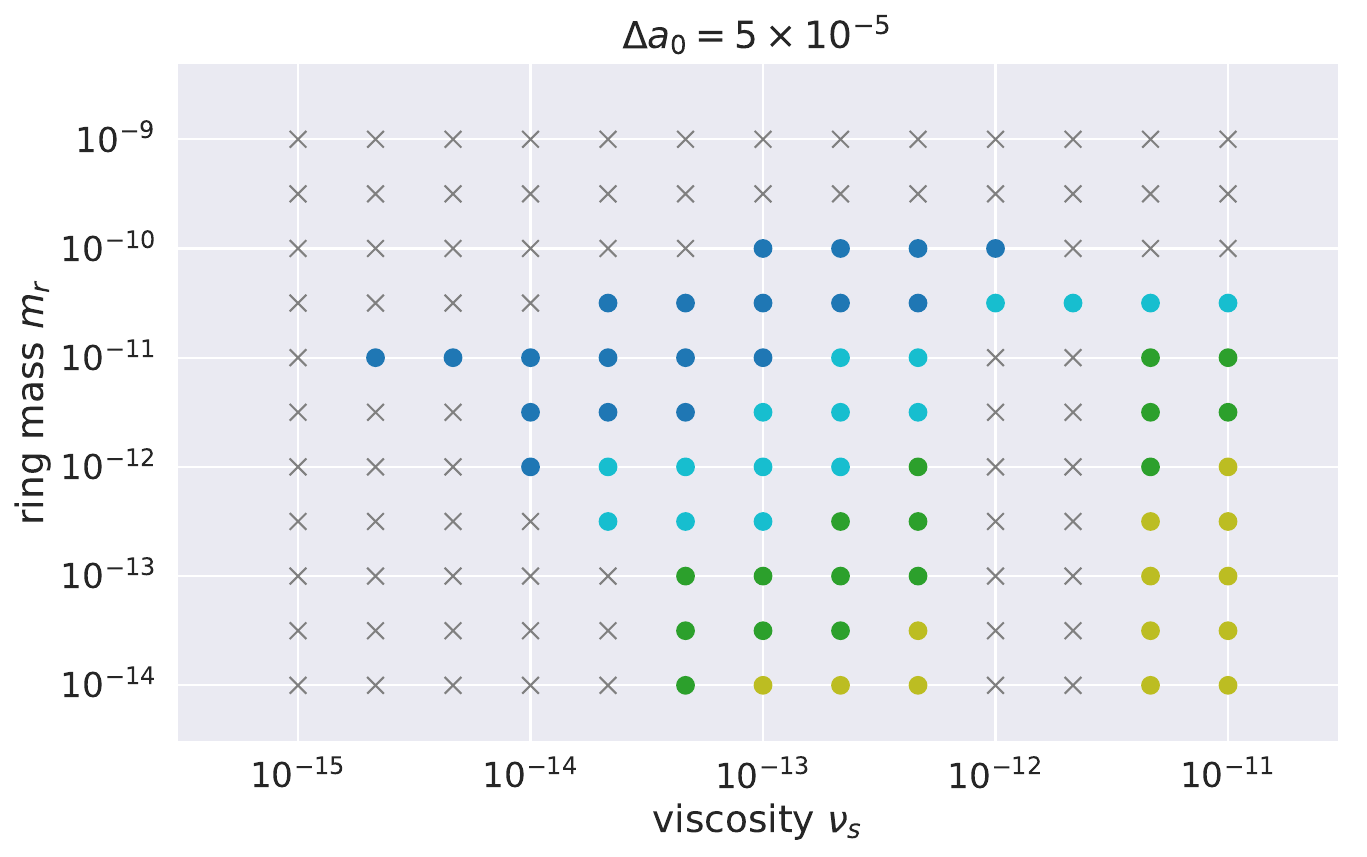}{0.5\textwidth}{}}
    \gridline{\fig{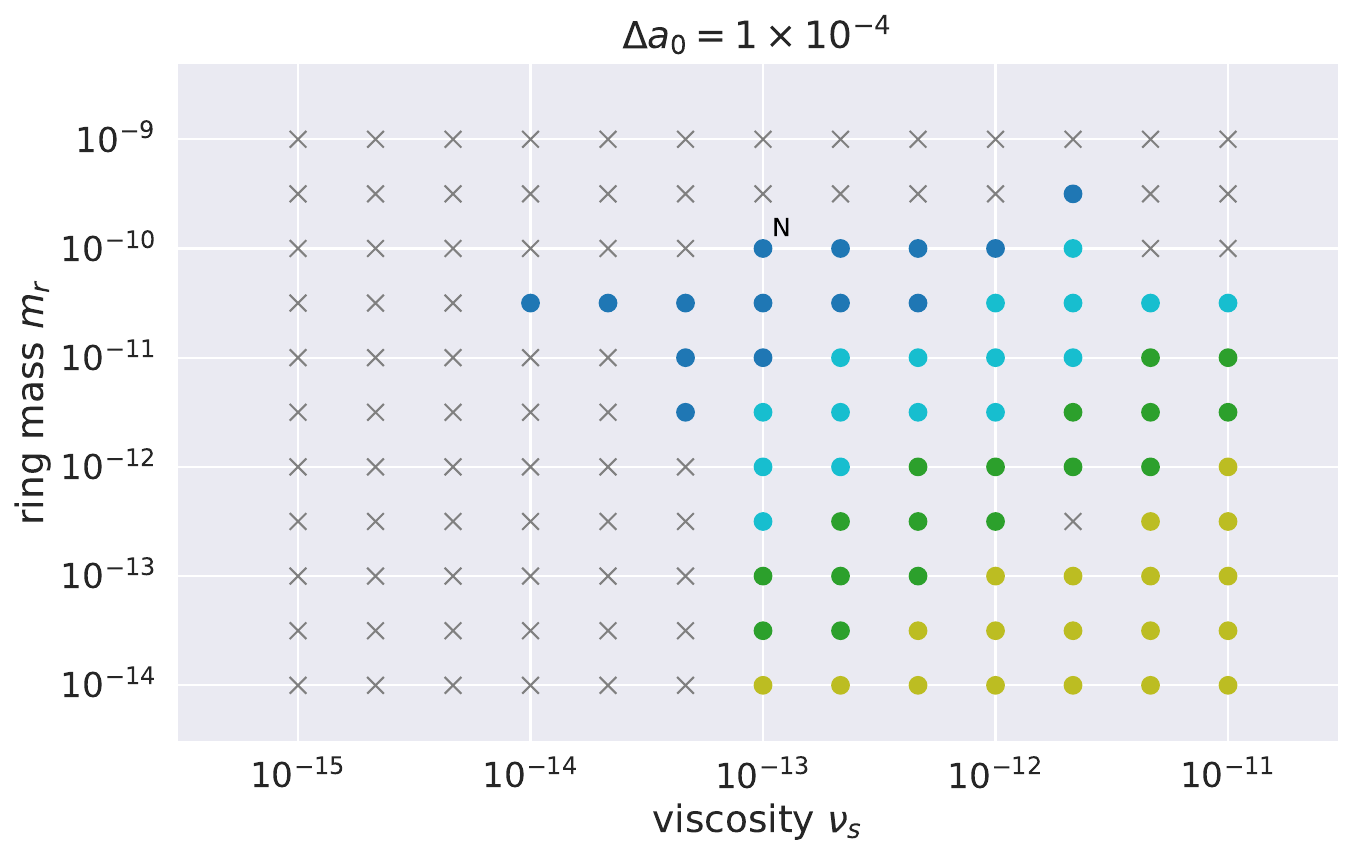}{0.5\textwidth}{}\fig{sim_grid_da_0p0001.pdf}{0.5\textwidth}{}}
    \gridline{\fig{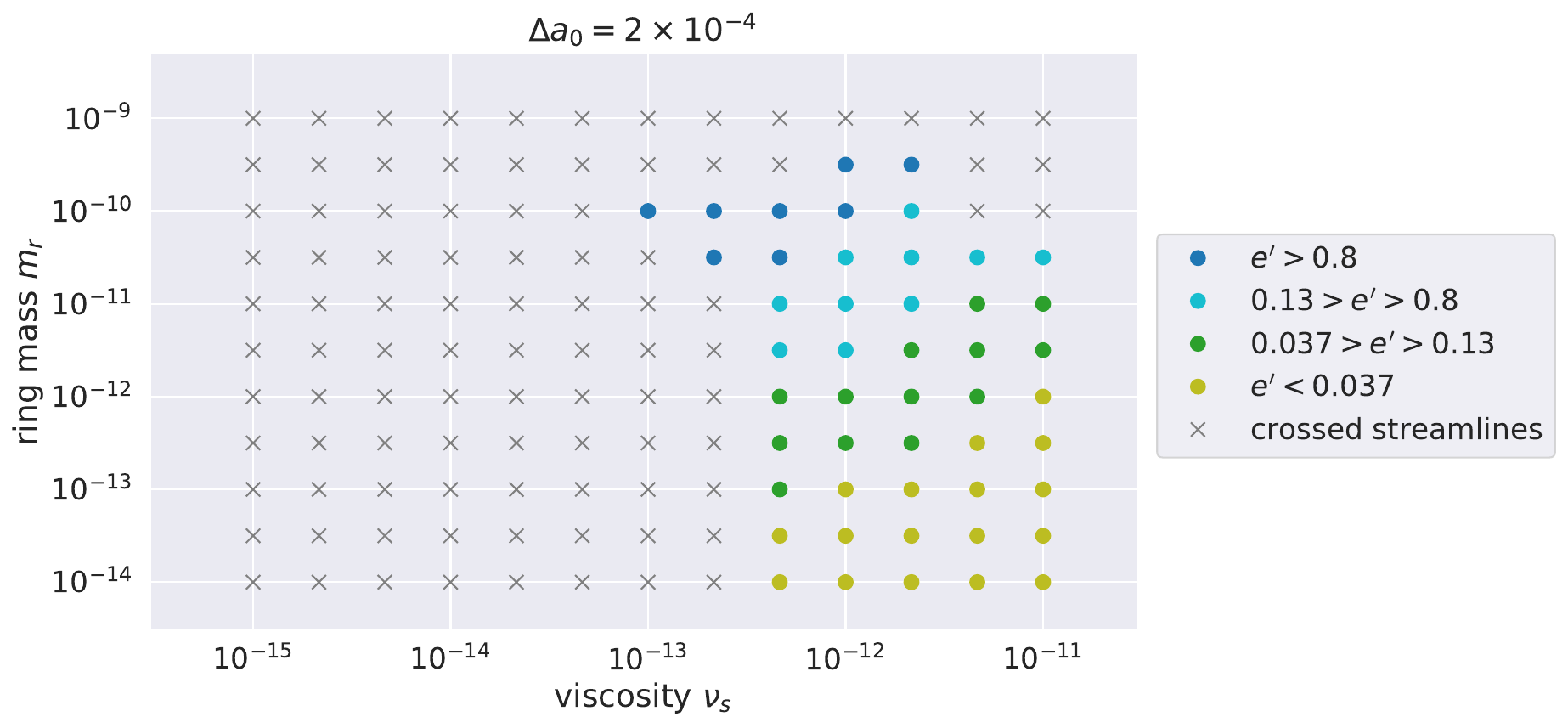}{0.7\textwidth}{}}
    \caption{
        \label{fig:sim_grid_da}
        Outcomes for 572 ringlet simulations having a variety of ringlet
        masses $m_r$ and shear viscosities $\nu_s$, with each panel showing results
        for those ringlets having the same initial radial width
        $\Delta a=2.5\times10^{-5}$, $5\times10^{-5}$, $1\times10^{-4}$, or $2\times10^{-4}$. Circles 
        indicate those ringlets that evolve into the self-confining state, while an $\times$ is
        for those simulations that terminate early when an epi\_int\_lite particle crossed an adjacent streamline,
        and N indicates the nominal ringlet simulation. The color of each self-confining ringlet shows
        whether the ringlet settled into a high--eccentricity gradient state with $e'> 0.8$ (blue),
        or smaller eccentricity gradients $0.13> e'>0.8$ (cyan), $0.037> e'>0.13$ (green), or $e<0.037$ (yellow).
        All simulations are evolved for the greater of ten dynamical timescales $\tau_{dyn}$
        or ten viscous timescales $10\tau_{\nu}$.
        The middle-right plot is redundant and is here to trick latex into rendering all plots on same page, please ignore.
    }
\end{figure}

To assess whether the nominal ringlet's evolution is typical of other ringlets
having alternate values of initial width $\Delta a$, total mass $m_r$, and shear viscosity $\nu_s$,
a survey of 571 additional ringlet simulations are executed. 
These survey ringlets are similar to the nominal ringlet 
with $N_s=2$ streamlines having $N_p=241$ particles per streamline, initial 
eccentricity $e=0.01$, initial eccentricity gradient $e'=0$, and viscosities $\nu_b=\nu_s$. 
But the survey ringlets instead have
total masses that are geometrically distributed between $10^{-14}\le m_r\le10^{-9}$,
shear viscosities distributed between $10^{-15}\le \nu_s\le 10^{-11}$,
and initial radial widths distributed between  
$2.5\times10^{-5}\le \Delta a\le2.0\times10^{-4}$. Survey results are summarized in Fig.\ \ref{fig:sim_grid_da}
where circles indicate those ringlets that do evolve into a self-confining state.
All simulations of self-confining ringlets are evolved in time until 
$10\tau_{dyn}$ where the so-called dynamical time $\tau_{dyn}$ is the moment when
the ringlet's nonlinearity parameter $q$ first exceeds $0.6$, since Fig.\ \ref{fig:de_prime_nominal}
suggests that $10\tau_{dyn}$ is sufficient time to assess whether the ringlet
has truly arrived at the self-confining state. 

The $\times$ simulations in Fig.\ \ref{fig:sim_grid_da}
terminated early when an epi\_int\_lite particle crossed a 
neighboring streamline. In reality, strong 
pressure forces would have developed as adjacent streamlines converged
and enhanced particle densities and particle collisions, with ring particles
possibly rebounding off this high-density region and/or splashing vertically,
none of which is accounted for with this version of epi\_int\_lite. 
So this survey simply terminates all such simulations and flags that occurrence with an 
$\times$ in Fig.\ \ref{fig:sim_grid_da}. Keep in mind though that this does not mean
that those particular ringlets would not have evolved into a self-confining state. Instead,
the streamlines in these ringlets have evolved so close to each other that a more sophisticated
and possibly nonlinear treatment of pressure effects would have been needed in order  
to accurately assess their fates.

Each circle in Fig.\ \ref{fig:sim_grid_da} represents a self-confining ringlet whose
nonlinearity parameter settles into a value that is close to the anticipated value, $q\simeq0.9$.
But these self-confining ringlets' final eccentricity gradients $e'$ also settle 
into a spectrum of values, $0\le e' \le 0.9$, with
outcomes indicated by circle color in Fig.\ \ref{fig:sim_grid_da}:
blue for a high--eccentricity gradient ringlets having 
$e'> 0.8$, cyan for smaller eccentricity gradients $0.13>e'>0.8$, green for $0.037>e'>0.13$,
and yellow for very low eccentricity gradients $e<0.037$,
with these intervals chosen so that there are approximately 60 simulations in each color-bin.
Inspection of Fig.\ \ref{fig:sim_grid_da} also shows that ringlets having lower $e'$ have
lower masses and higher viscosities than the higher $e'$ ringlets.

\begin{figure}
    \plotone{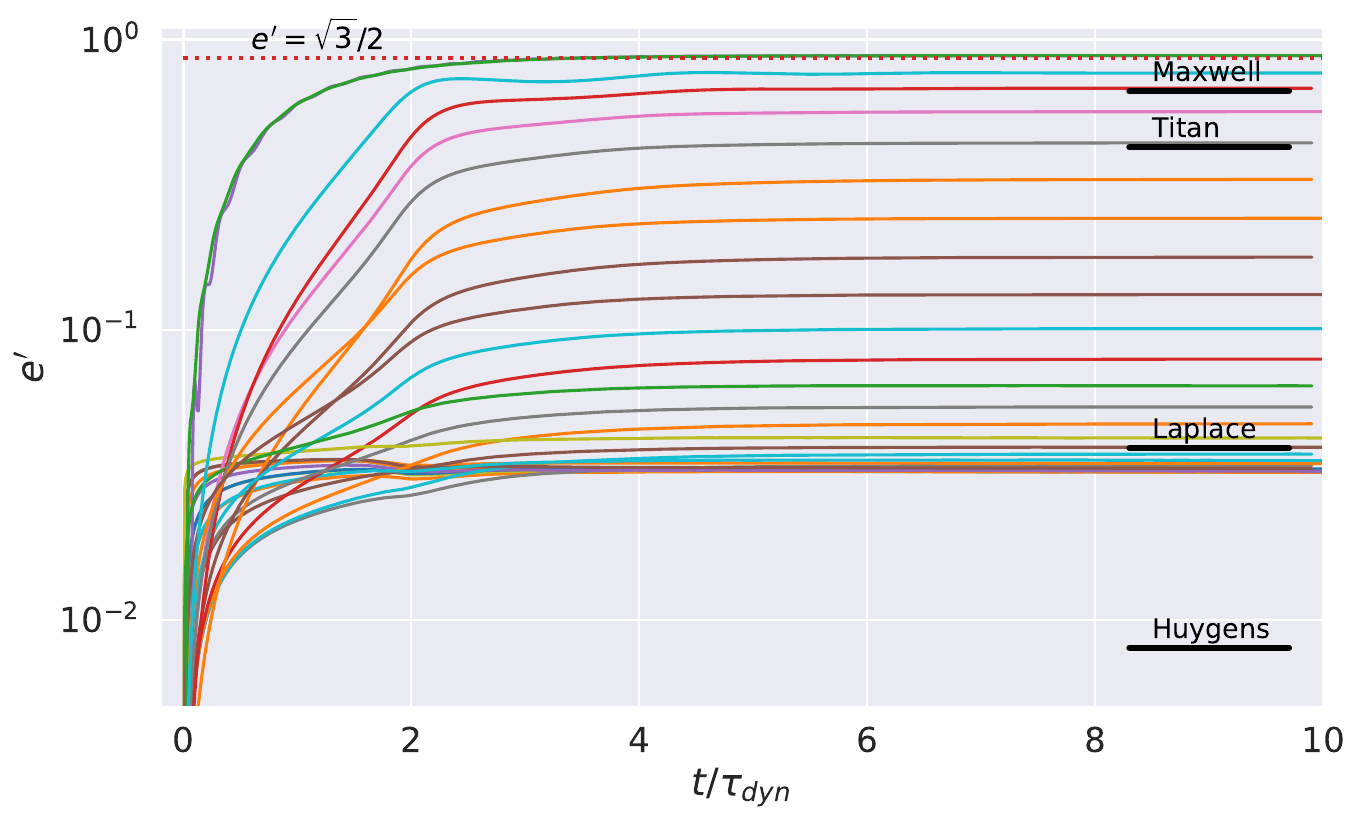}
    \caption{
        \label{fig:e_prime_q_vs_time}
        Eccentricity gradient $e'$ versus time $t/\tau_{dyn}$ for a sample of 28 self-confining
        ringlets such that their eccentricity gradients $e'$ span the range of values
        exhibited by the simulations reported in Fig.\ \ref{fig:sim_grid_da}. 
        The ringlet's dynamical timescale $\tau_{dyn}$ is time the when its nonlinearity parameter $q$ exceeds $0.6$.
        Sample ringlets have initial $2.5\times10^{-5}\le\Delta a\le2.0\times10^{-4}$,
        $10^{-14}\le m_r\le10^{-10}$, and $10^{-14}\le \nu_s\le10^{-11}$.
        Horizontal black lines indicated the observed eccentricity gradients 
        exhibited by the Maxwell, Titan, Laplace, and Huygens ringlets at Saturn 
        (\citealt{Netal14}, \citealt{Fetal16}, \citealt{SH16}). Dotted line is the $e'=\sqrt{3}/2$ threshold.
    }
\end{figure}

Figure \ref{fig:e_prime_q_vs_time} shows how eccentricity gradient varies over time $t/\tau_{dyn}$
for a sample of  self-confining ringlet simulations. That Figure also shows 
that the range of simulated $e'$ outcomes also agrees with the 
range of eccentricity gradients observed among several of Saturn's more well-studied narrow eccentric
ringlets, namely the Maxwell, Titan, and Laplace ringlets, whose $e'$ are
indicated by the black horizontal lines.
Figure \ref{fig:e_prime_q_vs_time} also shows there is a pileup of simulated 
outcomes near Laplace's eccentricity gradient $e'\simeq0.04$, as well as a dearth of simulations
at significantly lower values of $e'$. Which indicates that the Huygens ringlet, whose $e'\lesssim0.01$, 
can not be accounted for by the self-confining mechanism considered here.

Also keep in mind that Fig.\ \ref{fig:e_prime_q_vs_time} is not an apples-to-apples comparison 
of simulated ringlets to observed ringlets, since no attempt is made here to match
the observed ringlet's semimajor-axis width $\Delta a$ to the simulated ringlet's final $\Delta a$.
Rather,  Fig.\ \ref{fig:e_prime_q_vs_time}'s main point is that many of the observed narrow
eccentric ringlets that have $e'\gtrsim0.04$ are consistent with
this model of a viscous self-gravitating ringlet that is also self-confining.
And lastly, recall that $q=\sqrt{e'^2 + \tilde{\omega}'^2}$ which tells us that lower eccentricity gradient
ringlets have a larger periapse twist $|\tilde{\omega}'|$.

We also note the \cite{Fetal24} study of the Uranian ringlets (abbreviated here as F24)
which reports observed values for $e'$ and $\tilde{\omega}'$
for nine ringlets; see quantities $q_{em}=e'$ and $q_{\tilde{\omega}m}=\tilde{\omega}'$ for
the $m=1$ rows in Table 14 of F24. Nearly all of the ringlets monitored in F24 have 
nonlinearity parameters $q=\sqrt{q_{em}^2 + q_{\tilde{\omega}m}^2}<0.66$ that are significantly
smaller than the $q\simeq0.9$ achieved by the self-confining ringlets simulated 
here\footnote{The exception (F24) is Uranus' $\beta$ ringlet whose $q\simeq1.2$ is so large as to imply that
this ringlet is very disturbed, its steamlines are crossing (e.g.\ \citealt{BGT82}),
and that ring particles are crashing into each other.}. 
From this we conclude that the Uranian ringlets are unlike those simulated here.

\subsubsection{variations with ringlet viscosity}
\label{subsec:viscosity-variations}

\begin{figure}
    \plotone{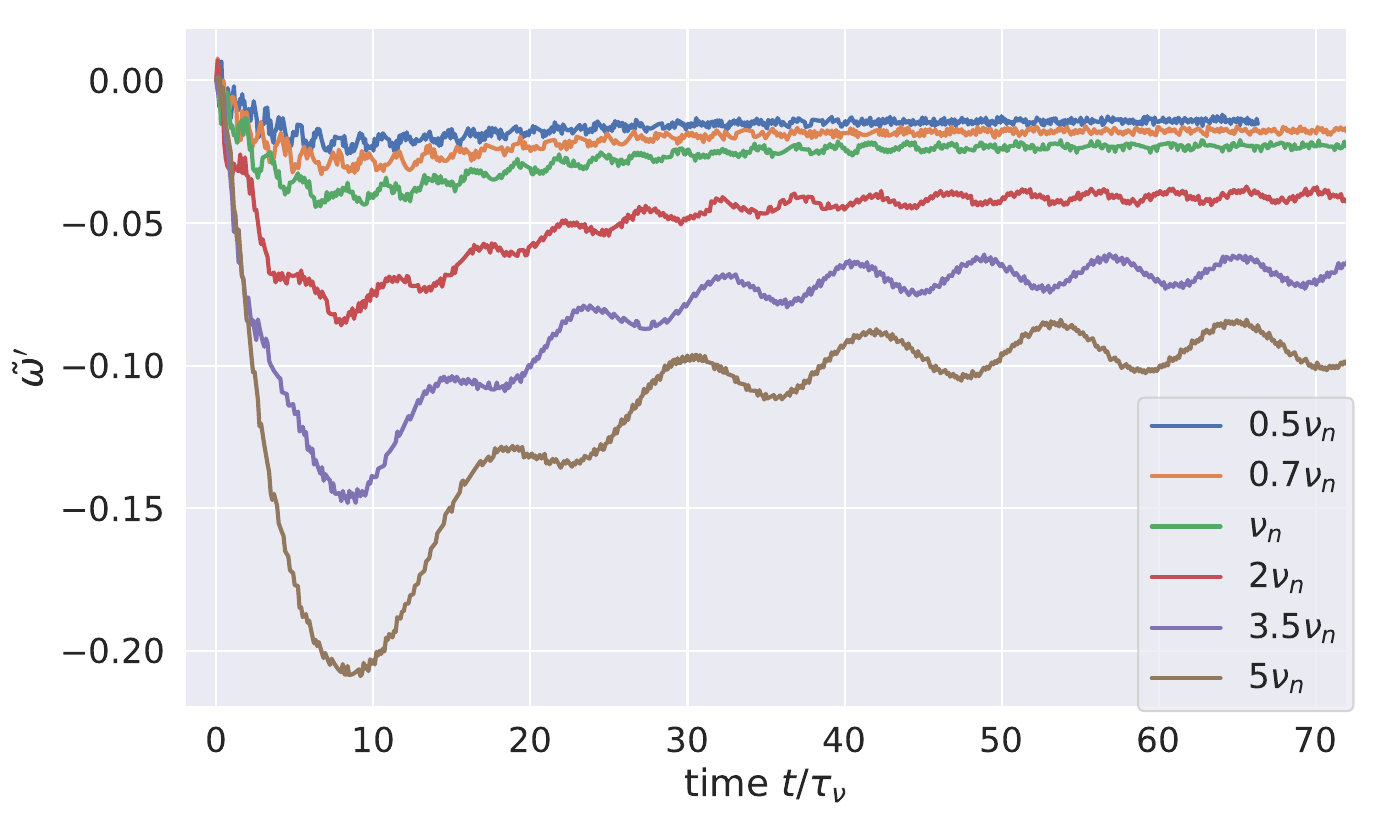}
    \caption{
        \label{fig:wt_prime_vs_time}
        Periapse twist $\tilde{\omega}'$ is plotted versus time $t/\tau_\nu$ for six ringlets
        having the same initial $e_0$, $\Delta a$, $m_r$ as the nominal ringlet
        but differing shear viscosities $\nu_s$ that range over $0.5\nu_n\le\nu_s\le5\nu_n$
        where $\nu_n=1.0\times10^{-13}$ is the nominal ringlet's shear visocity.
        The time-evolution of these ringlets' other orbit elements,  $\Delta a$, $e$, $e'$, and $q$,
        are very similar to that exhibited by the nominal ringlet, 
        Figs.\ \ref{fig:da_nominal}--\ref{fig:de_prime_nominal}. 
    }
\end{figure}

Figure \ref{fig:wt_prime_vs_time} shows the periapse twist $\tilde{\omega}' \simeq ea\Delta\tilde{\omega}/\Delta a$ 
versus time for six ringlets
having the same initial $e_0$, $\Delta a$, $m_r$ and $\nu_b$ as the nominal ringlet but with differing shear
viscosities $\nu_s$, and that plot shows that twist $|\tilde{\omega}'|$ varies with $\nu_s$. Which indicates that if 
the twist $|\tilde{\omega}'|$ could be observed in a self-confining ringlet, then 
the ringlet's viscosity could then be inferred. 

\subsubsection{variations with initial eccentricity}
\label{subsec:e-variations}

\begin{figure}
    \plotone{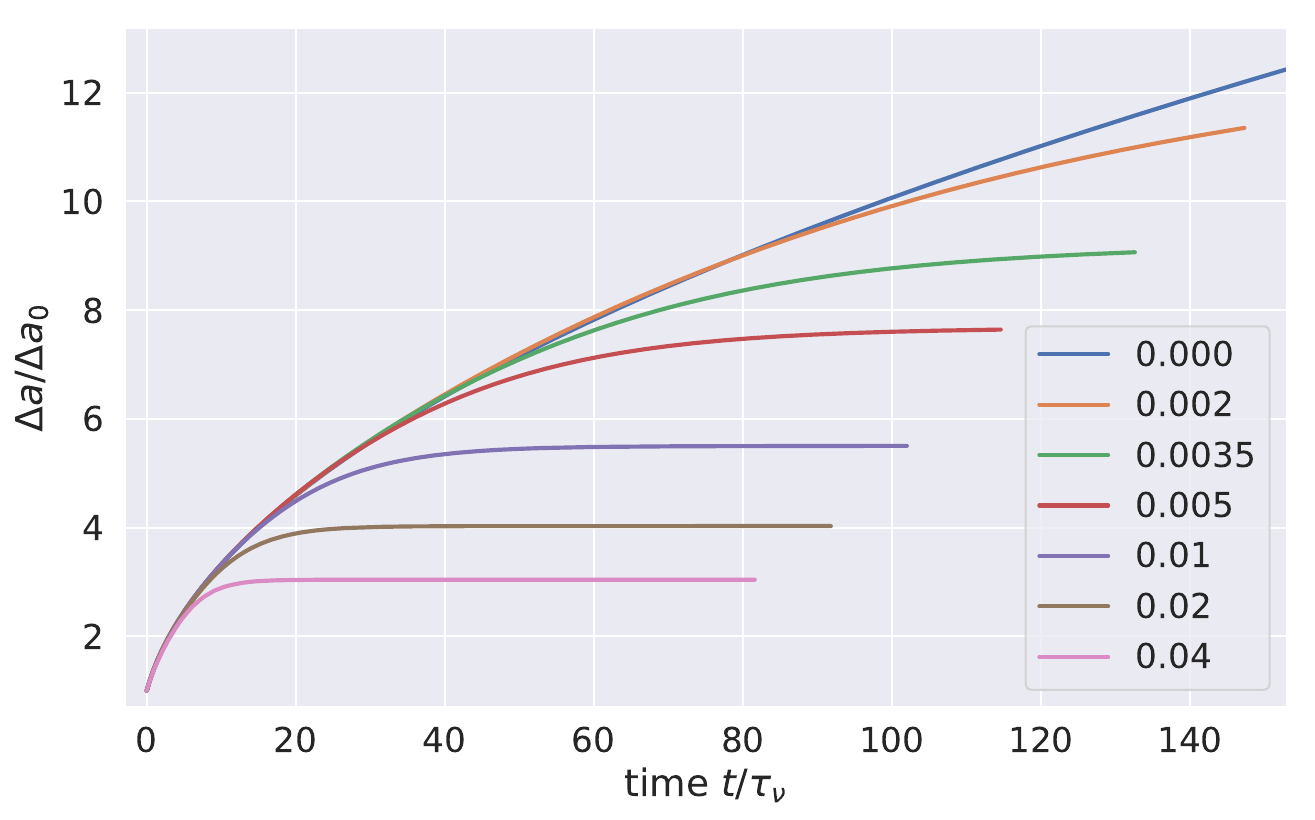}
    \caption{
        \label{fig:e0}
        Simulations of seven nominal ringlets having a variety of initial eccentricities $0\le e_0\le 0.04$.
        Curves show each ringlet's semimajor axis width $\Delta a$ in units of its initial $\Delta a_0$
        versus time $t/\tau_\nu$, and show that higher-$e$ ringlets settle into self-confinement
        sooner than the lower-$e$ ringlets.
    }
\end{figure}

Additional simulations illustrate how outcomes depend upon the ringlet's initial eccentricity $e_0$.
Figure \ref{fig:e0} shows seven simulations of the nominal ringlet that all have identical
physical properties (mass $m_r$, viscosity $\nu_s$, and initial width $\Delta a$) but differing initial $e_0$ ranging
over $0\le e_0\le 0.04$. That plot shows that higher-$e$ ringlets settle into the
self-confining state sooner than the lower-$e$ ringlets. This is because the higher-$e$ ringlet's
secular gravitational perturbation of itself drives its eccentricity gradient and hence $q$ towards 
$q\simeq\sqrt{3}/2$ faster than the lower-$e$ ringlets. Consequently, 
higher-$e$ ringlets tend to be narrower than lower-$e$ ringlets because they will have had
less time to spread before settling into self-confinement. Also note that the $e_0=0$ ringlet
(blue curve) experiences zero secular gravitational perturbations, so its $q$ is always zero
and is destined to spread forever.

The next Subsection will examine the rate at which
a self-confining ringlet's eccentricity $e$ decays over time due to viscosity, 
and all ringlets shown in Figs.\ \ref{fig:a_nominal}--\ref{fig:e_prime_q_vs_time}
have eccentricities $e$ that decay at the expected rates.

\subsection{eccentricity damping}
\label{subsec:eccentricity}

Viscous friction within the ringlet is a result of dissipative collisions among ringlet particles.
Particle collisions generate heat that is radiated into space, and the source of that radiated energy
is the ringlet's total energy $E_r=-m_rGM/2a + E_{sg}$ where $m_r$ is the ringlet's total
mass, $a$ its semimajor axis, and $E_{sg}$ is the ringlet's energy due to its self gravity
which is contant when the ringlet is self-confining. 
Collisions conserve angular momentum, so the ringlet's total angular momentum
$L_r=m_r\sqrt{GMa(1-e^2)}$ is constant so $dL_r/dt=0$ implies that
\begin{equation}
    \label{eqn:e2-dot}
    \frac{de^2}{dt} \simeq \frac{1}{a}\frac{da}{dt}
\end{equation}
to lowest order in the ringlet's small eccentricity $e$.
The ringlet's energy dissipation rate is $\dot{E}_r = dE_r/dt=m_rGM\dot{a}/2a^2$ so
$\dot{a}\simeq2\dot{E_r}/m_ra\Omega^2$ and 
\begin{equation}
    \label{eqn:de2/dt}
    \frac{de^2}{dt} \simeq \frac{2\dot{E_r}}{m_r a^2\Omega^2}
\end{equation}
where $GM\simeq a^3\Omega^2$ to lowest order in $J_2$. Also note that
the surface area of energy dissipation within a viscous disk is
\begin{equation}
    \delta = -\nu_s\sigma(r\omega')^2
\end{equation}
(\citealt{P81}) where $\omega=v_\theta/r$ is the angular velocity and 
$\omega'=\partial\omega/\partial r$ its radial gradient.

Now consider a small tangential segment within the ringlet whose length is $d\ell=rd\varphi$ where 
$\varphi$ is the segment's longitude measured from the ringlet's periapse and
$d\varphi$ is the small segment's angular extent. The segment's area is 
$dA=\Delta rd\ell=r\Delta r d\varphi$
where $\Delta r$ is the ringlet's radial width.
The rate at which that patch's viscosity dissipates orbital energy is $d\dot{E_r}=\delta dA$, so
the ringlet's total energy dissipation rate is 
$\dot{E}_r = \oint d\dot{E_r}$ when integrated about the ringlet's circumference, and so
$\dot{E}_r = -2\nu_s\lambda\int_0^\pi r^3\omega'^2d\varphi$
since the ringlet's linear density $\lambda=\sigma\Delta r\simeq m_r/2\pi a$. So the
total energy loss rate due to ringlet viscosity becomes
\begin{equation}
    \label{eqn:dE_r/dt}
    \dot{E}_r \simeq -\frac{9}{4}I(e')m_r\nu_s\Omega^2
\end{equation}
when Eqn.\ (\ref{eqn:domega-dr}) is used to replace $\omega'$, and the integral 
\begin{equation}
    I(e') =  \frac{1}{\pi}\int_0^\pi\left(\frac{1-\frac{4}{3}e'\cos\varphi}{1-e'\cos\varphi}\right)^2d\varphi .
\end{equation}
Note that  $I(e')$ is of order
unity except when $e'$ is very close to 1, and numerical evaluation shows that $I(e')\simeq0.889$
when $e'=\sqrt{3}/2$.

\begin{figure}
    \plotone{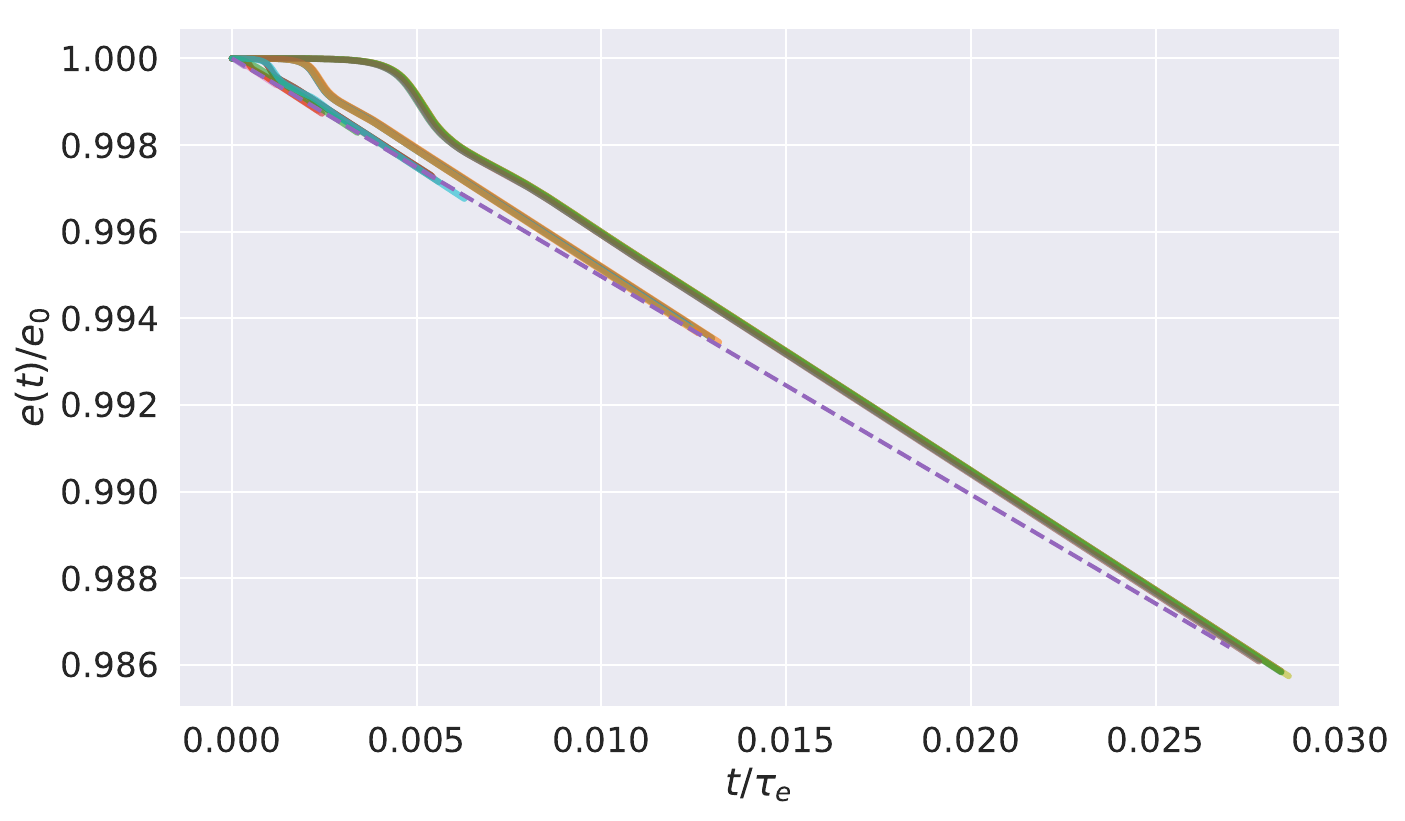}
    \caption{
        \label{fig:e_vs_tau_e}
        Plot of $e(t)/e_0$ versus time $t/\tau_e$ for the sample of 28
        simulations shown in Fig.\ \ref{fig:e_prime_q_vs_time},
        while the dashed curve is the expected behavior, Eqn.\ (\ref{eqn:e(t)}).
        This sample of simulations have $\tau_e$ ranging over 
        $2.5\times10^6 \le \tau_e \le 2.5\times10^9$.
    }
\end{figure}

Inserting Eqn.\ (\ref{eqn:dE_r/dt}) into (\ref{eqn:de2/dt}) then yields the rate at which $e^2$
is damped,
\begin{equation}
    \label{eqn:de2/dt_v2}
    \frac{de^2}{dt} = -\frac{9I\nu_s}{2a^2}
\end{equation}
which is easily integrated to obtain
\begin{equation}
    \label{eqn:e(t)}
    e(t) = e_0\sqrt{1-\frac{t}{\tau_e}}
\end{equation}
where $e_0$ is the ringlet's initial eccentricity and
\begin{equation}
    \label{eqn:tau_e}
    \tau_e = \frac{2a^2e_0^2}{9I\nu_s}
\end{equation}
is the ringlet's eccentricity damping timescale.
These expectations are also confirmed in Fig.\ \ref{fig:e_vs_tau_e}, which plots
$e(t)/e_0$ versus time $t/\tau_e$ for the sample of 28 simulations 
shown in Fig.\ \ref{fig:e_prime_q_vs_time}, with the dashed curve
indicating the theoretial predictions of Eqns.\ (\ref{eqn:e(t)}--\ref{eqn:tau_e}).
That all simulated curves have slopes similar to the dashed line tells us that
Eqns.\ (\ref{eqn:e(t)}--\ref{eqn:tau_e}) are a good indicator of outcomes across
a wide variety of ringlet parameters.

So viscosity circularizes the ringlet in time $\tau_e$, during which time the ringlet's
semimajor axis will have shrunk by $\Delta a=\dot{a}\tau_e=-e_0^2a$ by Eqns (\ref{eqn:e2-dot})
and (\ref{eqn:de2/dt_v2}), so the ringlet's
fractional drift inwards due to viscous damping is
\begin{equation}
    \label{eqn:delta-a}
    \frac{\Delta a}{a} = -e_0^2,
\end{equation}
which is small. And after the ringlet's inner edge damps to zero, its eccentricity gradient $e'$
will then shrink over time, angular momentum flux reversal will diminish, and the ringlet's 
viscous spreading will resume. So self-confinement of narrow eccentric ringlets is only temporary
after all, until time $\tau_e$ has elapsed, which is $\tau_e/2\pi\sim1.6\times10^6$ orbits for 
the nominal model considered here, which is only $\sim10^3$ years for a ringlet orbiting at 
$a\sim10^{10}$ cm about Saturn. Recall from Section \ref{subsec:nominal}
that the viscous lifetime of a non-self-confining
nominal ringlet is only $\tau_\nu/2\pi\sim500$ orbits, so self-confinement evidently
extends the lifetime of a narrow eccentric ringlet by an additional 
factor of $\sim3000$. But self-confinement
does not solve the ringlet's lifetime problem, because self-confinement is ultimately
defeated by viscous damping of the ringlet's eccentricity.

\subsection{number of streamlines $N_s$}
\label{subsec:num_streamlines}

When the simulated ringlet is composed of $N_s=2$ streamlines, the ringlet's
evolution is largely analytic ({\it c.f.}\ \citealt{BGT82, BGT83}), and the analytic
predictions provide excellent benchmark tests for the epi\_int\_lite integrator. This subsection 
assesses whether the results obtained for the simpler $N_s=2$ ringlet also apply to
more realistic ringlets having $N_s>2$.

\begin{figure}
    \plotone{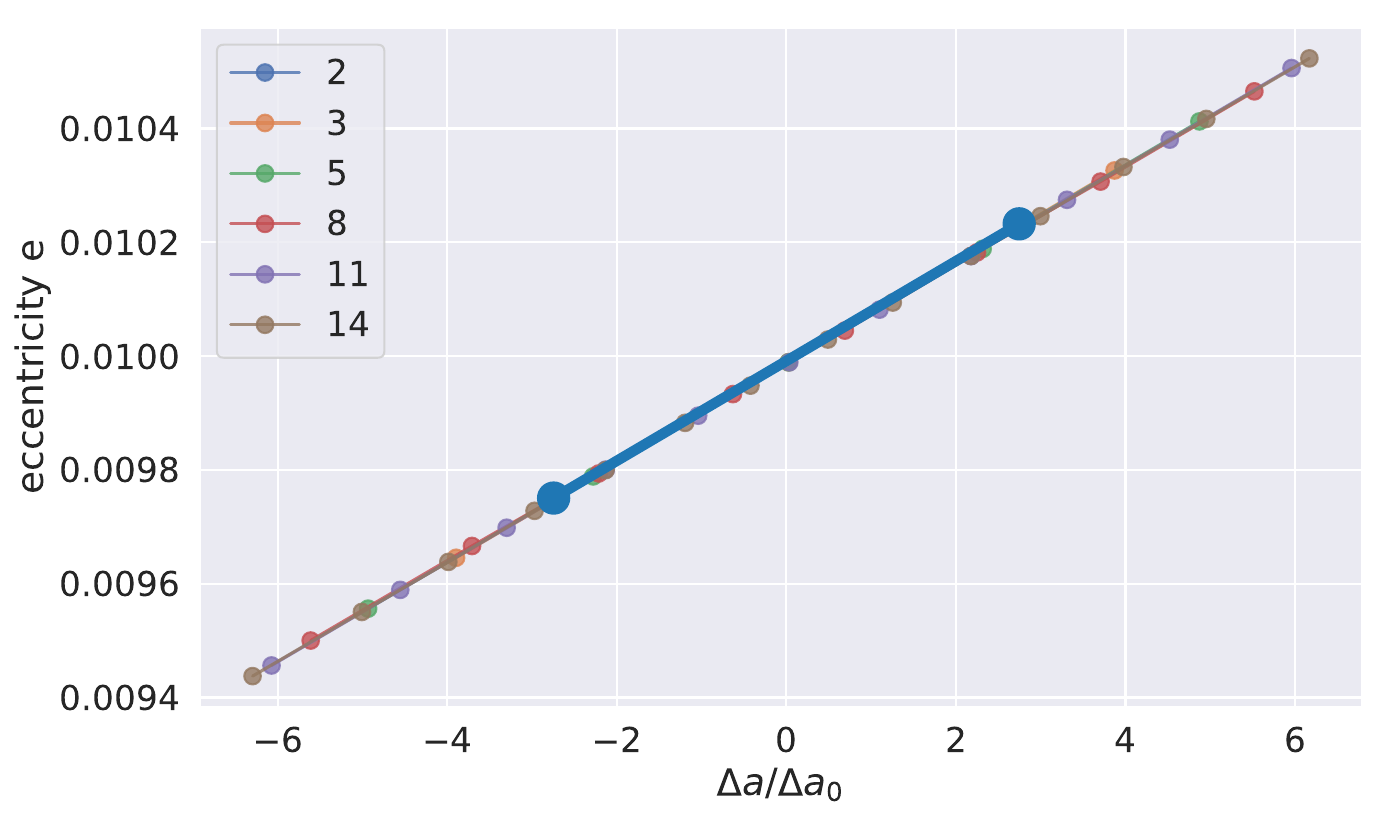}
    \caption{
        \label{fig:e_vs_da_streamlines}
        A suite of ringlets are evolved, these ringlets have the same properties as the
        nominal ringlet except they contain varied numbers of streamlines $2\le N_s\le14$,
        with $N_s$ indicated by the plot legend.
        Simulated ringlets' final eccentricities $e$ are plotted versus their relative
        semimajor axis $\Delta a=a-\bar{a}$ where $\bar{a}$ is the ringlet's mean semimajor axis. 
    }
\end{figure}

\begin{figure}
    \plotone{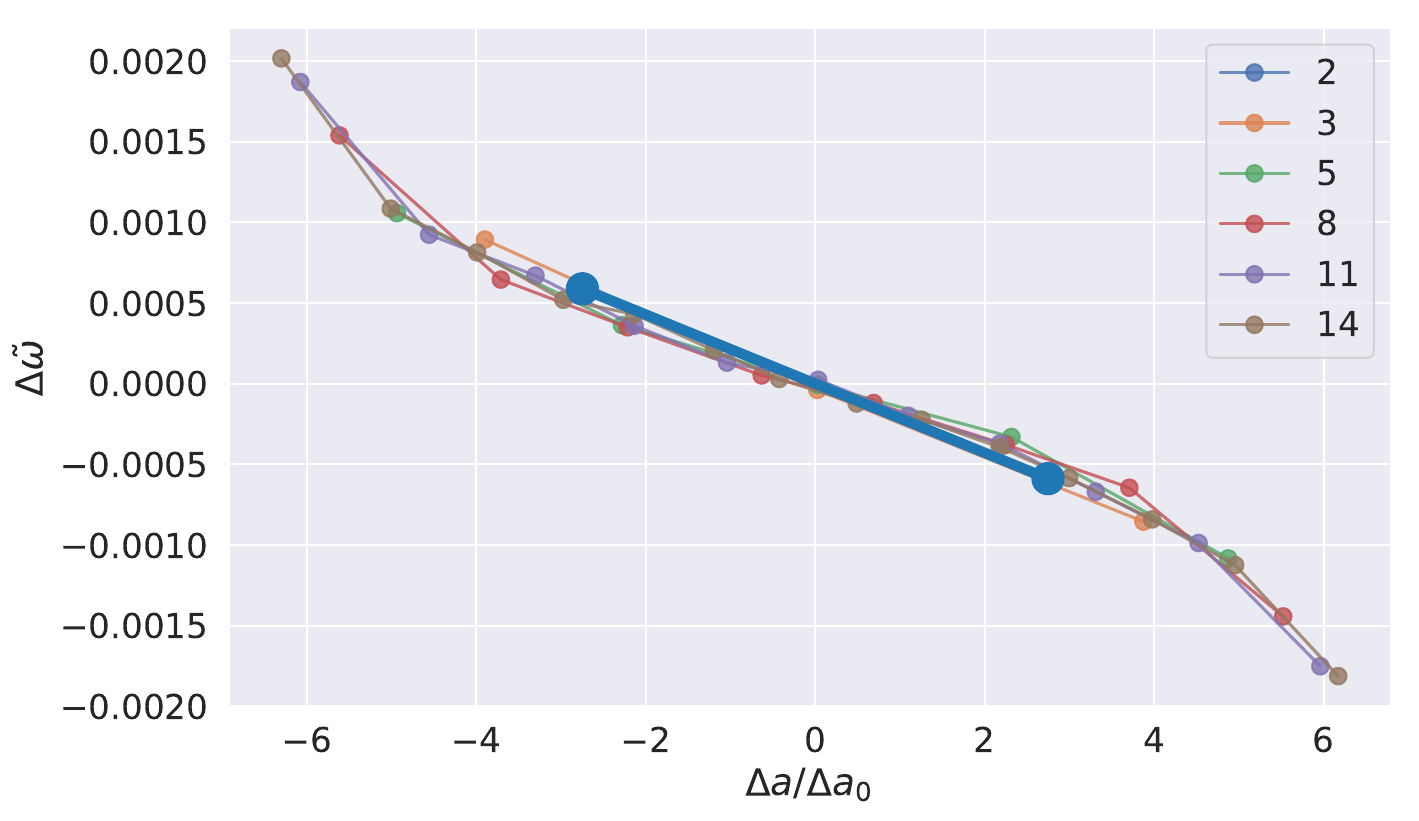}
    \caption{
        \label{fig:delta_wt_vs_da_streamlines}
        Simulated ringlets' final relative longitudes of periapse 
        $\Delta\tilde{\omega} = \tilde{\omega} - \bar{\omega}$ are plotted versus their relative
        semimajor axis $\Delta a=a-\bar{a}$, where $\bar{a}$ is the ringlet's mean semimajor axes 
        and $\bar{\omega}$ is their mean longitude of periapse, and number of streamlines $N_s$
        is indicated in the plot legend. 
    }
\end{figure}

Figures \ref{fig:e_vs_da_streamlines}--\ref{fig:delta_wt_vs_da_streamlines} recomputes the nominal
ringlet's evolution for ringlets having a range of streamlines, $2\le N_s\le14$, all of which
have $q$-evolution very similar to that exhibited by the $N_s=2$ simulation seen in 
Fig.\ \ref{fig:de_prime_nominal}.  Figure \ref{fig:e_vs_da_streamlines} 
plots each streamline's eccentricity $e$
versus their relative semimajor axis $\Delta a=a-\bar{a}$, which shows that
all simulated ringlet's have the same eccentricity gradient $e'$ regardless
of the number of streamlines $N_s$.
Ditto for the ringlets' relative longitudes of periapse, $\Delta\tilde{\omega}$ when
plotted versus $\Delta a$, Fig.\ \ref{fig:delta_wt_vs_da_streamlines},
which shows that all simulated ringlets have comparable gradients in $\tilde{\omega}$.
The only noteworthy difference seen here is that the smaller $N_s\le 3$ simulations
do not resolve the extra peripase twist that is seen at the edges of the higher resolution simulations.
Except for this one distinction,
the evolution of the $N_s>2$ ringlets are very similar to that exhibited by nominal
ringlet composed of $N_s=2$ streamlines.


\subsubsection{surface densities and sharp edges}
\label{subsubsec:sharp_edges}

\begin{figure}
    \plotone{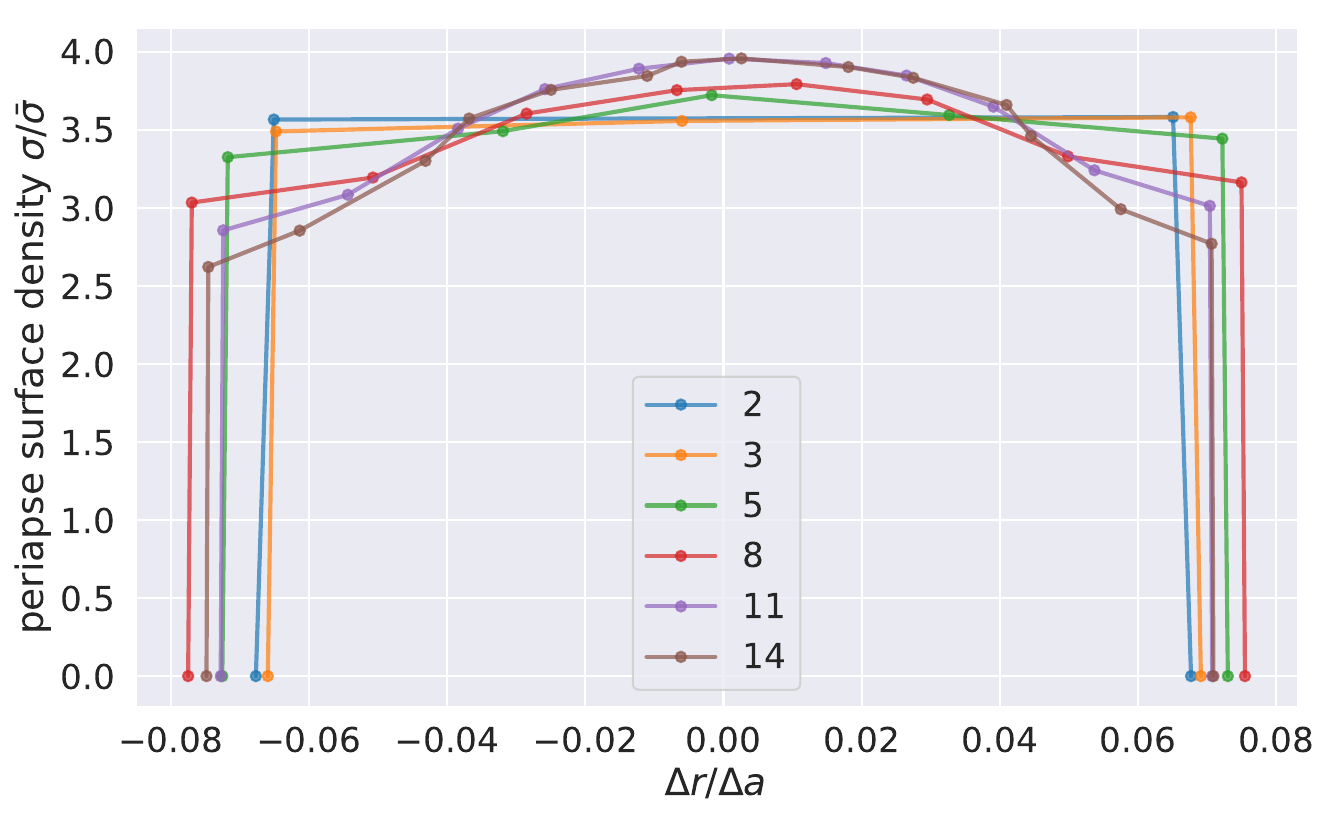}
    \caption{
        \label{fig:periapse_sigma_vs_r_streamlines.pdf}
        Radial surface density profiles $\sigma(\Delta r)$ 
        of the $2\le N_s\le14$ ringlets of Section \ref{subsec:num_streamlines}
        plotted along the ringlets' direction of periapse.
        Surface density $\sigma$ is shown in units of the ringlet's mean surface density $\bar{\sigma}$,
        and radial distance $\Delta r = r - r_{\text{mid}}$ is measured relative to chord's midpoint
        and in unit of the ringlet's semimajor axis width $\Delta a=a_{\text{outer}}-a_{\text{inner}}$. 
    }
\end{figure}

\begin{figure}
    \plotone{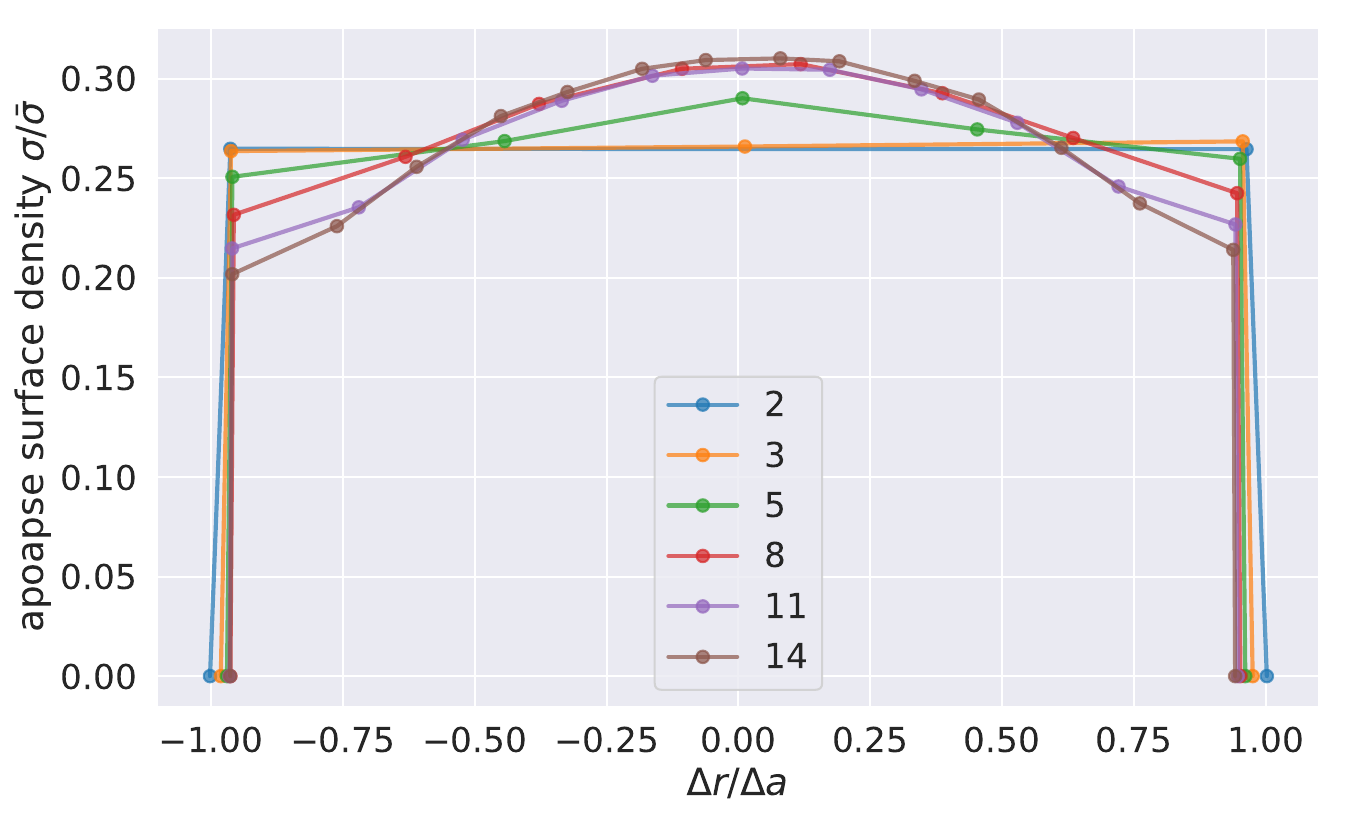}
    \caption{
        \label{fig:apoapse_sigma_vs_r_streamlines.pdf}
        Radial surface density profiles $\sigma(\Delta r)$ 
        of the $2\le N_s\le14$ simulations, plotted along the ringlets' direction of apoapse.
        Note that the ringlets' apoapse surface densities are about 13 times smaller
        than their periapse $\sigma$, Fig.\ \ref{fig:periapse_sigma_vs_r_streamlines.pdf},
        because they are also wider by that amount.
    }
\end{figure}

The main shortcoming of the $N_s=2$ simulation is that it reveals nothing about a ringlet's
possible sharp edges, since a two-streamline ringlet always has artificially sharp edges.
To examine this further, Fig.\ \ref{fig:periapse_sigma_vs_r_streamlines.pdf} shows the radial 
surface density profile along the periapse direction for the $2\le N_s\le14$ ringlets 
that were simulated in Section \ref{subsec:num_streamlines}. That plot shows that all ringlets'
edges are sharp after they arrive in the self-confining state, regardless of $N_s$. When self-confining,
each streamline is approximately equidistant (within $25\%$) from their neighbors, which causes
ringlet surface density $\sigma$ to be remain constant (within $25\%$) in the ringlet interior, 
which then plummets to zero beyond the ringlet's boundaries. Ditto for the surface density
profiles seen along the ringlets' apoapse direction, Fig.\ \ref{fig:apoapse_sigma_vs_r_streamlines.pdf}.
In summary, the self-confining ringlet examined here has a smooth radial surface density profile that is
concave--down, with surface density maxima at the ringlet' center, and has sharp edges.

Note that if a vicous ringlet were instead unconfined, then its positive angular momentum flux
would have repelled the edgemost streamlines away from the interior streamlines, 
which in turn would have caused the ringlet's concave--down surface density profile
to taper smoothly to zero at its edges, as is seen in Fig.\ 1 of \cite{P81}.
Meanwhile, the other ringlet models, those that rely on unseen shepherd satellites to maintain confinement, all
exhibit smooth surface densities having a variety of concavities, yet all have sharp edges (\citealt{GT79, CG00, ME02}).

In comparison, the Maxwell ringlet's radial optical depth profiles can be approximately 
described as concave--downish with a spiral density wave riding on top (\citealt{Netal14}). The Titan
ringlet is opaque so its concavity is unknown, but it does have sharp edges (\citealt{Netal14}).
The Huygens ringlet, which has a tiny eccentricity gradient, is concave--up with sharp edges 
(\citealt{Fetal16}), while the Laplace ringlet is possibly concave--down with sharp edges
and lots of internal structure. Main point: all well-observed narrow eccentric planetary ringlets
have a radial optical depth structure that is much more complicated than is exhibited by
published models to date.


\section{self--confining ringlet origin scenario}
\label{sec:origin}

Here we use the preceding results to speculate about the origin of narrow eccentric planetary ringlets.
If such ringlets are truly self-confining, then they are extremely young, only $\sim{\cal O}(1000)$ years 
old per Eqn.\ (\ref{eqn:tau_e}) and Section \ref{subsec:eccentricity}. Our preferred least-speculative origin scenario 
(which is still quite speculative) proposes that
the ringlet precursor was originally an especially large ring particle that was orbiting elsewhere such as
within a nearby dense planetary ring, presumably as an embedded moonlet aka propeller (\citealt{Tetal10}). 
If that embedded moonlet happened to 
form near the edge of the dense planetary ring, then the moonlet's reaction to
the shepherding torque that it exerts across the dense ring [which is $-1\times$ 
the radial integral of Eqn.\ (71) of \cite{GT82}] would
cause that moonlet to migrate towards and then beyond the ring's edge. And if a dense planetary ring can spawn
one such moonlet, it can also spawn additional moonlets. \cite{CSC10} also propose such 
for the origin for all the small satellites orbiting just beyond Saturn's A ring:
Atlas, Prometheus, Pandora, Epimethius, and Janus.

A moonlet's radial migration rate due to the ring-shepherding torque also
varies with moonlet mass, so a larger moonlet will eventually overtake a smaller moonlet that is orbiting beyond it.
Indeed, calculations by \cite{PS01} suggest that satellite Prometheus emerged from the outer edge of Saturn's A
ring $\sim10^7$ years ago, and that it will have a close encounter with the smaller satellite
Pandora that is currently orbiting just exterior to Prometheus in another $\sim10^7$ years.
Multiple close moonlet-moonlet encounters would then pump up moonlet eccentricities until they collide. If that collision
is sufficiently vigorous to disrupt one or both moonlets, then a narrow eccentric planetary ringlet composed of
collisional debris would result. Viscous particle-particle
collisions would then cause that young ringlet to spread radially (Fig.\ \ref{fig:da_nominal}) 
while its self-gravity would pump up its eccentricity gradient (Fig. \ref{fig:de_prime_nominal})
which, if sufficiently vigorous, would cause it to settle into the self-confining state that is
sharp-edged (Fig.\ \ref{fig:apoapse_sigma_vs_r_streamlines.pdf}) like most observed narrow eccentric ringlets. 
Nonetheless, self-confinement is temporary
and lasts only $\sim{\cal O}(1000)$ years due to the ringlet's viscosity damping its eccentricity until its inner edge gets circularized. Which would would cause the ringlet's $q$ to drop below the $q\simeq\sqrt{3}/2$ threshold
for self-confinement, the ringlet would again spread radially, and its edges would lose their sharpness.

So in summary, this purportedly least-speculative ringlet origin scenario
implies that dense planetary rings are frequently forming and emitting small embedded moonlets that then migrate away, 
due to their gravitational shepherding of that ring, towards and then beyond the nearby ring-edge
where they are later disrupted after colliding with other such moonlets. The resulting debris quickly
shears out into a low-eccentricity ringlet whose self-gravity excites its eccentricity gradient
until it settles temporarily into the self-confining state, which is sharp-edged. 
The next step in any assessment of the viability
of this least-speculative ringlet origin scenario would require estimating the various timescales that are relevant here,
namely, the moonlet's formation timescale, the moonlet's migration timescale, its lifetime versus collisional
disruption, and the ringlet's self-confinement lifetime, to confirm that none of these lifetimes are so
long as to cast doubt on this ringlet origin scenario. That more detailed analysis is deferred to a followup
study.

\section{summary of findings}
\label{sec:summary}

Main findings:

\begin{enumerate}

\item Simulations show that viscous self-gravitating
narrow eccentric ringlets having a wide variety of initial 
physical properties (mass, width, and viscosity) do evolve
into the self-confining state (Fig.\ \ref{fig:sim_grid_da}).

\item Self-gravity causes a self-confining ringlet's nonlinearity parameter $q$, 
which is the RMS sum of the ringlet's dimensionless eccentricity gradient $e'=ade/da$ 
and its dimensionless periapse twist $\tilde{\omega}' = ead\tilde{\omega}/da$,
to grow over time until it exceeds the $q\simeq\sqrt{3}/2\simeq0.866$ threshold where 
the ringlet's orbit-averaged angular momentum flux due to ringlet viscosity + self-gravity reversal is zero.

\item Gravitating ringlets that are self-confining have small but positive 
({\it i.e.}\ radially outwards) viscous angular momentum luminosity
that is counterbalanced by its negative gravitational angular momentum luminosity
(Section \ref{subsec:viscous_flux}). Ringlet self-gravity is also the reason
for a self-confining ringlet having nonlinearity parameter $q\simeq0.9$
that is slightly larger than the $q = \sqrt{3}/2 \simeq 0.866$
threshold that is expected of a viscous non-gravitating ringlet
\citep{BGT82}.

\item Simulated ringlets that do settle into self-confinement have final eccentricity
gradients $0.04\lesssim e' \lesssim 0.9$, Fig.\ \ref{fig:e_prime_q_vs_time}. Since all such ringlets 
have $q\simeq0.9$, this means that ringlets having smaller eccentricity gradients $e'$
also have larger periapse twist $|\tilde{\omega}'|$ that is likely due to a higher 
ringlet viscosity (Section \ref{subsec:viscosity-variations}).

\item The ringlet's total energy luminosity ${\cal L}_{E}$ is zero after the ringlet has settled into
the self-confining state (Section \ref{subsec:gravitational_flux}). 
The ringlet's gravitational energy luminosity ${\cal L}_{E,g}$ is also
zero at all times, consistent with secular theory. 

\item Self-confining ringlets have sharp edges (Section \ref{subsubsec:sharp_edges}).

\item Ringlet viscosity also circularizes the ringlet in time $\tau_e\sim10^6$ orbits $\sim1000$ years
(Section \ref{subsec:eccentricity}). After viscosity has reduced the eccentricity of the ringlet's 
inner edge to zero, the ringlet's  eccentricity gradient $e'$ will then shrink over time, its
angular momentum flux reversal will cease, and viscous spreading will resume as ringlet 
viscosity again transmits an  outwards flux of angular momentum. Ringlet circularization thus causes 
it to again resume spreading radially, and the ringlet would lose its sharp edges.

\item Self-confining narrow eccentric planetary ringlets are short lived unless there is a mechanism
that can pump up the ringlet's eccentricity.

\item We speculate that a sharp-edged self-confining narrow eccentric planetary ringlet first originated as
an exceptionally large ring particle orbiting within a nearby dense planetary ring. 
Possibly as a small embedded moonlet such as the presumed cause of the propeller structures
seen in Saturn's main A ring. If that embedded moonlet is orbiting near the dense ring's edge, then
its reaction to the gravitational shepherding torque that it exerts across the dense ring would
cause that moonlet to migrate towards and then beyond the dense ring's edge. If so, then 
a dense planetary ring can also birth multiple such moonlets. The ring-torque would also 
cause the largest moonlet to overtake any smaller migrating moonlets, which will encourage
moonlet-moonlet scattering and eventual collision. A disruptive moonlet collision
would generate debris that would quickly
shear out into a low-eccentricity ringlet whose self-gravity could excite its eccentricity gradient
until it settles into the self-confining state that is sharp-edged, albeit temporarily. 

\end{enumerate}

\section{additional followup studies}
\label{sec:followup}

Here we list possible followup studies that could also be pursued using the epi\_int\_lite streamline integrator:

\begin{enumerate}

\item {\it Nonlinear spiral density waves}: analytics theories exist, but they are complex and challenging to employ
\citep{SYL85, BGT86}, whereas some might regard the execution of an epi\_int\_lite simulation 
as the simpler/faster way forward. Spiral density waves are nonlinear
when the surface density variation due to the wave is comparable to the ring's undisturbed surface density
{\it i.e.}\ $|\Delta\sigma|\sim\sigma_0$. Since $\sigma=\lambda/\Delta r$, this means that 
streamline separation $\Delta r$ need only shrink by a factor of two to make a density wave nonlinear.
Note that eccentricities associated with the wave are still small, 
which means that the streamline-integrator approach is well suited for simulating 
nonlinear density waves. See for example the \cite{Fetal16B} simulation
of a marginally nonlinear spiral density wave using the epi\_int code,
which is the predecessor of epi\_int\_lite.

\item {\it Partly incompressible nonlinear spiral density waves}: one can imagine that a sufficiently vigorous nonlinear
density wave could drive particle densities high enough such that ring particles in the affected region are packed 
shoulder-to-shoulder so that further compaction is impossible
because that region has becomes incompressible. We suspect that these incompressible
patches could also result in shocks and/or vertical splashing as the wave drives 
additional ring particles into the affected regions. Which surely would alter the wave's dynamics and may
lead to new and possibly observable phenomena ({\it e.g.}\ \citealt{BGT85}). Note that the equation of state (EOS) employed by
epi\_int\_lite assumes that the ring is a compressible particle gas, but that EOS could easily be adapted
to account for the additional forces that would result as particles enter into or recoil
from the incompressible patches generated by the wave. 

\item {\it Nonlinear spiral bending waves}: although theories for nonlinear spiral density waves do exist, an
analytic theory for a nonlinear spiral bending wave does not. The fundamental assumption
of linearized bending wave theory is that the radial forces associated with the wave are negligible
compared to the wave's vertical forcing of the ring \citep{S84}. Which is true provided the bending wavelength
$\lambda_b$ is large compared to the bending wave amplitude $A$, which is the ring plane's maximal
vertical displacement due to the bending wave, {\it i.e.}\ $\lambda_b\gg A$.  
Although the epi\_int\_lite streamline integrator is 
currently a two dimensional code, adapting it to the 3$^{rd}$ dimension should be straightforward,
which means that the revised code would then be well positioned to simulate a nonlinear spiral bending wave having
$\lambda_b\lesssim A$. Also note that, when in this regime, the gravitational forces exerted by the
warped ring plane are no longer purely vertical {\it i.e.}\ the wave's forcing of the warped ring-plane
within distances $\sim\lambda_b$ will also have an in-plane component. And that radial forcing of a
self-gravitating ring tends to beget spiral density waves. Which suggests that
a spiral bending wave, whose wavelength $\lambda_b$ shrinks as it propagates, will
eventually go nonlinear in a way that might spawn a spiral density wave
that could be observable in Cassini observations of Saturn's rings.

\item {\it Galactic spiral structure}: the main difference between planetary and galactic trajectories
is in their precession rates. Planetary orbits precess slowly, 
$|\dot{\tilde{\omega}}|=|\Omega-\kappa|\ll\Omega$, whereas galactic trajectories,
which can also be described by equations like Eqn.\ (\ref{eqn:revised_linearized_soln}),
precess rapidly, $|\dot{\tilde{\omega}}|\sim\cal{O}($$\Omega)$. Which suggests
that epi\_int\_lite could also be used to simulate galactic spiral stucture
when the code is provided with appropriate functions for $\Omega(r)$ and $\kappa(r)$.
Epi\_int\_lite is also quite general in that it can simulate waves in both 
gravity and pressure dominated disks, and
additional perturbations from a central bar would be straightforward to include.

\item {\it Asymmetric circumstellar debris disk}: a debris disk is a dusty circumstellar disk that 
is often found
in orbit about younger stars, $\sim10^{7-9}$ years \citep{Matthews18}, 
and is a possible signature of ongoing planet formation. Many such star-disk systems are suspected of hosting an unseen
planetesimal `birth ring', wherein collisions among those planetesimals generates dust whose smaller grains are 
launched into very wide eccentric orbits due to stellar radiation pressure \citep{SC06}.
Those eccentric dust grains populate a very broad disk that can be observed out to stellar-centric
distances of $r\sim1000$ AU when imaged via scattered starlight, which also tends to favor the discovery of 
edge-on debris disks. These edge-on dust disks' ansae
routinely exhibit surface-brightness asymmetries as well other structures that
may be due to gravitational forcing by unseen protoplanets and/or giant impacts into those protoplanets
(\citealt{Jetal23}). Surface-brightness asymmetries can also be due to the birth ring having an
eccentricity \citep{H09}, which again implicates gravitational forcing by unseen planets or protoplanets, but differential
precession due to the protoplanets' gravities will eventually defeat the birth ring's eccentricity
unless that is resisted by the ring's self-gravity. 
The longevity of these possibly eccentric birth rings and their debris disks, which is inferred from their 
host stars' $10^{7+}$ yr ages, suggests that these possibly eccentric
birth rings can also be very long-lived. Which causes us to wonder whether
self gravity may play a role in preserving a birth ring's structure
via the self-confinement mechanism considered here, which epi\_int\_lite would be well-suited to
investigate.

\end{enumerate}

The epi\_int\_lite streamline integrator source code is available at https://github.com/joehahn/epi\_int\_lite,
and readers are encouraged to reach out if they wish to use that code in their research.

\acknowledgments
\section{Acknowledgments}
\label{sec:Acknowledgments}

This research was supported by the National Science Foundation via Grant No.\ AST-1313013.
A portion of this paper as well as the epi\_int\_lite integrator
were composed and developed while at the Water Tank karaoke bar in northwest Austin TX, 
and the authors thank proprietor James Stryker for his hospitality.
The authors also thank Joe Spitale and an anonymous review for their comments about this work.

\appendix

\section{Appendix A}
\label{sec:Appendix A}

The drift step implemented in the original epi\_int code utilized the epicyclic equations of \cite{BL94}
to advance an unperturbed particle along its trajectory about an oblate planet. Those equations
convert a particle's spatial coordinates $r, \theta, v_r, v_\theta$
into epicyclic orbit elements $a,e,M,\tilde{\omega}$ that are easily advanced by timestep
$\Delta t$, after which the epicyclic elements are converted back to spatial coordinates
with an accuracy to ${\cal O}(e^2)$. Note though that those epicyclic equations are not reversible;
every conversion from spatial to epicyclic coordinates or vice-versa also introduces an 
error of order ${\cal O}(e^3)$, and the accumulation of those errors causes the orbits of the epi\_int particles to 
slowly drift over time.  But that slow numerical drift
was not problematic for the relatively short epi\_int simulations reported in \cite{HS13} that evolved
Saturn's B ring for typically $\sim10^4$ orbit periods. However the viscous self-gravitating
ringlets considered here must be evolved for $10^5$ to $10^6$ orbit periods in order for self-confinement to occur,
and simulations having those longer execution times were being defeated by this orbital drift. 
The remedy is to derive an alternate set of epicyclic equations that
{\em are} reversible such that the epi\_int\_lite's drift step will not be a significant
source of numerical error.

Begin with the equation of motion for an unperturbed particle in orbit about the central planet, 
\begin{equation}
    \label{eqn:EOM}
    \ddot{\mathbf{r}}=-\nabla\Phi
\end{equation}
where the planet's gravitational potential is
\begin{equation}
    \label{eqn:Phi}
    \Phi = -\frac{GM}{r} - \frac{J_2GMR^2}{2r^3}
\end{equation}
where $M$ is the planet's mass, $R$ its radius, $J_2$ its second zonal harmonic due to oblateness,
and with higher-order $J_{n>2}$ terms ignored here. The angular part of Eqn.\ (\ref{eqn:EOM}) is
\begin{equation}
    \label{eqn:angular_EOM}
    \frac{1}{r}\frac{d}{dt}(r^2\dot{\theta}) = -\frac{\partial\Phi}{\partial\theta} = 0
\end{equation}
so the particle's specific angular momentum $h=r^2\dot{\theta}=r v_\theta$ is conserved {\it i.e.}
\begin{equation}
    \label{eqn:theta-dot}
    \dot{\theta} = \frac{h}{r^2},
\end{equation}
and inserting that into the radial part of Eqn.\ (\ref{eqn:EOM}) yields
\begin{equation}
    \label{eqn:radial_EOM}
    \ddot{r} = \frac{h^2}{r^3} -\frac{GM}{r^2} - \frac{3J_2GMR^2}{2r^4}.
\end{equation}
Solving the above to first-order accuracy ${\cal O}(e^1)$ ordinarily yields
\begin{eqnarray}
    \label{eqn:linearized_soln}
    r &\simeq& a(1-e\cos M) \nonumber\\
    \theta &\simeq& \frac{\Omega}{\kappa}(M+2e\sin M) + \tilde{\omega} \nonumber\\
    v_r = \dot{r} &\simeq& ea\kappa\sin M \nonumber\\
    v_\theta = r\dot{\theta} &\simeq& a\Omega(1 + e\cos M)
\end{eqnarray}
where
\begin{equation}
    \label{eqn:omega}
    \Omega(a) = \sqrt{\frac{GM}{a^3}}\left[1 + \frac{3}{2}J_2\left(\frac{R}{a}\right)^2\right]^{1/2}
\end{equation}
is the particle's mean orbital frequency and 
\begin{equation}
    \label{eqn:kappa}
    \kappa(a) = \sqrt{\frac{GM}{a^3}}\left[1 - \frac{3}{2}J_2\left(\frac{R}{a}\right)^2\right]^{1/2}
\end{equation}
its epicyclic frequency. In the above,  $a$, $e$, $\tilde{\omega}$ and $M=\kappa t$ are the epicyclic orbit elements
that have errors of order ${\cal O}(e^2)$. Equations (\ref{eqn:linearized_soln})
are easily inverted, which would then provide epicyclic
orbit elements $a$, $e$, $\tilde{\omega}$, $M$ as functions of the particle's 
spatial coordinates $r$, $\theta$, $v_r$, $v_\theta$, but applying them
here would be even more problematic since the conversion from spatial coordinates to
orbit elements would accrue ${\cal O}(e^2)$ errors during every drift step, which is even
worse that the rate at which epi\_int's drift step accrues errors.

The remedy is to choose an alternate set of equations (\ref{eqn:linearized_soln}) that also
solve the equation of motion (\ref{eqn:EOM}) to the same accuracy while  
satisfying Eqn.\ (\ref{eqn:theta-dot}) exactly. Begin with the above expression for the
angular coordinate, $\theta=(\Omega/\kappa)(M+2e\sin M) + \tilde{\omega}$,
so that $\dot{\theta} =\Omega(1+2e\cos M)$. Then require the constant $h=r^2\dot{\theta}=a^2\Omega$
be satisfied exactly, which provides the revised expression for $r$ as a function of orbit elements,
$r = a/\sqrt{1+2e\cos M}$, noting that this expression differs from that in Eqn.\ (\ref{eqn:linearized_soln}) 
by an amount of order ${\cal O}(e^2)$ which is this solution's allowed error.
It is then straightforward to complete the revisions to Eqn.\ (\ref{eqn:linearized_soln}):
\begin{eqnarray}
    \label{eqn:revised_linearized_soln}
    r &=& \frac{a}{\sqrt{1+2e\cos M}} \nonumber\\
    \theta &=& \frac{\Omega}{\kappa}(M+2e\sin M) + \tilde{\omega} \nonumber\\
    v_r = \dot{r} &=& \frac{ea\kappa\sin M}{(1 + 2e\cos M)^{3/2}}  = ea\kappa\sin M\left(\frac{r}{a}\right)^3\nonumber\\
    v_\theta = r\dot{\theta} &=& a\Omega\sqrt{1 + 2e\cos M} = \frac{a^2\Omega}{r}
\end{eqnarray}
where $\Omega$ and $\kappa$ are Eqns.\ (\ref{eqn:omega}--\ref{eqn:kappa}) and $M=\kappa t$.

Eqns.\ (\ref{eqn:revised_linearized_soln}) must also be inverted so
that the particle's epicyclic orbit elements $a, e, \tilde{\omega}, M$ can be obtained
from the particle's spatial coordinates $r, \theta, v_r, v_\theta$, which is done via
the following:

\begin{enumerate}

\item Calculate the particle's specific angular momentum $h=r v_\theta$ and then solve 
$h^2 = a^4\Omega^2 = GMa[1 + (3J_2/2)(R/a)^2]$
for the particle's semimajor axis $a$, which is
\begin{equation}
    \label{eqn:a}
    a = R\left(C + \sqrt{C^2 - 3J_2/2}\right)
\end{equation}
where constant $C=h^2/2GMR$.

\item Calculate $\Omega$ and $\kappa$ using Eqns.\ (\ref{eqn:omega}--\ref{eqn:kappa}).

\item Note that $e\sin M = (v_r/a\kappa)(a/r)^3$ and $e\cos M = \onehalf[(a/r)^2 - 1]$, which 
provides the particle's eccentricity via $e = \sqrt{(e\sin M)^2 + (e\cos M)^2}$
and its mean anomaly $M$ inferred from $\tan M = e\sin M/e\cos M$.

\item The particle's longitude of periapse is 
$\tilde{\omega} = \theta - (\Omega/\kappa)(M + 2e\sin M)$.

\end{enumerate}
These equations satisfy the equation of motion (\ref{eqn:EOM}) to accuracy ${\cal O}(e)$ and
with errors of order ${\cal O}(e^2)$. To advance a particle during epi\_int\_lite's drift step from 
time $t$ to time $t + \Delta t$, use the above steps 1-4 to convert the particle's spatial
coordinates $r, \theta, v_r, v_\theta$ to epicyclic orbit elements $a,e,M,\tilde{\omega}$,
update the particle's mean anomaly via $M\rightarrow\kappa(t + \Delta t)$, and then use
Eqns.\ (\ref{eqn:revised_linearized_soln}) to compute the particle's revised spatial coordinates.
And because steps 1-4 convert coordinates to elements exactly, epi\_int\_lite's drift step 
is not a significant source of numerical error.\\

Lastly we derive the streamlines' angular shear $\partial\omega/\partial r$.
The angular velocity $\omega=v_\theta/r=\Omega(a/r)^2$ by Eqn.\ (\ref{eqn:revised_linearized_soln}) so
$\omega \simeq\Omega(1+2e\cos\varphi)$
to first order in $e$ where true anomaly $\varphi=\theta -\tilde{\omega}\simeq M$.
Thus $\partial\omega/\partial a \simeq -(3\Omega/2a)(1-\frac{4}{3}e'\cos\varphi)$
since $J_2$ and $e$ are small though
the eccentricity gradient $e'=a(\partial e/\partial a)$ might not be small. Likewise
$r\simeq a(1-e\cos\varphi)$ so $\partial r/\partial a\simeq 1-e'\cos\varphi$ and so
\begin{equation}
    \label{eqn:domega-dr}
    \omega' = \frac{\partial\omega}{\partial r} = 
        \frac{\partial\omega}{\partial a}/\frac{\partial r}{\partial a}
        \simeq -\left(\frac{3\Omega}{2a}\right) \frac{1-\frac{4}{3}e'\cos \varphi}{1-e'\cos \varphi},
\end{equation}
which interestingly changes sign near periapse, $\varphi\simeq0$, when the eccentricity gradient is large enough, 
$e'\gtrsim3/4$.

\section{Appendix B}
\label{sec:Appendix B}

Write the gravitational angular momentum flux ${\cal F}_{L,g}=\lambda r A^1_{g,\theta}$
in terms of orbit elements, for a ringlet composed of two streamlines. The one-sided gravitational
acceleration that the inner streamline exerts on a particle in the outer streamline is
$A^1_g=-2G\lambda/\Delta$. The perturbing streamline's orientation relative to the particle
is illustrated Figure 3 of \cite{HS13}, which shows that the particle's tangential acceleration
is $A^1_{g,\theta}\simeq-(v_r/v_\theta)A^1_g$ where $v_r$ and $v_\theta$ are the perturbing 
streamline's radial and azimuthal velocities at the point of minumum separation
where $v_r/v_\theta\simeq e_0\sin\varphi$ by Eqns.\ (\ref{eqn:revised_linearized_soln}) 
to first order in the inner streamline's 
eccentricity $e_0$ and assuming $\kappa\simeq\Omega$. 
The inner streamline's orbit elements 
are designated $a_0$, $e_0$, $\tilde{\omega}_0$ while the outer streamline has
$a_1 = a_0 + \Delta a$, $e_1 = e_0 + \Delta e$, 
$\tilde{\omega}_1 = \tilde{\omega}_0 + \Delta \tilde{\omega}$, and
the particle's  longitude relative to the inner streamline's periapse is 
$\varphi\simeq\theta - \tilde{\omega}_0$.
The streamline-particle separation is then
\begin{equation}
    \label{eqn:Delta}
    \Delta = r_1 - r_0 \simeq (1 - e'\cos\varphi - \tilde{\omega}'\sin\varphi)\Delta a
\end{equation}
to first order in the small quantities $\Delta a$, $\Delta e$, and $\Delta\tilde{\omega}$
and assuming $e_0\ll e'$, so the gravitation flux of angular momentum
at the particle is 
 \begin{equation}
    \label{eqn:F_L_grav_approx}
    {\cal F}_{L,g} \simeq 2eG\lambda^2\left(\frac{a}{\Delta a}\right)
        \frac{\sin\varphi}{1 - e'\cos\varphi - \tilde{\omega}'\sin\varphi}
\end{equation}
where the streamline's linear density $\lambda=m_1/2\pi a = m_r/4\pi a$ for 
a streamline whose mass $m_1=m_r/2$ is half the total ringlet mass $m_r$.

\section{Appendix C}
\label{sec:Appendix C}

This Appendix compares the evolution of epi\_int\_lite simulations to theoretical predictions for 
narrow viscous ringlets, and for narrow gravitating ringlets.\\

\subsubsection{viscous evolution}
\label{subsubsec:viscous evolution}

\begin{figure}
    \plotone{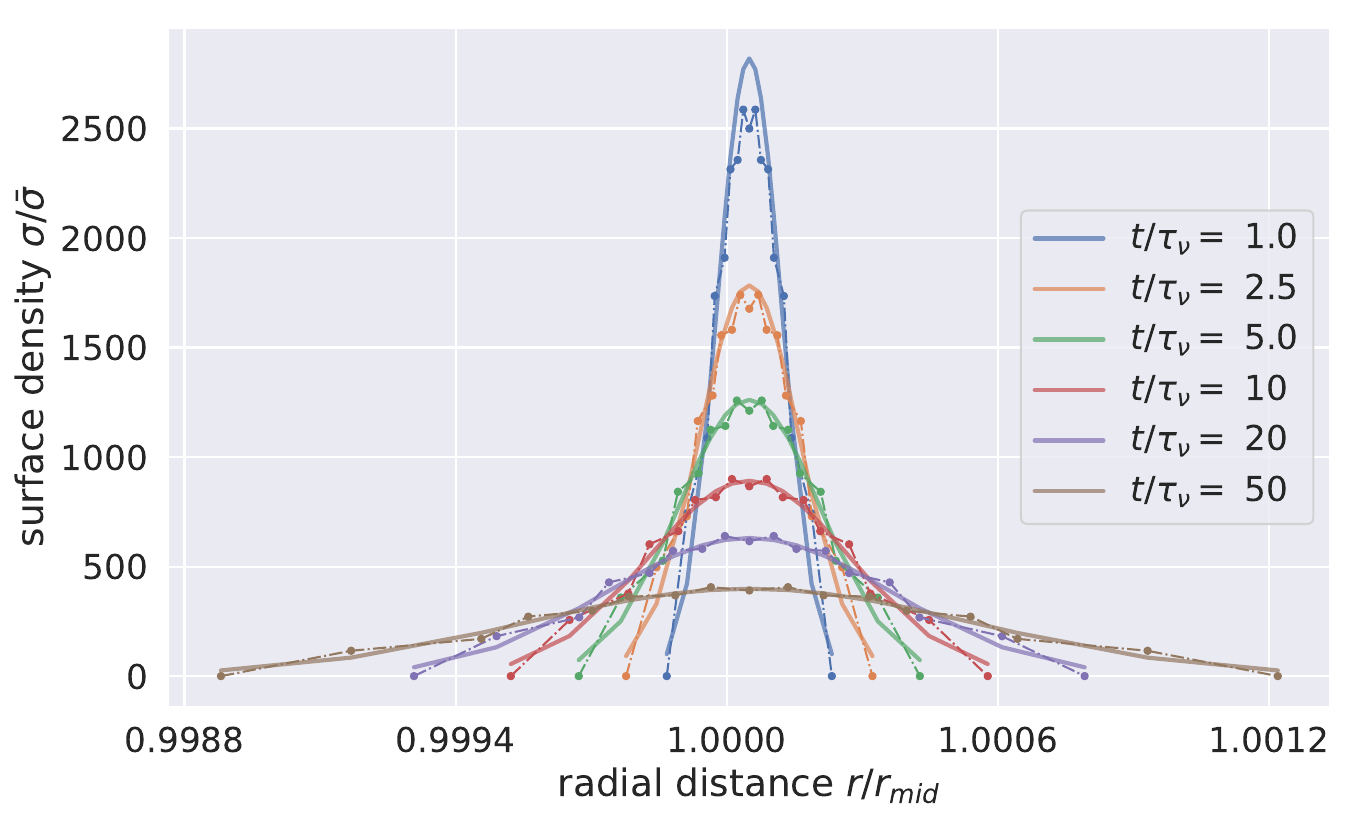}
    \caption{
        \label{fig:viscous_spreading}
        Radial spreading of a initially narrow circular viscous non-gravitating ringlet,
        simulation (dashed curves with dots) versus theory (solid curves).
        The simulated ringlet has the same physical properties as the nominal ringlet:
        total mass $m_r=10^{-10}$, initial semimajor axis width
        $\Delta a_0=10^{-4}$, shear viscosity $\nu_s=10^{-13}$, and bulk viscosity $\nu_b=\nu_s$.
        However this simulated ringlet is composed of $N_s=13$ streamlines instead of two, 
        and all streamlines are initially circular, $e=0$. And because a circular ringlet
        is always axisymmetric only $N_p=3$ particles per streamline need be used here. 
        Each colored curve indicates the ringlet's radial surface density profile
        $\sigma(r)/\bar{\sigma}$ versus radial distance $r/r_{mid}$ at various times $t/\tau_\nu$,
        where $\tau_\nu$ is the viscous timescale and $r_{mid}$ is the radius of the ringlet's middle.
        Solid theoretical curves are from Eqn.\ (2.13) of \cite{P81}
        where the constant $\bar{\sigma}=m_r/\pi r_{mid}^2$.
    }
\end{figure}

The test described in Fig.\ \ref{fig:viscous_spreading}
examines the radial spreading of a narrow viscous non-gravitating ringlet.
The simulated ringlet has the same physical properties as the nominal ringlet except that it is circular and 
it has many more streamlines than the $N_s=2$ simulation of Section \ref{subsec:nominal}.
Each colored curve indicates the ringlet's radial surface density profile $\sigma(r)$ at 
various times $t$. Note that the simulated surface density profiles (dashed curves with dots)
track nicely with the theoretical prediction (solid curves) of \cite{P81}, which is a good indicator that
epi\_int\_lite is computing the ringlet's viscous evolution correctly.

\subsubsection{self gravitating ringlets}
\label{subsubsec:self gravitating ringlets}

\begin{figure}
    \plotone{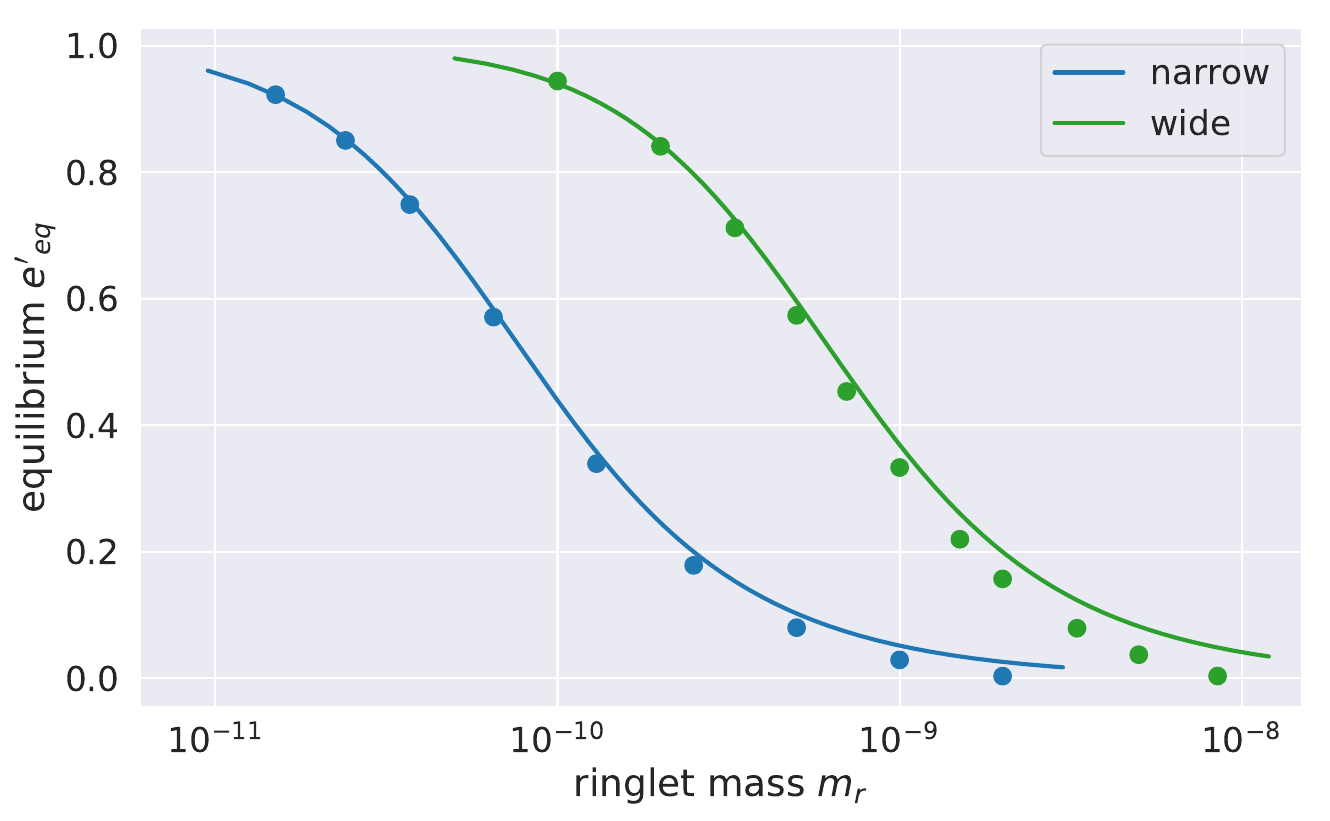}
    \caption{
        \label{fig:equilibrium}
        Equilibrium $e'_{eq}$ for two kinds of inviscid $\nu=0$ gravitating ringlets. Blue curve is for narrower lower-mass
        ringlets that all have the same semimajor axis width $\Delta a=5\times10^{-4}$, eccentricity $e=0.0025$, and the indicated
        range of ringlet masses $m_r$. Green curve is for wider higher-mass ringlets whose $\Delta a$ and $e$ are 
        larger by $\times2$. 
        Dots indicate outcomes from epi\_int\_lite simulations while continuous curves are the theoretical
        predictions from \cite{BGT83AJpaper}.
    }
\end{figure}

\cite{BGT83AJpaper} show that a gravitating ringlet has an equilibrium eccentricity difference 
$\Delta e^0=e_\text{outer} - e_\text{inner}$ that is stationary. 
Which means that if a ringlet is in equilibrium,
it will precess as rigid body with its streamlines experiencing zero relative motions and with steady $\Delta e$ over time.
Multiplying their expression for the equilibrium $\Delta e^0$ by $a/\Delta a$
then provides the ringlet's equilibrium dimensionless eccentricity gradient:
\begin{equation}
    \label{eqn:e_prime_eq}
    e'_{eq} = a\frac{\Delta e^0}{\Delta a} = \frac{21\pi eJ_2}{4H(q)}\left(\frac{M}{m_r}\right)\left(\frac{R_p}{a}\right)^2
        \left(\frac{\Delta a}{a}\right)^2
\end{equation}
where the nonlinearity parameter $q=|e'|$ since $\tilde{\omega}'=0$ for an inviscid ringlet in equilibrium, and the function
\begin{equation}
    \label{eqn:H(q)}
    H(q) = \frac{1 - \sqrt{1-q^2}}{q^2\sqrt{1-q^2}}. 
\end{equation}
\cite{BGT83AJpaper} also show that when the ringlet is displaced slightly from equilibrium, $e'$ will librate about
the equilibrium point, Eqn\ (\ref{eqn:e_prime_eq}). Which makes it straightforward to iteratively determine
a ringlet's equilibrium $e'_{eq}$ numerically via a handful of short epi\_int\_lite simulations, the results of which are 
summarized in Fig.\ \ref{fig:equilibrium} which compares simulated ringlet 
equilibria (dots) with theory (continuous curves) 
for a variety of inviscid gravitating ringlets. That Figure shows that the agreement between the epi\_int\_lite simulations
and theoretical expectations, Eqn.\ (\ref{eqn:e_prime_eq}), is overall very good, 
though modest discrepancies do exist for the higher-mass ringlets whose equilibrium $e'_{eq}$ are small.

\begin{figure}
    \plotone{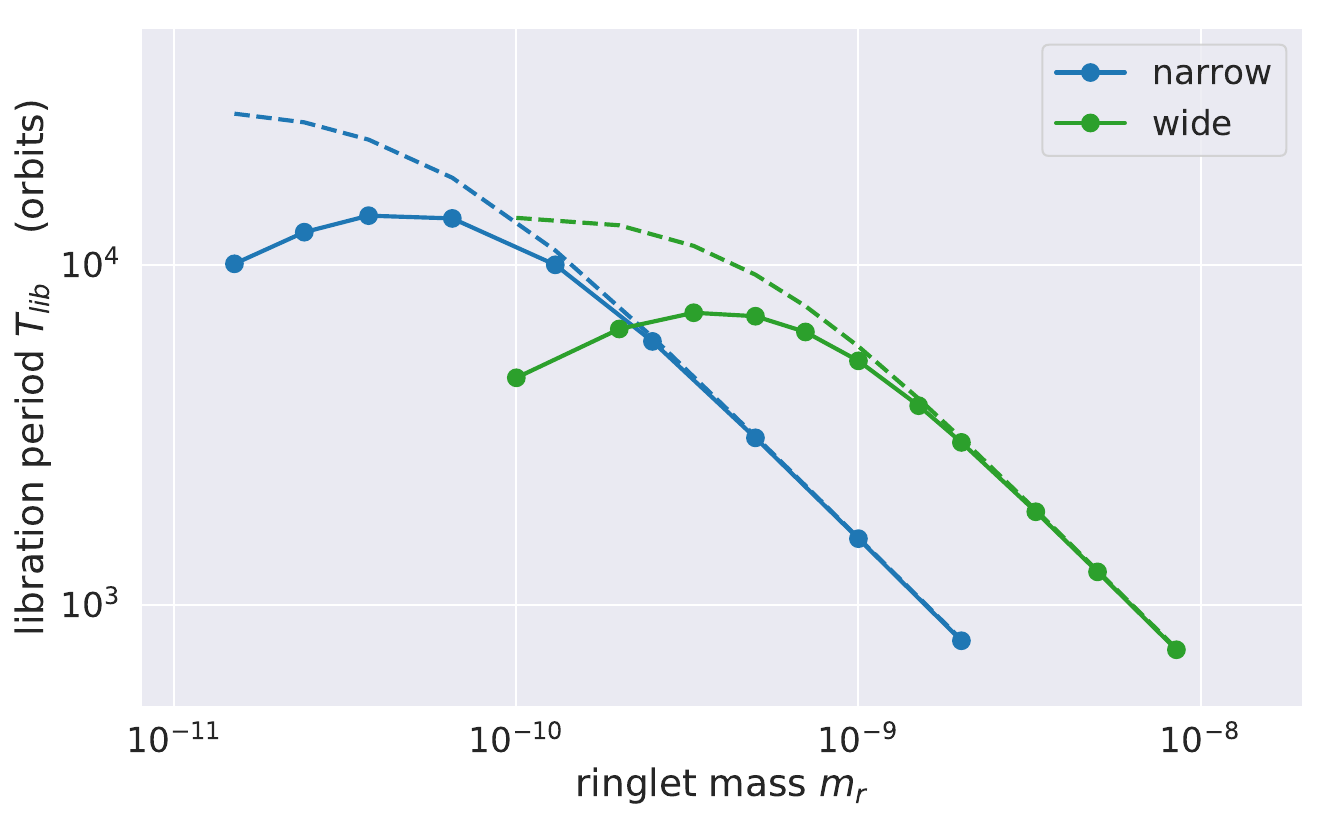}
    \caption{
        \label{fig:libration}
        Libration period $T_\text{lib}$ versus ringlet mass $m_r$ for the suite of narrow and wide ringlets
        also simulated in Fig.\ \ref{fig:equilibrium}. Dotted curves are the simulated outcomes while dashed
        curves are from Eqn.\ (\ref{eqn:T_lib}) derived in \citep{BGT83AJpaper}. Note that outcomes for the lower-mass ringlet
        simulations do disagree with theory, but they are exploring a parameter space where the 
        assumptions utilized in the derivation do not apply. 
    }
\end{figure}

And when a ringlet is displaced slightly from equilibrium, it will
librate about the equilibrium point (\ref{eqn:e_prime_eq}) 
with angular frequency $\Omega_\text{lib}$ given by Eqn.\ (30) of \cite{BGT83AJpaper}.
So a ringlet's libration period is
\begin{equation}
    \label{eqn:T_lib}
    T_\text{lib} = \frac{2\pi}{\Omega_\text{lib}} = 
        \frac{\pi}{H(q)}\left(\frac{M}{m_r}\right)\left(\frac{\Delta a}{a}\right)^2 T_\text{orb}
\end{equation}
where $T_\text{orb}$ is the ringlet's orbital period, which is shown in Fig.\ \ref{fig:libration} for the 
simulated ringlets of Fig \ref{fig:equilibrium}. 
That Figure compares predicted libration periods (dashed curves) versus simulated $T_\text{lib}$
(dotted curves) for the suite of narrow and wide ringlets. The rightmost portions of the blue and green curves show that the
higher-mass simulated ringlets' have libration periods $T_\text{lib}$ that are in excellent agreement with theory, 
Eqn.\ (\ref{eqn:T_lib}). However significant disagreement exists in the leftmost portions
of these curves, which corresponds to the lower-mass narrow \& wide ringlets. 
This disagreement is a consequence of the simulated ringlets' evolution
violating a key assumption of the \cite{BGT83AJpaper} derivation: 
that the ringlet's libration amplitude $A$ and its libration frequency $\Omega_\text{lib}$ are constant during a libration
cycle. In all of the simulations reported on in Fig.\ \ref{fig:libration}, the ringlet's libration amplitude and
frequency do vary over time, but those simulations show the their fractional variations $\Delta A/A$ and 
$\Delta\Omega_\text{lib}/\Omega_\text{lib}$ 
are progressively smaller in the higher-mass ringlets. For instance the three lowest-mass blue and green
simulations in Fig.\ \ref{fig:libration} have $\Delta A/A>0.5$ while that quantity is significantly smaller
for the higher-mass ringlets. From this we conclude that the lower-mass simulations 
are exploring parameter space where the theoretical predictions do not apply. And that the higher-mass simulations 
are in excellent agreement with theory.

\section{Appendix D}
\label{sec:Appendix D}

\begin{figure}
    \plotone{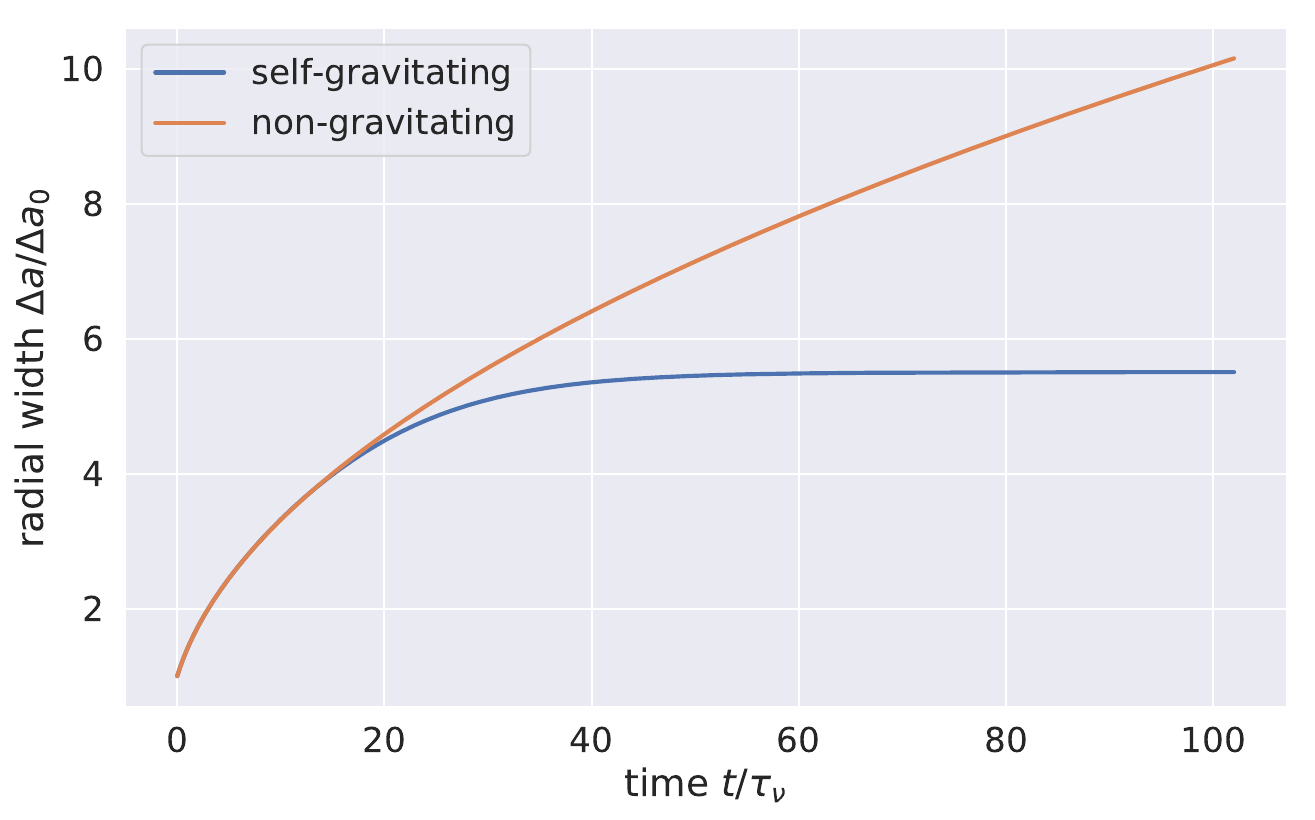}
    \caption{
        \label{fig:da_nogravity}
        Blue curve is the nominal ringlet's semimajor axis width $\Delta a$ versus time $t$,
        and this ringlet's radial spreading ceases by time $t\sim40\tau_\nu$ when it's self-gravity
        has excited the ringlet's eccentricity gradient $e'$ sufficiently; 
        see blue curve in Fig.\ \ref{fig:de_prime_nogravity}. Orange curve shows that
        the non-gravitating ringlet's $\Delta a$ grows without limit due to the ringlet's
        much lower eccentricity gradient. Note that planetary oblateness would
        cause the non-gravitating streamlines to precess differentially and eventually cross
        when $J_2>0$, so the non-gravitating simulation also sets $J_2=0$ 
        to avoid differential precession.
    }
\end{figure}

\begin{figure}
    \plotone{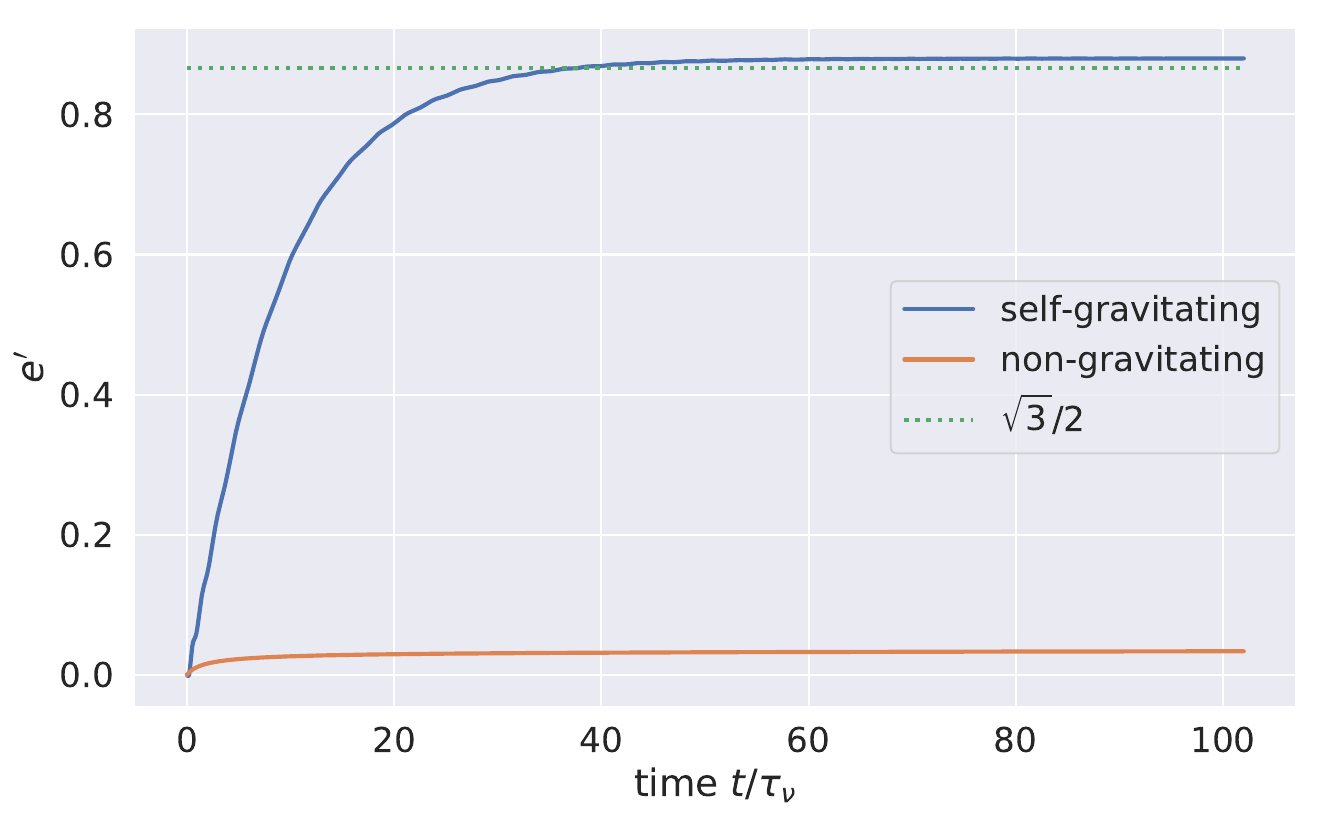}
    \caption{
        \label{fig:de_prime_nogravity}
        Eccentricity gradient $e'$ versus time $t$ for self-gravitating nominal ringlet (blue curve),
        and non-gravitating ringlet (orange).
    }
\end{figure}

This examines the viscous evolution of a narrow eccentric non-gravitating
ringlet that is identical to the nominal ringlet of Section \ref{subsec:nominal} but
with ringlet self-gravity neglected and $J_2=0$.
As the orange curve in Fig.\ \ref{fig:da_nogravity} shows, the non-gravitating ringlet's
radial width $\Delta a$ grows over time due to ringlet viscosity, 
long after the nominal self-gravitating ringlet (blue curve)
has settled into the self-confining state by time $t\sim40\tau_\nu$. This is due to the
ringlet's secular gravitational perturbations of itself,
which tends to excites the ringlet's outer streamline's eccentricity at the expense
of the inner streamline (see Fig.\ \ref{fig:e_nominal}) until the ringlet eccentricity gradient $e'$
(blue curve in Fig.\ \ref{fig:de_prime_nogravity}) grows beyond the
limit required for complete angular momentum flux reversal 
that results in the ringlet's radial confinement (dotted line). 
Note that viscosity also excites the non-gravitating
ringlet's eccentricity gradient some (orange curve), but not sufficiently to halt the ringlet's 
viscous spreading.

\bibliography{jmh_bibliography}

\begin{thebibliography}{}
\expandafter\ifx\csname natexlab\endcsname\relax\def\natexlab#1{#1}\fi

\bibitem[{{Borderies} {et~al.}(1982){Borderies}, {Goldreich}, \&
  {Tremaine}}]{BGT82}
{Borderies}, N., {Goldreich}, P., \& {Tremaine}, S. 1982, \nat, 299, 209

\bibitem[{{Borderies} {et~al.}(1983{\natexlab{a}}){Borderies}, {Goldreich}, \&
  {Tremaine}}]{BGT83}
---. 1983{\natexlab{a}}, Icarus, 55, 124

\bibitem[{{Borderies} {et~al.}(1983{\natexlab{b}}){Borderies}, {Goldreich}, \&
  {Tremaine}}]{BGT83AJpaper}
---. 1983{\natexlab{b}}, \aj, 88, 1560

\bibitem[{{Borderies} {et~al.}(1985){Borderies}, {Goldreich}, \&
  {Tremaine}}]{BGT85}
---. 1985, Icarus, 63, 406

\bibitem[{{Borderies} {et~al.}(1986){Borderies}, {Goldreich}, \&
  {Tremaine}}]{BGT86}
---. 1986, Icarus, 68, 522

\bibitem[{{Borderies-Rappaport} \& {Longaretti}(1994)}]{BL94}
{Borderies-Rappaport}, N., \& {Longaretti}, P.-Y. 1994, \icarus, 107, 129

\bibitem[{{Brouwer} \& {Clemence}(1961)}]{BC61}
{Brouwer}, D., \& {Clemence}, G.~M. 1961, {Methods of celestial mechanics} (New
  York: Academic Press, 1961)

\bibitem[{{Chambers}(1999)}]{C99}
{Chambers}, J.~E. 1999, \mnras, 304, 793

\bibitem[{{Charnoz} {et~al.}(2010){Charnoz}, {Salmon}, \& {Crida}}]{CSC10}
{Charnoz}, S., {Salmon}, J., \& {Crida}, A. 2010, \nat, 465, 752

\bibitem[{{Chiang} \& {Goldreich}(2000)}]{CG00}
{Chiang}, E.~I., \& {Goldreich}, P. 2000, \apj, 540, 1084

\bibitem[{{French} {et~al.}(2024){French}, {Hedman}, {Nicholson}, {Longaretti},
  \& {McGhee-French}}]{Fetal24}
{French}, R.~G., {Hedman}, M.~M., {Nicholson}, P.~D., {Longaretti}, P.-Y., \&
  {McGhee-French}, C.~A. 2024, \icarus, 411, 115957

\bibitem[{{French} {et~al.}(2016{\natexlab{a}}){French}, {Nicholson}, {Hedman},
  {Hahn}, {McGhee-French}, {Colwell}, {Marouf}, \& {Rappaport}}]{Fetal16B}
{French}, R.~G., {Nicholson}, P.~D., {Hedman}, M.~M., {et~al.}
  2016{\natexlab{a}}, \icarus, 279, 62

\bibitem[{{French} {et~al.}(2016{\natexlab{b}}){French}, {Nicholson},
  {McGhee-French}, {Lonergan}, {Sepersky}, {Hedman}, {Marouf}, \&
  {Colwell}}]{Fetal16}
{French}, R.~G., {Nicholson}, P.~D., {McGhee-French}, C.~A., {et~al.}
  2016{\natexlab{b}}, \icarus, 274, 131

\bibitem[{{Goldreich} {et~al.}(1995){Goldreich}, {Rappaport}, \&
  {Sicardy}}]{Getal95}
{Goldreich}, P., {Rappaport}, N., \& {Sicardy}, B. 1995, \icarus, 118, 414

\bibitem[{{Goldreich} \& {Tremaine}(1979{\natexlab{a}})}]{GT79}
{Goldreich}, P., \& {Tremaine}, S. 1979{\natexlab{a}}, \aj, 84, 1638

\bibitem[{{Goldreich} \& {Tremaine}(1979{\natexlab{b}})}]{GT79c}
---. 1979{\natexlab{b}}, \nat, 277, 97

\bibitem[{{Goldreich} \& {Tremaine}(1981)}]{GT81}
---. 1981, \apj, 243, 1062

\bibitem[{{Goldreich} \& {Tremaine}(1982)}]{GT82}
---. 1982, \araa, 20, 249

\bibitem[{{Hahn}(2009)}]{H09}
{Hahn}, J.~M. 2009, in AAS/Division of Dynamical Astronomy Meeting, Vol.~40,
  AAS/Division of Dynamical Astronomy Meeting, \#06.12--+

\bibitem[{{Hahn} \& {Spitale}(2013)}]{HS13}
{Hahn}, J.~M., \& {Spitale}, J.~N. 2013, \apj, 772, 122

\bibitem[{{Jones} {et~al.}(2023){Jones}, {Chiang}, {Duch{\^e}ne}, {Kalas}, \&
  {Esposito}}]{Jetal23}
{Jones}, J.~W., {Chiang}, E., {Duch{\^e}ne}, G., {Kalas}, P., \& {Esposito},
  T.~M. 2023, \apj, 948, 102

\bibitem[{{Longaretti}(2018)}]{L18}
{Longaretti}, P.~Y. 2018, {Theory of Narrow Rings and Sharp Edges} (Cambridge
  University Press), 225--275

\bibitem[{{Matthews} {et~al.}(2018){Matthews}, {Greaves}, {Kennedy},
  {Matr{\`a}}, {Wilner}, \& {Wyatt}}]{Matthews18}
{Matthews}, B., {Greaves}, J., {Kennedy}, G., {et~al.} 2018, in Astronomical
  Society of the Pacific Conference Series, Vol. 517, Science with a Next
  Generation Very Large Array, ed. E.~{Murphy}, 161

\bibitem[{{Mosqueira} \& {Estrada}(2002)}]{ME02}
{Mosqueira}, I., \& {Estrada}, P.~R. 2002, Icarus, 158, 545

\bibitem[{{Murray} {et~al.}(2005){Murray}, {Chavez}, {Beurle}, {Cooper},
  {Evans}, {Burns}, \& {Porco}}]{Metal05}
{Murray}, C.~D., {Chavez}, C., {Beurle}, K., {et~al.} 2005, \nat, 437, 1326

\bibitem[{{Nicholson} {et~al.}(2014){Nicholson}, {French}, {McGhee-French},
  {Hedman}, {Marouf}, {Colwell}, {Lonergan}, \& {Sepersky}}]{Netal14}
{Nicholson}, P.~D., {French}, R.~G., {McGhee-French}, C.~A., {et~al.} 2014,
  \icarus, 241, 373

\bibitem[{{Poulet} \& {Sicardy}(2001)}]{PS01}
{Poulet}, F., \& {Sicardy}, B. 2001, \mnras, 322, 343

\bibitem[{{Pringle}(1981)}]{P81}
{Pringle}, J.~E. 1981, \araa, 19, 137

\bibitem[{{Rimlinger} {et~al.}(2016){Rimlinger}, {Hamilton}, \& {Hahn}}]{RHH16}
{Rimlinger}, T., {Hamilton}, D., \& {Hahn}, J.~M. 2016, in AAS/Division of
  Dynamical Astronomy Meeting, Vol.~47, AAS/Division of Dynamical Astronomy
  Meeting, \#47, id.400.02

\bibitem[{{Shu}(1984)}]{S84}
{Shu}, F.~H. 1984, in IAU Colloq. 75: Planetary Rings, ed. R.~{Greenberg} \&
  A.~{Brahic}, 513--561

\bibitem[{{Shu} {et~al.}(1985){Shu}, {Yuan}, \& {Lissauer}}]{SYL85}
{Shu}, F.~H., {Yuan}, C., \& {Lissauer}, J.~J. 1985, \apj, 291, 356

\bibitem[{{Spitale} \& {Hahn}(2016)}]{SH16}
{Spitale}, J.~N., \& {Hahn}, J.~M. 2016, \icarus, 279, 141

\bibitem[{{Strubbe} \& {Chiang}(2006)}]{SC06}
{Strubbe}, L.~E., \& {Chiang}, E.~I. 2006, \apj, 648, 652

\bibitem[{{Tiscareno} {et~al.}(2010){Tiscareno}, {Burns},
  {Srem{\v{c}}evi{\'c}}, {Beurle}, {Hedman}, {Cooper}, {Milano}, {Evans},
  {Porco}, {Spitale}, \& {Weiss}}]{Tetal10}
{Tiscareno}, M.~S., {Burns}, J.~A., {Srem{\v{c}}evi{\'c}}, M., {et~al.} 2010,
  \apjl, 718, L92

\bibitem[{{Weiss} {et~al.}(2009){Weiss}, {Porco}, \& {Tiscareno}}]{Wetal09}
{Weiss}, J.~W., {Porco}, C.~C., \& {Tiscareno}, M.~S. 2009, \aj, 138, 272

\end{thebibliography}
\end{document}